\documentclass{aa}  
\usepackage{graphicx}
\usepackage[figuresright]{rotating}
\usepackage{txfonts}
\begin{document}
   \title{The stellar host in blue compact dwarf galaxies:}

   \subtitle{the need for a two-dimensional fit}

   \author{Ricardo O. Amor\'in\inst{1}, Casiana Mu\~noz-Tu\~n\'on\inst{1},
	  J. Alfonso L. Aguerri\inst{1}, 
	  Luz M. Cair\'os\inst{2} \\
          \and
          Nicola Caon\inst{1}
          }
   \authorrunning{R. O. Amor\'in et al.}
   \titlerunning{Two-dimensional fits of the stellar hosts in BCD galaxies}
   \offprints{R. Amor\'in}

   \institute{Instituto de Astrof\'isica de Canarias (IAC), 
     V\'ia L\'actea, E-38200 La Laguna, Tenerife, Spain\\
            \email{ricardo.amorin@iac.es}, 
             {casiana@iac.es}, 
	     {jalfonso@iac.es}
             {nicola.caon@iac.es} 
	 \and
             Astrophysikalisches Institut Potsdam, An der Sternwarte 16, 
     D-14482 Potsdam, Germany\\ 
             \email{luzma@aip.de}
             }
   \date{Received, 2006 August 1; accepted 2007 February 12}

 
  \abstract
   {The structural properties of the low surface brightness stellar host in 
   blue compact dwarf galaxies are often studied by fitting $r^{1/n}$ models 
   to the outer regions of their radial profiles. The limitations imposed by 
   the presence of a large starburst emission overlapping the underlying 
   component makes this kind of analysis a difficult task.}
   {We propose a two-dimensional fitting methodology in order to improve the 
   extraction of the structural parameters of the LSB host. We discuss its 
   advantages and weaknesses by using a set of simulated galaxies and 
   compare the results for a sample of eight 
   objects with those already obtained using a one-dimensional technique.}
   {We fit a PSF convolved \mbox{S\'ersic} model to synthetic galaxies, and 
   to real galaxy images in the $B$, $V$, $R$ filters. 
   We restrict the fit to the stellar host by masking out the starburst 
   region and take special care to minimize the sky-subtraction uncertainties. 
   In order to test the robustness and flexibility of the method, we carry out
   a set of fits with synthetic galaxies. Furthermore consistency checks are 
   performed to assess the reliability and accuracy 
   of the derived structural parameters.}
   {The more accurate isolation of the starburst emission is the most 
   important advantage and strength of the method. 
   Thus, we fit the host galaxy in a range of surface brightness and 
   in a portion of area larger than in previous published 1D fits with the same dataset.
   We obtain robust fits for all the sample galaxies, all of which, except 
   one, show \mbox{S\'ersic} indices $n$ very close to 1, with good 
   agreement in the three bands. 
   These findings suggest that the stellar hosts in BCDs have near-exponential 
   profiles, a result that will help us to understand the mechanisms that form and 
   shape BCD galaxies, and how they relate to the other dwarf galaxy classes.}
   {}


   \maketitle
%

\section{Introduction}

Blue compact dwarf (BCD) galaxies have been recognized for some time 
as playing an important role in our understanding of star formation (SF) 
processes, and their study provides relevant information for our knowledge of 
the evolution of galaxies.
BCDs are gas-rich (H{\sc i} mass fraction typically higher than 30\%) and 
metal-deficient ($Z_{\odot}/50 \leq Z \leq Z_{\odot}/2$) extragalactic 
systems, which display very intense and narrow emission lines in the optical 
(similar to those seen in H{\sc ii} regions), due to the intense 
SF activity distributed in one to several star-forming bursts 
(Cair\'os \cite{C01b}, hereafter C01b). 
Stars are formed at high rates (0.1--1$ M_{\odot}$, 
Fanelli \cite{Fanelli88}), exhausting their gas content in much shorter times 
than the age of the Universe. The mechanisms triggering the cloud collapse in 
these low mass systems is still unknown since, unlike in normal spirals, 
density waves are inhibited. Moreover, their characteristics are believed 
to have been common among unevolved low-mass galaxies at high to intermediate 
redshift. 
Local BCDs are therefore the most suitable nearby laboratories for studying  
the spectrophotometric and chemodynamic properties of 
distant and faint galaxies counterparts at high spatial resolution.\\
\indent The nature and genesis of BCDs have been studied and are now better understood. 
The idea of BCDs being genuinely young galaxies forming stars for the very 
first time (Sargent \& Searle \cite{SyS70}; Kunth \cite{Kunth88}) is now 
discarded, at least for the vast majority of them. 
Deep photometric studies in the optical (Loose \& Thuan \cite{LT}; 
\cite{Telles}; Papaderos \cite{P96b}; Doublier \cite{Doublier1},1999;
Cair\'os \cite{C00}; Cair\'os \cite{C01a}, hereafter C01a; C01b; 
Bergvall \& {\"O}stlin \cite{ByO02}) and in the near-infrared (Noeske 
\cite{Noeske05}, and reference therein) have in fact revealed that 
virtually all BCDs present a low-surface brightness (LSB) stellar host. 
This stellar component, underlying the SF regions, is an extended 
envelope that generally shows elliptical isophotes, displays red 
colours, indicative of an evolved stellar population (Papaderos \cite{P96b}; 
C01a; CO1b; Cair\'os \cite{C02}; Cair\'os \cite{C03}; 
Bergvall \& {\" O}stlin \cite{ByO02}), and is therefore a ``witness'' of
former events of star formation. 

LSB hosts are typically  found to dominate the intensity and colour 
distribution of BCDs for $\mu_{\rm B}$$\gtrsim$24 mag/arcsec$^{-2}$ 
(Papaderos \cite{P96b}; C01b). This evolved stellar population is observed in 
all types of BCDs, except for the extremely rare i0 type (from the 
classification scheme of Loose \& Thuan \cite{LT}).        

In order to establish the luminosities, structures and evolutionary 
status, as well as the star-forming history 
of BCD galaxies, it is indispensable to characterize the LSB component 
underlying the starburst regions. 
The first step in this process is to derive the LSB host ages and chemical abundances.
Also, a comprehensive spectrophotometric study of the starburst in BCDs
needs a precise determination of the structural properties of the LSB stellar
host in order to remove its contribution. Moreover, the structural,
kinematic and dynamical properties of the host galaxy in BCDs 
are important issues in dwarf galaxy research, as they help to understand 
such processes as the regulation of the SF activity, and the possible evolutionary 
connections between different dwarf galaxy types.

\cite{2006AJ....131.1318V} show, from $NIR$ surface photometry,
 that the starburst is usually a small fraction of the total mass.
Furthermore, provided that dark matter does not dominate the mass
within the Holmberg radius (Papaderos \cite{P96b}), the LSB component is,
together with the H{\sc i} halo, mainly responsible for the global
gravitational potential within which the starburst phenomenon takes place.
Thus, the analysis of the projected luminosity distribution is fundamental
for modelling the gravitational potential and the dynamics of BCDs, as well as 
the effects of starburst events in their interstellar medium, such as galactic 
winds. Whether the processed stellar material will be released, causing the 
contamination of the intergalactic medium, depends strongly on the galaxy 
structure (Tenorio-Tagle \cite{TT03}), being easier in disc-like objects 
than in spheroids (Silich \& Tenorio-Tagle \cite{ST01}, see their Fig.3).

The comparison of the properties of the LSB stellar component (e.g.\ 
structural parameters, average colours, and colour gradients) with those of 
other dwarf galaxy classes (dwarf irregulars, dIs; dwarf elliptical, dEs), 
and low-surface brightness galaxies is crucial for testing those 
evolutionary scenarios that link these galaxies with BCDs 
(Thuan \cite{T85}; Davies \& Phillipps \cite{DyPh88}; Papaderos
\cite{P96a}; Marlowe \cite{M97}, \cite{M99}; Cair\'os \cite{C00}). 

The faint surface brightness of the LSB component and the contamination 
caused by the starburst emission make the derivation of the structural 
parameters of the underlying host in BCDs a complicated task. 
Thus, the derived structural parameters strongly depend 
on how well the starburst has been excluded from the fit, on the extent of the 
fitted LSB radial profile, on the quality of the dataset, and on the method 
and model used to parameterize the surface brightness profile. 

Some previous studies, in which one-dimensional models of the BCDs
LSB component were fitted, have shown discrepant results even for the same 
galaxies, especially when a \mbox{S\'ersic} law was applied 
(see examples in Cair\'os \cite{C03} and Caon \cite{Caon05}). 
Moreover, during the extraction of a radial profile one can find serious 
limitations and ambiguities. Each of the different procedures has its own 
drawbacks, resulting in information loss from the image (Baggett \cite{Baggett}).\\ 
\indent At the present time, several well-tested two-dimensional algorithms 
are available. There are also several examples in the literature of studies showing 
that the two-dimensional method is generally more reliable than one-dimensional 
methods in bulge--disc decompositions (de Jong \cite{dJong}), as it is able
to retrieve more accurate structural parameters. 
Particularly in BCDs, a two-dimensional fit to the image, by using a $\chi^{2}$ 
minimization technique, may provide significant advantages over the fit to
some averaged one-dimensional profile. \\
\indent In Caon \cite{Caon05} (hereafter Paper~I) we presented a very detailed 
one-dimensional \mbox{S\'ersic} fitting of a sample of eight BCDs, showing that by 
carefully selecting the fitted radial range and by performing consistency 
checks on the fits we can derive reliable structural parameters. 
We also analysed in detail the problems and uncertainties involved in the 
fitting of a 1D \mbox{S\'ersic} model. \\
\indent As a natural continuation of the work in Paper~I, here we explore the LSB stellar 
host structural characterization by carrying out a two-dimensional fitting 
technique to the same galaxies. 
We propose a consistent methodology, explain its advantages and discuss its
possible drawbacks and limitations.

\indent
One of the crucial issues to be investigated is whether or not there 
is an evolutionary link among the different classes of the dwarf galaxies population.
Although detailed studies of individual objects are important and can provide tests, 
examples, or counterexamples, which are undoubtfully useful, statistical studies are 
the key for extending or establishing general paths. 
To do so, ``semiautomatic'' procedures able to analyse a large data base are necessary. 
In particular, to classify the LSB host into structures with low or high $n$ \mbox{S\'ersic} 
index with well-controlled uncertanties would be an important first step.
{
\sloppy
In Section~2 we present the sample of galaxies; in Section~3 we describe the
two-dimensional fitting methodology. Simulations with synthetic galaxies are
described in Section~4. In Section~5 we discuss the observational
uncertainties involving the fitting of a \mbox{S\'ersic} model to the BCDs
host galaxy. Section~6 presents the results of the 2D fitting to the sample of
galaxies. Summary and conclusions are the subject of Section~7.
}
\section{The Sample}
Eight galaxies were selected from a larger sample of 28 BCDs already studied 
by our group. A complete description of this sample, as well as the image 
reduction and calibration procedures, were given in C01a and C01b. 
The complete collection of broad-band and H$\alpha$ images of the 
galaxies can be found on-line
\footnote{see http://www.iac.es/proyect/GEFE/BCDs/BCDframe.html}.

We have selected the same objects already analysed in Paper~I. 
Such galaxies are those with the deepest images, located in areas of 
the sky free from nearby bright stars, and with few overlapping background 
or foreground objects. The basic data on the galaxies are presented in 
Table~\ref{T1}.  

\section{The Two-Dimensional Fitting Method}

It is clear that many BCDs present light profiles with
significant deviations from a pure exponential (Cair\'os \cite{C03}; Paper~I).
Following our previous works (Cair\'os \cite{C00}, C01a, Cair\'os \cite{C03}, and Paper~I) 
we have adopted the S\'ersic law (\mbox{S\'ersic} 1968) to fit our data. 
This paper does not concern itself with which mathematical function best suits the 
surface brightness profiles of most BCDs host galaxies.
The applicability of the \mbox{S\'ersic} profile and the drawbacks associated with it 
were the subject of Paper~I.
This profile has been shown to describe the light distribution in ellipticals 
(from dwarfs to brightest cluster members, \cite{1993MNRAS.265.1013C}, 
\cite{1994MNRAS.268L..11Y}, \cite{1996ApJ...465..534G}, \cite{2003AJ....125.2936G}, 
\cite{2005AJ....130..475A}) and spiral bulges (\cite{2001A&A...367..405P}).
Here, for the first time, a new methodology for fitting BCD hosts by using a 
two-dimensional technique is developed.   
We used the publicly available software called GALFIT v2.0.3b  
(\cite{2002AJ....124..266P}; hereafter P02). Detailed information on GALFIT and
how it is implemented can be found in P02, and 
on-line.\footnote{http://zwicky.as.arizona.edu/~cyp/work/galfit/galfit.html} 
GALFIT has been used in recent years in several works where  modelling of the
light distribution of galaxies was required, e.g.\ in disc galaxies 
(\cite{2005ApJ...635..959B}), field spheroidal and bulge-dominated galaxies
(\cite{2005ApJ...633..174T}), active galaxies (\cite{2004ApJ...614..586S};
\cite{2006AJ.131.1236D}) and luminous blue compact galaxies (\cite{2006ApJ...640L.143N}),
among others. 
However, it has not yet been applied to blue compact and irregular dwarfs. 
We give here a brief introduction on how GALFIT works and describe
the methodology we implemented and applied to fit our galaxies.
GALFIT was designed to extract structural components from galaxy images, 
as it is capable of fitting a galaxy with an arbitrary number of 
components simultaneously, whose geometry is described by axially symmetric generalized 
ellipses (see \cite{1990MNRAS.245..130A}), and whose profile
can be one of several analytic functions such as \mbox{S\'ersic}, Gaussian, 
exponential, or a constant or tilted background. 
The program has the option of convolving the models with the point-spread 
function (PSF) to simulate the seeing. 
GALFIT minimizes the $\chi^{2}_{\nu}$ (normalized $\chi^{2}$, see P02) residuals using a 
downhill gradient/parabolic expansion method, called Levernberg-Marquardt 
(Press \cite{Press}), by iteratively creating model images, convolving them with the PSF and 
subtracting them from the data. Also, GALFIT is able to use the readout noise
and gain parameters of the image to build a Poissonian noise model in 
order to weigh the pixels used in the fit.  
In our case, we fitted a single \mbox{S\'ersic} model:
\begin{equation}
I(r) = I_{e} e^{-b_{n}[(r/r_{e})^{1/n} - 1]},
\end{equation}
where $I_{\rm e}$ is the intensity at the effective radius \mbox{$r_{\rm e}$} that 
encloses half of the total light from the model 
(\cite{1993MNRAS.265.1013C}), while the constant $b_n$ is coupled with the 
\mbox{S\'ersic} shape parameter, $n$. 
The generalized ellipses:
\begin{equation}
r = \left( {\arrowvert x-x_c \arrowvert}^{2+c} + {\arrowvert \frac{y-y_c}{q} \arrowvert}^{2+c}\right)^{1/(2+c)}
\end{equation}
can be rotated to any position angle (PA). The parameter $q$ is the axis 
ratio of the model. The parameter $c$ is positive for boxy isophotes,
and negative for discy ones. 

The flux, integrated over all radii for an elliptical \mbox{S\'ersic} profile with 
an axis ratio $q$, is given (following the notation by P02) by:
\begin{equation}
F_{tot} = 2 \pi {r_{e}}^2\Sigma_{e} e^{k_{n}}n{k_{n}}^{-2n}\Gamma(2n)q/R(c),
\end{equation}
where $\Gamma(2n)$ is the Gamma function, and
\begin{equation}
R(c)=\frac{\pi (c+2)}{4\beta (1/(c+2), 1+1/(c+2))},
\end{equation}
where $\beta(1/(c+2), 1+1/(c+2))$ is the beta function with two arguments.
$R(c)$ is a function that accounts for the area ratio between a perfect 
ellipse and a generalized ellipse of discyness/boxiness parameter $c$ (P02).

In total, we have 8 free parameters:  
$x_c$, $y_c$, $m_{\rm tot}$, $r_e$, $n$, $q$, $PA$, and $c$. Note 
that the same number of parameters must be determined 
in a one-dimensional S\'ersic fit.

\subsection{Implementation}

We prepare the images (galaxy and PSF) and the GALFIT input 
files. Here we select the appropriate model we want to fit. 
Then, we identify the starburst emission from the image and 
make a set of elliptical (preliminary) masks in order to 
mask-out that emission from the fit. 
Next, we perform a set of fits and plot the output parameters 
versus the size of the masks. From that and the residual images 
we identify both the range of mask sizes where the parameters 
fit data with the minimum scatter, and the mask outside which 
the starburst emission is practically absent. 
Afterwards, we use the residual image to refine the size and 
shape of the mask and we fit again with the new mask in an 
iterative process. Finally, the procedure stops when the best 
solution is obtained. A sketch of the procedure is shown in 
Fig.\ref{F2}. The method is described in detail as follows:\\ 

1 - {\em Preparing the images and the input files\/} 

\noindent Reduced and calibrated images were cleaned for bad pixels, cosmic 
ray events, and other artefacts. 
In some cases, the edge of the images were trimmed (Paper~I) 
in order to eliminate spurious sources and vignetting. 
Background and foreground objects were masked out by using $DS9$ and the IRAF/PROS 
task {\sl plcreate}. 
Images were corrected (Paper~I) for small large-scale variations in the sky 
background by fitting and subtracting out a low-order polynomial, using the 
IRAF task {\sl imsurfit}.\\
GALFIT requires a first-guess set of parameters to be given in an input file.
We used the same $m_{\rm tot}$, $r_e$, $n$ values published in Paper I;
for $x_c$, $y_c$, $PA$  $q$ we used mean values averaged outside the
starburst region from the isophote fitting carried out in Paper I, while
$c$ was set to 0.

Finally, GALFIT allows us to optionally fit also the sky-background as a 
plane that can tilt in $x$ and $y$:
\begin{equation}
sky(x,y)=sky(x_0,y_0)+(x-x_0)\frac{dsky}{dx}+(y-y_0)\frac{dsky}{dy}
\end{equation}

We discuss the application of this option in \S~4.\\

2 - {\em The input PSF\/}

\noindent As we already mentioned, GALFIT uses a PSF image in order to simulate the 
seeing and convolve it with models. When fitting the central pixels of a 
light distribution, the convolution can be very important since the effects 
of seeing are more significant (\cite{2001MNRAS.321..269T},\cite{2001MNRAS.328..977T}). 
Since it is the most accurate way to do it, we extracted a PSF image 
directly from the data, finding an unsaturated star with high S/N.\\

3 - {\em Elaboration of input masks\/} 

\noindent When fitting the radial surface profiles of the LSB component in BCDs, the 
determination of the radial interval is critical (Paper~I). 
Whatever the technique employed, we must make sure that we fit only those 
regions free from the starburst (young stars+gas). 
So, when fitting an image, we need to mask out the area of the galaxy 
occupied by the starburst, whose identification is not always straightforward.

Colour maps and H$\alpha$ images (when available) could help to identify 
and delimit the starburst emission, and thus define a ``first order'' size 
and shape of the masks.
However, a more accurate determination of the region where the starburst 
is practically absent is needed when fitting a S\'ersic function.
In order to do this, an iterative procedure described below, has been applied.
The input masks for fitting consist of a set of elliptical masks of different sizes. 
We have used the IRAF tasks {\sl plcreate} and {\sl imreplace} in order to 
create masks of the same size as the input image, with value 0 in those regions to be 
fitted (``starburst-free region'') and value 1 in the region we want 
to mask out from the fit (``starburst region''). The size of these masks go from 
the small ones containing the inner isophotes, i.e. masking out the starburst peak(s), 
up to the larger ones, following the shape of the outer isophotes, ensuring that we mask out 
the entire starburst emission.\\

4 - {\em Searching for Stability\/}

\noindent The derived parameters may strongly depend on the masked area of the galaxy. 
Small masks that do not contain all of the starburst emission or very large 
masks extending far into the starburst-free region are both expected to produce 
unreliable and/or large uncertainties in the fitted parameters. 
In Paper~I we used two variables: the transition radius, $R_{\rm tran}$ as 
the radius beyond which starburst emission is absent, and $R_{\rm max}$, 
the outermost point fitted. 
The quantity $R_{\rm max}-R_{\rm tran}$ then indicates the fitted radial interval 
where the fit is stable, i.e. the parameters fits with the minimum scatter.

Successive fits with elliptical input masks of different sizes is a possible  
way of estimating both $R_{\rm tran}$ and the stability of a given parameter,
as the range of the light distribution that gives the minimum scatter. 
Since GALFIT fits the entire image, weighting each 
pixel with its own noise, we cannot set an outer radius (i.e. $R_{\rm max}$) for the fit, 
so it is difficult to estimate $R_{\rm max}-R_{\rm tran}$. 

Instead, we prefer to quantify the percentage of galaxy pixels 
we actually mask out from fit, $P_{\rm area}$, defined as 
\begin{equation}
P_{\rm area}=100\left(\frac{N_{\rm mask}}{N_{\rm gal}}\right) = 100\left(\frac{R_{\rm mask}}{R_{\rm gal}}\right)^{2},
\end{equation}
where $N_{\rm mask}$ is the number of pixels masked out, contained in a circle of 
radius $R_{\rm mask}$, and $N_{\rm gal}$ is the total number of pixels with $S/N>1$, 
contained in a circle of radius $R_{\rm gal}$. 
The quantity $100-P_{\rm area}$, when $R_{\rm mask}=R_{\rm tran}$, quantifies the fraction 
of the galaxy (with $S/N>1$) free of starburst emission we actually fit.   

In the fitting procedure, we progressively increase the mask size, 
starting from a mask somewhat smaller than the starburst and ending with a 
mask significantly larger than the starburst. In this way, we aim to detect sudden 
changes in the structural parameters, which could indicate a significant 
contamination by the starburst light, i.e. a rough estimation of the transition radius.

The number of masks, and thus of independent fits, varies depending upon 
the galaxy size, but between six and ten is usually enough to generate 
a ``stability plot'' in which the structural parameters are represented 
versus $P_{\rm area}$ (and/or $R_{\rm mask}$, 
see Figs.~\ref{F4a}-\ref{F5}). 
The step between masks, i.e.\ the step in $R_{\rm mask}$, 
depends on the extension of the starburst in each filter for 
a given galaxy. 
Each fit with a given mask is carried out twice with a fixed sky value and 
a sky value free to vary in the fit.
As we shall discuss later, the difference between the values for the \mbox{S\'ersic} 
parameters in the two cases gives us a further estimation of the 
uncertainties on fitting the LSB component.
The procedure can be automatized up to this point.\\

5 - {\em The Final Mask\/}

\noindent From the stability plots we identify the best model, i.e. the one in which the 
starburst has been completely masked out (see asterisks in Figs.~\ref{F4a}-\ref{F5}). 
From that model we determine the $P_{\rm area}$ range where the parameters do not 
change significantly (e.g. see grey band in Fig.~\ref{F7}). 
We inspect the residual images (galaxy $-$ model) in  
order to check the goodness of the model outside the mask. Note that
the size and shape of the starburst and possible extra features overlapping the galaxy 
are generally not axisymmetric, so
the size and shape of the mask for each galaxy image can be refined 
after inspecting the residuals, as explained below. 
Any signs of non-axisymmetric features, such as spiral arms, distortions or 
possible dust lanes, can be detected in the residual images. The model 
should be fainter than the galaxy in the starburst region and therefore, 
the residuals within the masked inner region should be positive.

From the image of residuals, we use {\sl imreplace} to create a new mask by 
setting to 1 all those pixels whose residual is larger than three times the local rms. 
In this way, the new mask will generally be smaller and will follow  
more precisely the shape of the starburst region.
We call these new masks {\em irregular masks}. Then, we fit the image again.
This iterative process aims at optimizing $P_{\rm area}$ by adapting the mask to
the actual extension and shape of the starburst region.  
When no significant changes of the parameters are obtained by
refining the masks (typically $<$1\%), and the residuals do not change appreciably,
the iterative procedure stops.

\begin{figure}
     \centering
     \includegraphics[width=8.0cm]{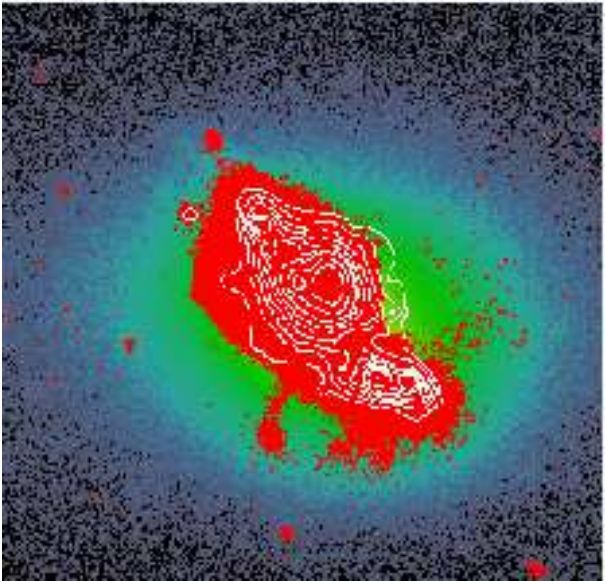}
     \caption{Mrk~35, $B$ band. Overlapped in red, the final mask. In white, the H$\alpha$ contours.}
     \label{F1}
   \end{figure}

\begin{figure}
      \begin{tabular}{l}
	\includegraphics[width=6.25cm,angle=270]{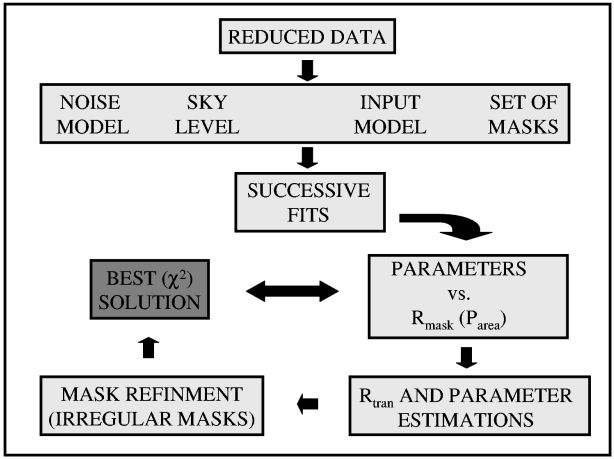}
      \end{tabular}
      \caption{Scheme showing the main steps of the method.}
      \label{F2}
    \end{figure}

In Fig.~\ref{F1} we show a comparison between the final mask derived from 
the iterative fits and the continuum-subtracted H$\alpha$ 
contours ($>$~3$\sigma$), overlapping the $B$ band image of Mrk~35. 
The H$\alpha$ emission was useful to give us a first guess mask but it turned out to be 
somewhat smaller than the whole starburst region in the $B$ band after the complete 
mask refinement. 
Thus, relying only on the H$\alpha$ image may lead to an overestimation (albeit small) 
of the $n$ shape parameter.\\ 

6- {\em Definition of Solution}

\noindent The final solution is obtained by using the irregular masks.
Once we determine the stability range as well as refine the mask in shape and size, 
we choose the best solution as the one with the minimum $\chi^2_{\nu}$ residuals. 
This solution is constrained by the dispersion of the stability range and the limits 
imposed by the possible error sources, as we will discuss later. 

\section{Simulations}  

In order to test the reliability of the method and to determine its 
robustness and flexibility, we carried out a set of fits with synthetic 
galaxies. 
Our interest in this exercise is to determine the typical errors induced by 
the limited number of pixels used in the fit and to study the stability of the fitted 
parameters as a function of $P_{\rm area}$. This way, we are able to define a data range 
within which the procedure provides reliable solutions. 
We have built a set of model galaxies following the criteria:

\begin{itemize}

\item The surface brightness profile was assumed to be a \mbox{S\'ersic} law, 
with two possible values for $n$: 1 (disk-like objects), and 4 (spheroidal 
objects). In both cases the ellipticity was set to 0, assuming in all 
cases a face-on configuration ($q=1$).
\item We simulated the starburst emission as one single burst having a 
Gaussian profile, concentric with the \mbox{S\'ersic} distribution.
\item The difference in magnitude between the whole galaxy (host + starburst) 
and the LSB component was assumed to vary between 0.5 and 1. 
\item The simulated burst contaminates (by 0.01 mag or more) the radial 
\mbox{S\'ersic} profile between $\sim0.5-2.5$ times $R_e$.  
\item We introduce a Poissonian noise and a background sky, 
generated by the {\sl mknoise} task in IRAF. We used the same
read-noise and gain parameters of the real galaxy images in our sample. 
The synthetic galaxies have then surface brightness limits and signal-to-noise 
ratios similar to those of the real images. 
\end{itemize} 

\begin{figure}
      \centering
      \includegraphics[width=8.0cm]{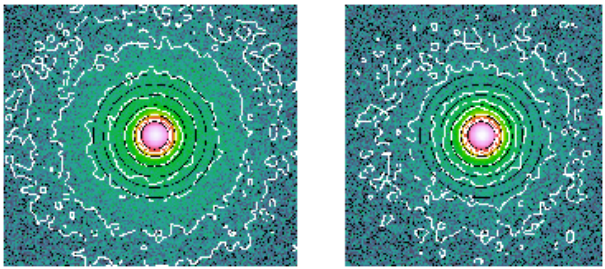}
      \caption{Two synthetic galaxies with \mbox{S\'ersic} index $n=1$ (left) and 
	$n=4$ (right). The outermost isophote corresponds to $1\sigma_{\rm sky}$ (in 
	white), $1\sigma_{\rm sky}$ being the {\em rms} of the sky-background. 
	Circular isocontours, overlapped in black, represent the set of masks used 
	in successive fits.}
      \label{F3}
    \end{figure}

As an example, two synthetic galaxies, a \mbox{S\'ersic} profile with $n=4$ on the 
right, and with $n=1$ on the left, are presented in Fig.~\ref{F3}. 
Both  include a synthetic Gaussian burst overlapping their light 
distribution up to $P_{\rm area}=9$. The circular contours (in black) show 
the size of some of the applied masks. The outermost isophote (in white) 
correspond to $1\sigma_{sky}$, where $\sigma_{sky}$ is the {\em rms} 
background of the image. 

\begin{figure*}
     \begin{tabular}{l l l l}
       \includegraphics[width=8.0cm]{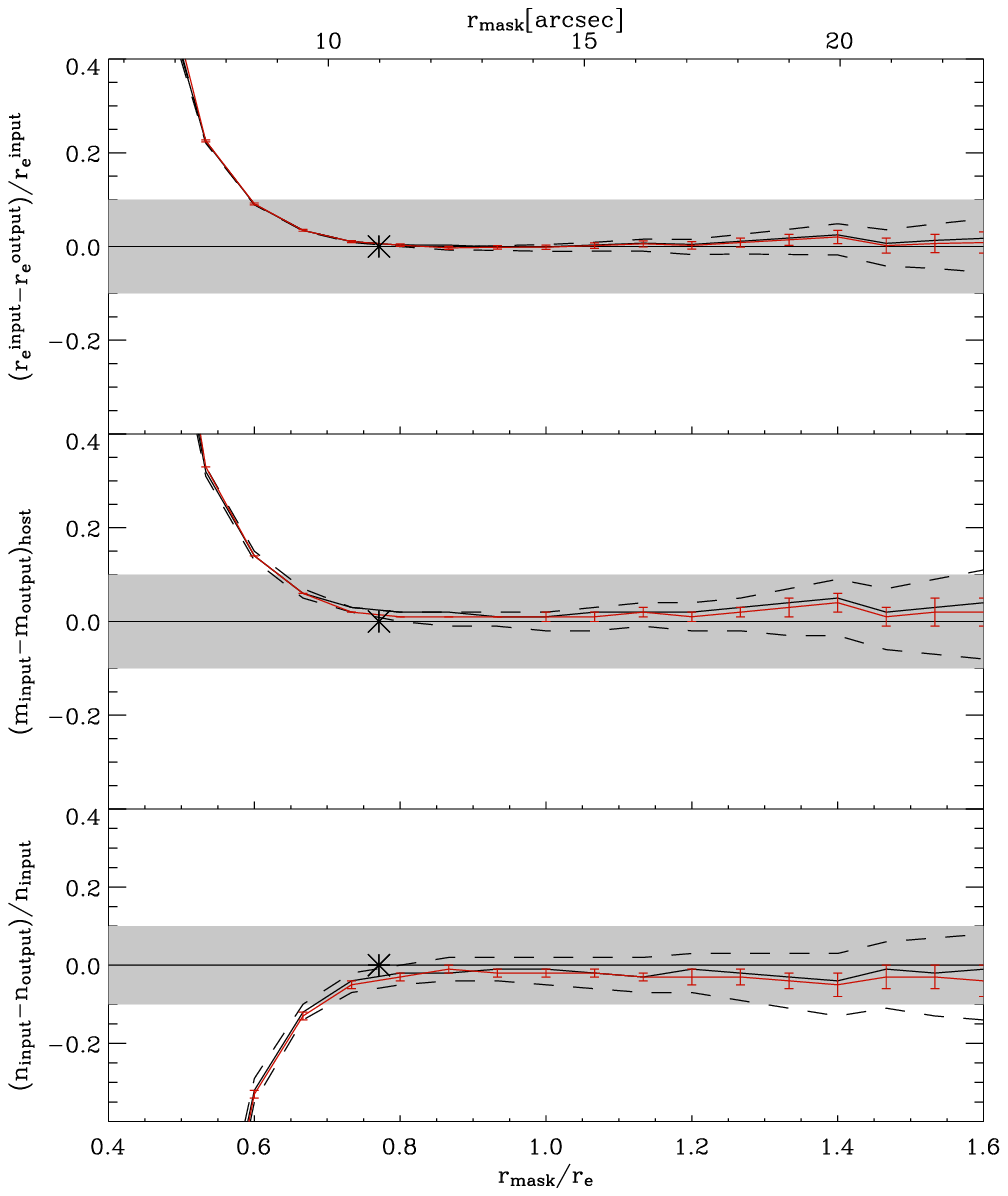}&
       \includegraphics[width=8.0cm]{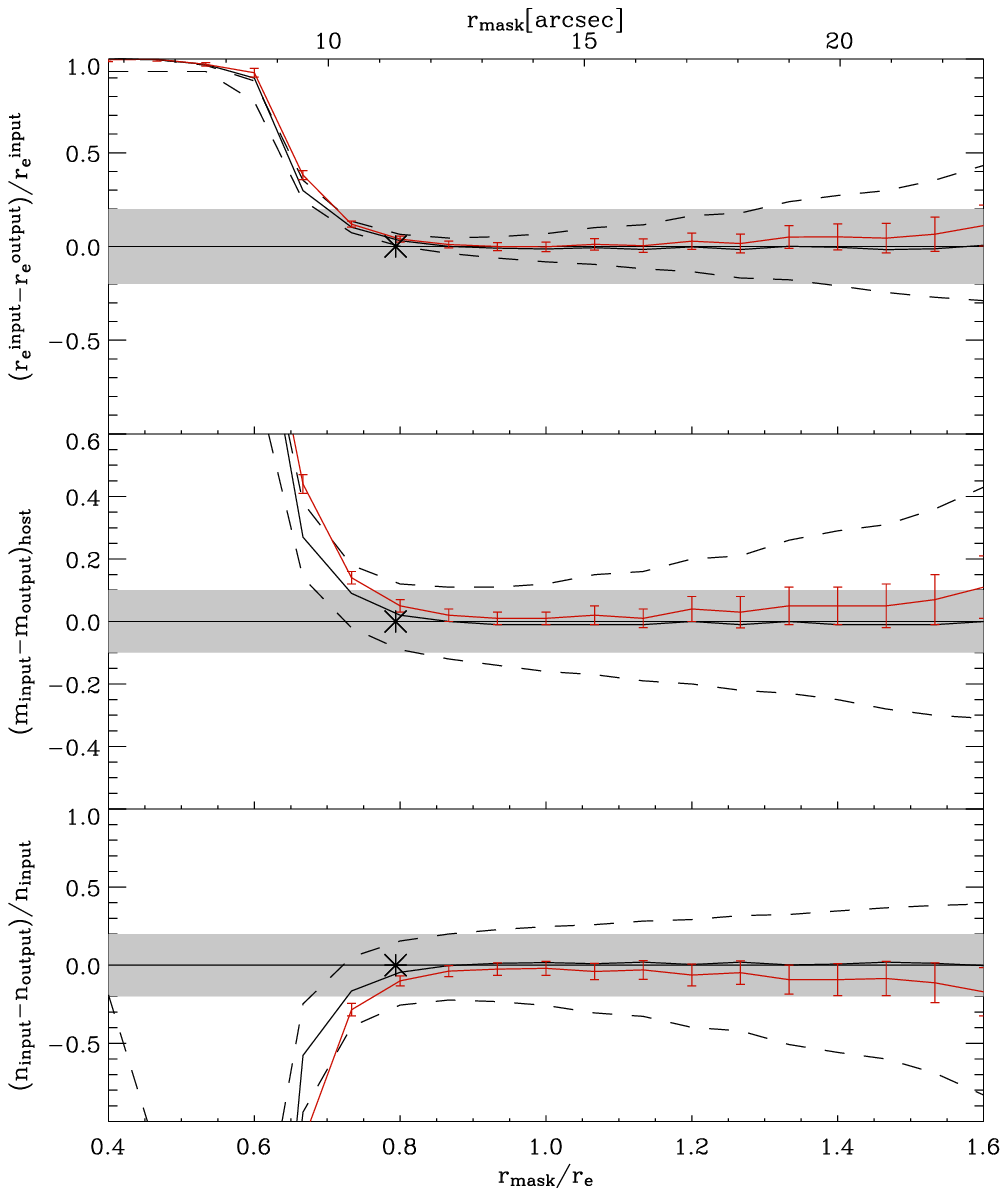}
     \end{tabular}
     \caption{Relative deviations in the three model parameters ($R_{\rm e}$, $n$ and 
       $m_{\rm tot}$) as a function of $R_{mask}$ (upper scale) and $R_{\rm mask}/R_{\rm e}$ 
       (bottom scale). 
       To the left we show results for n=1 models, while n=4 models are presented to the right.  
       Models fitted with a sky value free to vary are represented by solid red lines, 
       while {models fitted with a fixed sky value} are shown by black solid lines. 
       Lower and higher dashed lines show those models for which the sky was fixed 
       to $<sky> \pm \sigma$, where $<sky>$ is the best sky estimation and $\sigma$ the sky 
       uncertainty. The grey band indicates deviations of the 10\% and 20\% for recovered 
       $R_{\rm e}$ and $n$ parameters in the $n=1$ and $n=4$ cases respectively, while 0.1 mag 
       deviations for $m_{\rm tot}$ in both cases. 
       Asterisks in all cases show an estimation of the transition radius ($R_{\rm tran}$).}
     \label{F4a}
   \end{figure*}

\begin{figure*}
     \begin{tabular}{l l l l}
       \includegraphics[width=8.0cm]{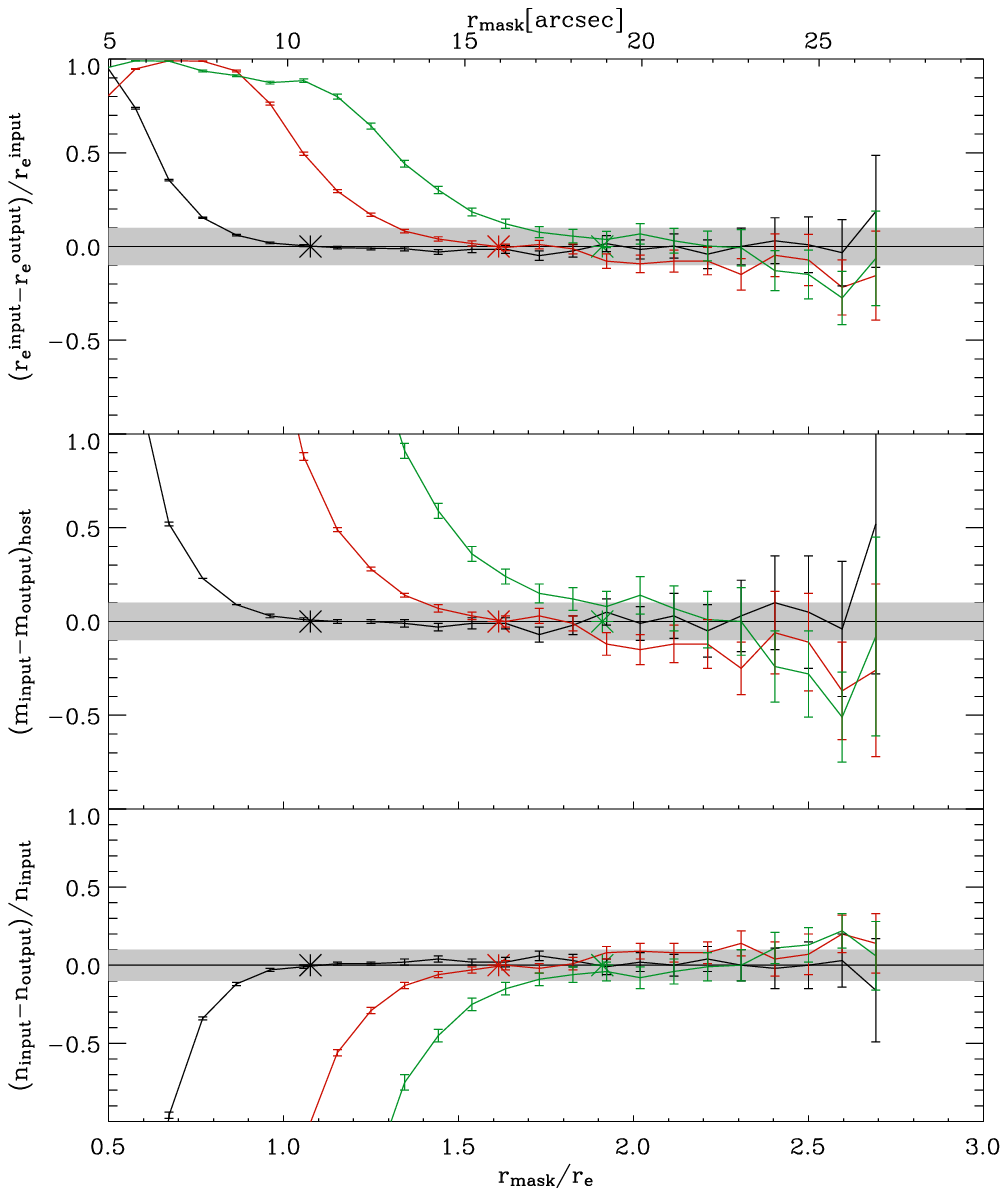}&
       \includegraphics[width=8.0cm]{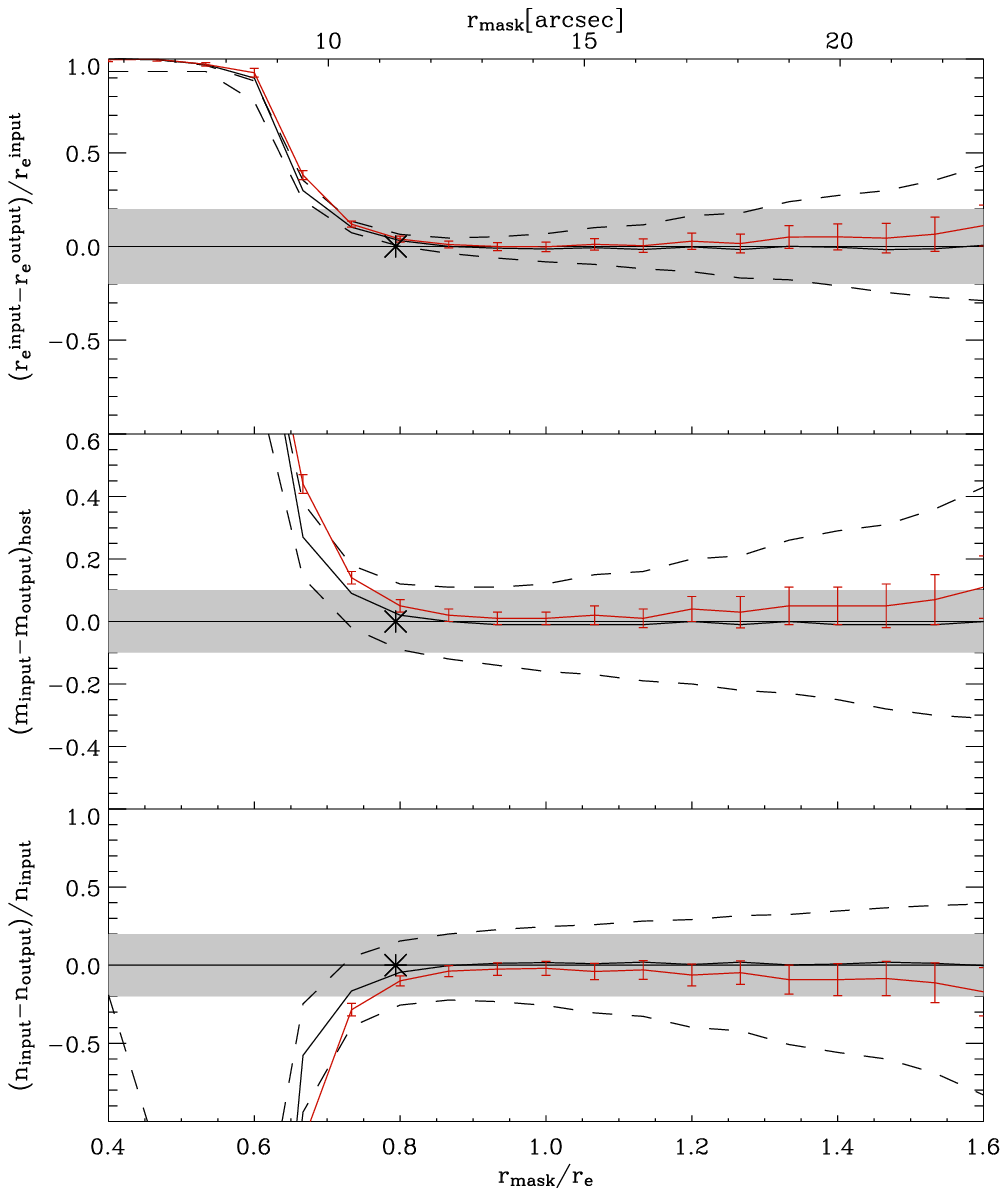}
     \end{tabular}
     \caption{Relative deviations in $R_e$, $n$ and $m_{\rm tot}$ as a function of 
       $R_{\rm mask}$ ({\em top}) and $R_{\rm mask}/R_{\rm e}$ ({\em bottom}), and the 
       simulated starburst size. Lines show \mbox{S\'ersic} recovered models in $n=1$ (left panel) 
       and $n=4$ (right panel) in synthetic galaxies for which the starburst size is 
       $\sim R_{\rm e}$ (black), $\sim1.5 R_{\rm e}$ (red), and $\sim2 R_{\rm e}$ (green). 
       Asterisks in all cases show an estimation of the transition radius ($R_{\rm tran}$).}
     \label{F4b}
   \end{figure*}

\begin{figure*}
     \centering
     \includegraphics[width=8.0cm]{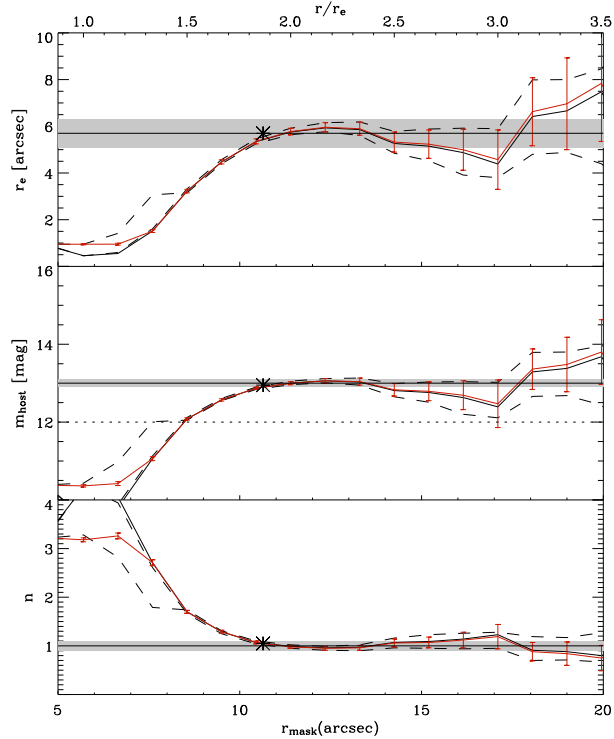}
     \caption{Example of stability plots: $R_{\rm e}$, $n$ and $m_{\rm tot}$, as a function 
       of $R_{\rm mask}$ and $R_{\rm mask}/R_{\rm e}$, for a $n=$1 synthetic galaxy.
       Models fitted with a sky value free to vary are represented by solid red lines, 
       while models fitted with a fixed sky value are shown by black solid lines. 
       Lower and higher dashed lines show those models for which the sky was fixed to 
       $<sky> \pm \sigma$, where $<sky>$ is the best sky estimation and $\sigma$ the sky uncertainty.
       The grey band indicates deviations of the 10\% on $R_{\rm e}$ and $n$, 
       while 0.1 mag deviations for $m_{\rm tot}$.
       Asterisks show an estimation of the transition radius ($R_{\rm tran}$).}
     \label{F5}
   \end{figure*}

We applied the method described in \S3 to fit and recover the 
\mbox{S\'ersic} parameters. We ran GALFIT with a first-guess estimate as an initial input. 
In Figures~\ref{F4a},~\ref{F4b} and~\ref{F5} we present examples of the results.

We show in Fig.~\ref{F4a} the relative deviation (i.e., the difference between 
output and input values divided by the input value) in $R_{\rm e}$ and $n$, as well as 
the deviation (i.e., the difference between output and input values) in $m_{\rm tot}$ 
as a function of $R_{\rm mask}$ (upper scale) and $R_{\rm mask}/R_{\rm e}$ (bottom scale) 
for two galaxies described by $n=1$ (left) and $n=4$ (right) \mbox{S\'ersic} profiles.
In this synthetic galaxy, the starburst component contaminates $\sim$35\%. 
The last plotted model ($R_{\rm mask}\sim$23$''$) corresponds to $P_{\rm area}\sim$75.  
Red solid lines are models with the sky left as a free parameter. 
Black solid lines are models with a fixed sky value. 
The error bars correspond to the statistical uncertainties estimated by the GALFIT 
$\chi^{2}_{\nu}$ minimization. 
The grey bands indicate deviations of $10\%$ and $20\%$ in the output $R_{\rm e}$ 
and $n$ parameters for $n=1$ and $n=4$ respectively, and 0.1 mag deviations for 
$m_{\rm tot}$ in both cases.
Lower and upper dashed-lines are models for which the sky was fixed 
to $<sky>\pm \sigma$, where $\sigma$ is the sky uncertainty.  
We obtain reasonable estimations of the sky level ($\sigma \sim$ 0.2-0.5\% $sky$) 
when the portion of sky in the frame is large enough, typically $\geq3$ times the diameter 
of the $1\sigma_{sky}$ isophote.  
Two independent methods were applied in order to estimate a value of the sky. 
In the first, we use the {\sl ellipse} task in IRAF to calculate an extended profile 
of the galaxy to the edge of the frame. 
By plotting the flux as a function of radius we can estimate a region where the contribution 
of the galaxy is negligible and the background can be considered flat. Thus, the median  
and the standard deviation of the outer points are used (e.g. see \cite{2006A&A...454..759P}). 
The second method is easier and faster; in addition to stars and other background and/or 
foreground objects, the entire galaxy is masked out to an ellipse with a flux smaller 
than the {\em rms} of the background.
Thus, we fit with GALFIT only the sky function to the image obtaining $<sky>\pm\sigma$. 
Both estimations agree for our simulations within $\pm0.05$ counts ($\sim0.2\% sky$).

Fig.~\ref{F4b} shows the comparison between the deviations of the output $m_{\rm tot}$, 
$R_{\rm e}$ and $n$ parameters as a function of $R_{\rm mask}$ (upper scale) and 
$R_{\rm mask}/R_{\rm e}$ (bottom scale), for three galaxies with different starburst size. 
Galaxy with $n=1$ is to the left while the $n=4$ galaxy is to the right. 
Different line colours were used to indicate galaxies with different synthetic starburst size, 
varying between $\sim$1$R_{\rm e}$ and 2$R_{\rm e}$ 
($P_{\rm area}\sim$17\% in black, $\sim$35\% in red, and $\sim$50\% in green).
The error bars and the grey bands have the same meaning as in Fig.~\ref{F4a}.

Finally, in Fig.~\ref{F5} we show the main output parameters ($m_{\rm tot}$, $R_{\rm e}$, 
and $n$) versus $R_{\rm mask}$ and $R_{\rm mask}/R_{\rm e}$ for a compact synthetic 
disc-like object with a starburst occupying $\sim$50\% ($\sim$1.9$R_{e}$) of the pixels 
until the $1\sigma_{\rm sky}$ level. The last model ($R_{\rm mask}=$20) has 
$P_{\rm area}\sim$95. 
The horizontal dotted line in the $m_{\rm tot}$ plot corresponds to the total luminosity 
of the galaxy (host+starburst). Lines as well as deviations and uncertainties have the 
same meaning as in Fig.~\ref{F4a}.

The examples shown in Figs.~\ref{F4a}~-~\ref{F5} illustrate the existence of two features:  
a sudden change of the parameters, i.e. a significant change of the slope (an elbow), 
which indicates the region where the starburst becomes practically absent ($R_{\rm tran}$, 
indicated by asterisks in the figures), and a range of models (with $R_{\rm mask} 
\geq R_{\rm tran}$) for which the \mbox{S\'ersic} parameters are stable within the 
uncertainties showed by the grey bands.

We resume and discuss our results as follows:

$\bullet$ 
Our simulations show that the starburst contamination has been clearly discriminated by 
plotting the successive models with different masks, i.e. {\em parameters} $vs.$ 
$R_{\rm mask}$. 
The sudden change in the main parameters ($m_{\rm tot}$, $R_{\rm e}$ and $n$) followed 
by a more or less extended stability region give an estimation of the $R_{\rm tran}$ value. 
That change is well determined by the elbow in the plot, as shown in the 
Figures~\ref{F4a}~-~\ref{F5}. The \mbox{S\'ersic} parameters are recovered within the 
formal errors from $R_{\rm mask}=R_{\rm tran}$. The parameter uncertainties grow 
up when $R_{\rm mask}>R_{\rm tran}$, showing no systematic effects.
Models with $R_{\rm mask}<R_{\rm tran}$ (i.e, with too small a mask) produce large 
systematic overestimations in $n$ and luminosity, which also depends on the 
starburst luminosity. The condition $m_{galaxy}<m_{host}$ is a logical
(physical) constraint to the solution, as shown in the Fig.~\ref{F5} (dotted line); 
in a narrow range of fits, where $R_{\rm mask}<R_{\rm tran}$, the luminosity of the host 
can be brighter than the whole galaxy.

$\bullet$
When we mask the entire starburst, the amount of galaxy pixels with  $S/N>1$ used for 
the fit is critical in order to recover the \mbox{S\'ersic} parameters with small 
uncertainties (see Figs.~\ref{F4a}-\ref{F5}). 
The errors induced by sky subtraction as well as the random uncertainties estimated 
from $\chi^{2}_{\nu}$ minimization in the model parameters increase with $P_{\rm area}$, 
those for  $n=1$ models being smaller than those of $n=4$ models. 

$\bullet$
Besides the starburst contamination, the other possible and critical systematic error source 
in fitting \mbox{S\'ersic} profiles at very faint levels is the measurement of the correct 
sky value.
Relative deviations caused by an erroneous sky subtraction in low $n$ objects are smaller 
than in higher $n$ objects. The most affected \mbox{S\'ersic} parameter is $n$. 
For low $n$ objects, we estimate uncertainties within 10\% in a significant range of 
$P_{\rm area}$ for a given combination of model parameters. For high $n$ galaxies that 
range could be narrower, and errors within 20\% to 30\% are expected. 
We estimate ranges of, at least, $\sim$25-30\% $R_{\rm e}$ in which the parameters are 
stable within these relative deviations. 

$\bullet$
In order to weight the pixels in the fit (with their own noise) GALFIT makes a model from 
the data information, so accurate solutions require accurate noise information, 
basically: readnoise, gain, the number of images combined, and sky level. 
In our simulations we have shown that fits with both fixed and free sky value are 
in good agreement if the portion of sky is large enough, with uncertainties growing up 
towards low $S/N$ levels. The $\chi^{2}_{\nu}$ minimization uncertainties in 
fits with a free sky value are greater than in those with a fixed one. 

$\bullet$ 
In our simulations all the parameters behave in a similar way for all the starburst sizes 
and luminosities considered; e.g, a downward slope followed by a flatter curve. 
As shown in Fig.~\ref{F4b} the $R_{\rm tran}$ value is well determined in all cases.
The range where parameters are stable (grey bands in Figs.~\ref{F4a}-\ref{F5}) depends 
on $R_{\rm mask}$ and on the $S/N$ of the fitted portion of the galaxy.
An important result is that our solutions show an acceptable stability in their 
parameters in the range $R_{\rm mask}/R_{\rm e}\leq 2$, with relative deviations growing as
the mean $S/N$ ratio of the fitted portion of galaxy decreases.
Even when the percentage of masked galaxy is large ($\gtrsim$1.5$R_{\rm e}$),
we can still find a range of stability, $\gtrsim$0.4-0.5$R_{\rm e}$, with uncertainties 
between 10\% and 20\%. For low $n$ galaxies the  portion of galaxy required for a good fit 
is smaller than for high $n$ galaxies, which generally show larger random uncertainties 
(see Figs.~\ref{F4a}-\ref{F5}). 

\section{On Fitting the LSB Host Galaxy in BCDs}

In Paper~I we analysed the benefits and drawbacks of fitting a \mbox{S\'ersic} law 
to the starburst-free region of BCDs radial profiles. 
Here we apply our previous research to fit two-dimensional images with \mbox{S\'ersic} 
profiles. Although it was argued that the \mbox{S\'ersic} law provides a reliable 
description of the host galaxy, we must assess how sensitive the derived two-dimensional
\mbox{S\'ersic} parameters are to the observational uncertainties. 

\subsection{Sensitivity to the sampled area}

Regardless of the specific technique used to fit a \mbox{S\'ersic} model to BCD host
galaxies, we must be careful about the choice of the fitted data points
(in a radial profile) or number of pixels (in an image), since the \mbox{S\'ersic} 
parameters could be very sensitive to that choice.
In Paper~I we explored the effect of a limited radial interval in a 1D fit and
concluded that accuracy in its selection is crucial 
for small radial ($R_{\rm tran} \geq 1-2$ scale lengths)
 and surface brightness intervals (typically less than 4 mag). A broad overview of
this problem in the simulations of ideal galaxies has already been presented in
\S4. As was already observed by Makino et al.\ (1990) for 1D fits, 
we find that in 2D fits the shape parameter $n$ is also the 
most sensitive and least constrained parameter, especially when the fit is done
on a restricted surface brightness range.

The inner limit of the fitted interval is given by the size and shape
of the mask, as parameterized by \mbox{$R_{\rm tran}$}.
Since the starburst emission generally has a steeper light distribution, a small
underestimation of $R_{\rm tran}$ will provide significantly higher $n$ values. 

Uncertainties in this case can be quantified by means of the dispersion of the output 
parameters of fits with masks of different sizes. 
We have tested the procedure through our simulations, taking into 
account different settings of the input models. 
In particular, the relationship between $R_{\rm tran}$, $R_{\rm mask}$ and $R_{\rm e}$ 
was studied (see \S4).
In our sample we found that the stability in the parameters is generally 
reached when \mbox{$R_{\rm tran}$}$\gtrsim$\mbox{$R_{\rm e}$} (see Table~\ref{T2}). 
This indicates that these galaxies have extended starburst emission, and thus we have 
to take special care with the size and shape of the masks. 
The accurate isolation of the starburst emission is the most important strength of the 
technique.

As shown in the simulations (see \S4) the more concentrated the starburst is, 
the easier it is to find reliable and stable solutions 
(e.g. Mrk~370, Mrk~36, Tol~127 and Mrk~35).

\begin{figure}
     \centering
     \includegraphics[width=8.2cm]{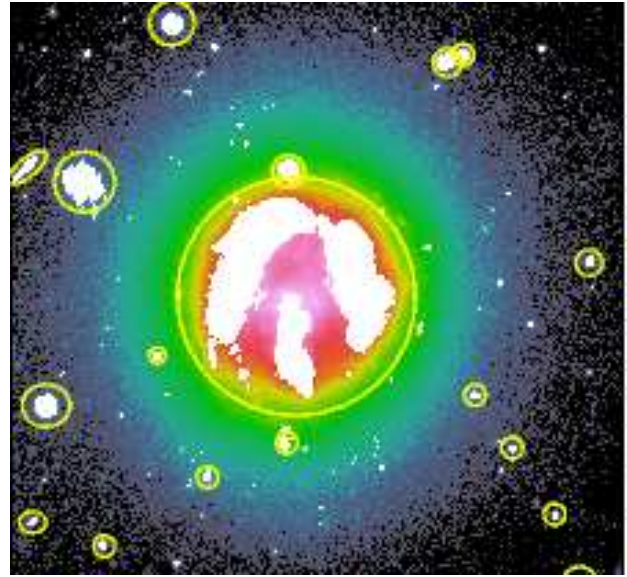}
     \caption{Mrk~86, $R$ band. In white, the mask covering the starburst peaks. 
       In yellow, the contours of the elliptical mask containing most of the 
       starburst emission.}
     \label{F6}
   \end{figure}

\begin{figure*}
\centering
\includegraphics[width=15.cm, angle=90]{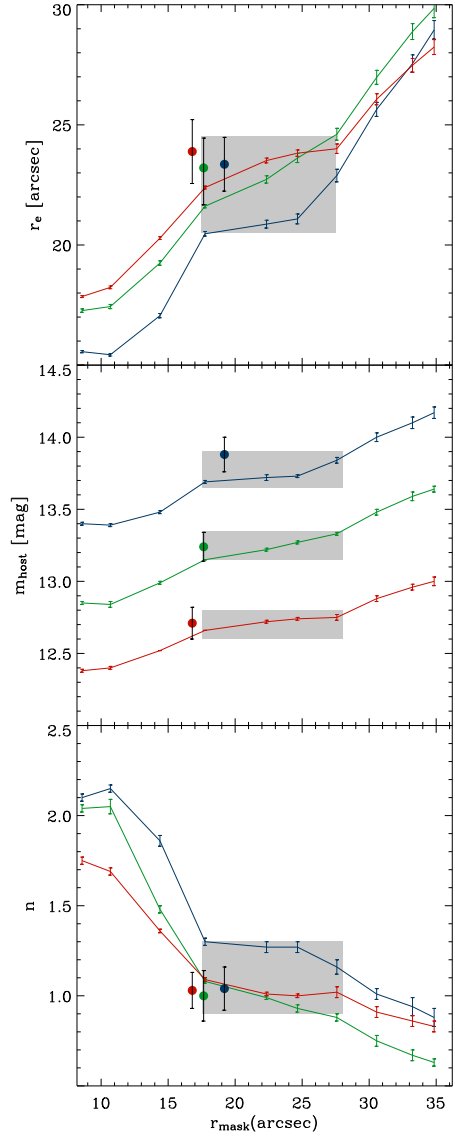}
\caption{Stability in the 2D fit of Mrk~370: the free parameters $R_{\rm e}$, $m_{\rm tot}$ 
and $n$, are plotted as a function of $R_{\rm mask}$ in the three bands (blue=$B$, 
green=$V$, and red=$R$). Dots in colour are the final best fits. 
Their error bars take into account uncertainties from the sky subtraction. 
Grey bands indicates the stability region considered.}
\label{F7}
\end{figure*}

In those galaxies where the star-forming knots are spread out
over the whole galaxy, we could recover information from those inter-knot pixels, 
increasing the number of pixels being fitted. Although this could be useful to check
the stability of solutions, those pixels could have a small starburst light contribution. 
Thus, an accurate and more reliable solution is obtained when the whole starburst 
contamination is completely masked out.
Fig.~\ref{F6} displays an example, using the $R$-band image of Mrk~86. 
An irregular mask derived by our procedure (shown in 
white), masks out only the peaks of the main knots. The fitted parameters are 
$n=1.07$, $m_{\rm tot}=11.60$ and \mbox{$r_{\rm e}$}$=25.86$ arcsec.
The contour of an elliptical mask (shown in yellow) including 
almost the whole starburst emission is also shown.  
Fitting the host galaxy with that mask, $n=1.28$, $m_{\rm tot}=11.85$ and 
\mbox{$r_{\rm e}$}$=31.56$ arcsec are obtained. 
Notice that when fitting a larger fraction of pixels (i.e.\ when using the irregular 
mask) the $n$ value is lower than in the opposite case, but the luminosity is 0.25 mag higher. 
Although there seems to be still a small amount of starburst contamination, it is 
evident that the host galaxy of Mrk~86 has a disc-like object structure.
Our best fit in the $R$-band was $n=1.02$, $m_{\rm tot}=11.73$ and 
\mbox{$r_{\rm e}$}$=25.49$ arcsec, by using a bigger irregular mask (derived from the 
$B$-band frame) which isolates the whole starburst emission following its shape.

\subsection{Sensitivity to sky subtraction errors}

As we have already shown in our simulations the
uncertainties resulting from noise fluctuation and sky background subtraction are 
propagated into the structural parameters when we fit a model to the outer,
low signal-to-noise regions of a galaxy. 
In particular, the \mbox{S\'ersic} profile has more extended and shallower tails 
for high values of $n$. Thus, an over-subtracted sky background will produce 
an underestimated $n$ value, while an under-subtraction will overestimate 
$n$. We expect that the larger the intrinsic $n$ value, the more 
sensitive it becomes to this error source (see Paper~I and \S4). 

We analysed the case of Mrk~5 in order to illustrate this error source
and compare it with our previous simulations and with a similar example given in Paper~I.
Results are shown in Table~\ref{T4}. Three cases were examined. 
First we considered the best sky background subtraction. 
In the second, we over-subtracted the sky background by about 3 times 
the sky uncertainty, and in the third we under-subtracted it by the same amount. 
This is equivalent to fitting two profiles that follow the envelopes 
of the  error bars considered in Fig.~5 of Paper~I.
For each of the three cases we computed the \mbox{S\'ersic} parameters for 
the final irregular mask ($R_{\rm tran}=10.97$ arcsec).
The $B$-band image of Mrk~5 has its best-fit $n=0.99$ for 
$R_{\rm tran}=10.97$ arcsec (bold row in the Table~\ref{T4}). 
By under-subtracting the sky background $n$ goes up to 1.17, \mbox{$r_{\rm e}$} 
increases by 9\%, while the total luminosity does not change by more than 0.05 
mag. Over-subtracting the sky background by the same amount, $n$ falls to 0.88, 
\mbox{$r_{\rm e}$} decreases by 4\% and the total luminosity decreases 0.06 mag. 

\subsection{Other error sources}

In the fitting procedure, GALFIT estimates the statistical uncertainties by using the 
covariance matrix of the parameters (see P02 for a detailed description). 
Those uncertainties are random errors (Poissonian fluctuations) since GALFIT 
assumes that the fitted profile is correct, so these estimated errors are 
generally very small compared to non-random errors from other sources. 
For our galaxies, these random errors are $<$1$\%$.

Galaxies are not perfectly axially symmetric objects. 
Since GALFIT does not allow for radial changes in axis ratio and position angle 
of a given component, we expect to obtain larger residuals when fitting 
regions with twisted or irregular isophotes. 
This is a remarkable weaknesses of the method.
Mrk~35 (see Fig.~\ref{F9}) is useful to illustrate this case.
This galaxy shows negative and positive residuals in the border of the mask, 
just in the region where isophotes twist.
However, this does not seem to have a great effect on $n$, \mbox{$r_{\rm e}$}, 
or $m_{\rm tot}$ values. 
If we observed systematic residuals produced by these effects, we could check whether 
they are produced by a real feature (e.g. dust, spiral arms, bars, isophote twists, etc...), 
and quantify the changes in the parameters by playing with the ellipticity and the 
position angle, i.e., fixing different values as input in the fits, and comparing 
its residuals.

\subsection{Consistency checks}

In order to overcome the above problems and to assess the reliability of the
fit, we need to carry out some consistency checks.

First, the reliability of the derived \mbox{S\'ersic} parameters are assessed by examining 
the stability plots. Given their sensitivity to the $P_{area}$ range we explore 
whether there exists a range in the light profile where the fit is stable  
and does not depend on the exact choice of $R_{\rm tran}$. 
The stability criterion is the same as that applied to the simulations in \S4.
Second, as we expect that the old stellar host galaxy has negligible color gradients 
(as shown by the observed behaviour in the outer parts, see for example 
C01b, \cite{2005ApJS..156..345G}), both $n$ and $R_{\rm e}$ should be the same in 
all passbands, while the differences in $\mu_{\rm e}$ and $m_{\rm tot}$ reflect the LSB colors.
Finally, the sky background analysis implemented in our simulations (see \S4) 
is systematically applied.

In Fig.~\ref{F7} we show the three model parameters 
($m_{\rm tot}$, $R_{\rm e}$ and $n$) of Mrk~370 for three 
bands: $B$ (blue line), $V$ (green line) and $R$ (red line), versus $R_{\rm mask}$. 
Each point in the lines corresponds 
to one fit by using an elliptical mask with a given $R_{\rm mask}$ (and thus, $P_{\rm area}$) 
value. 
We notice some features: first, there exists a sudden change in the slope of the curves 
by increasing the size of the mask. This ``elbow'', as shown in the previous simulations, 
indicates the region where the starburst becomes practically absent. The second,
these free parameters are quite stable (grey bands) 
in the range between $R_{\rm mask}\sim$17 and 28 arcsec. 
The grey bands correspond to dispersions $\sigma_{\rm n}=0.20$, 
$\sigma_{\rm m_{\rm tot}}=0.10$, and $\sigma_{\rm R_{\rm e}}=2''$. 
In the case of $n$ and $R_{\rm e}$ these bands take into account the dispersion between 
filters, which can be explained by differences in image depth and quality.
We refine the mask by using the positive residuals ($>3\sigma$) in this range of stability 
thus obtaining the final irregular masks (see \S3). 
With it, we obtain a better fit, i.e. $\chi^{2}_{\nu}$, constraining the starburst mask 
in size and shape.
We also perform various fits constraining parameters such as $c$, $PA$, $q$, or the 
centre coordinates, by varying the input values, and playing with small variations of 
the sky background in order to check the stability and find the best $\chi^{2}_{\nu}$ and 
residuals. We indicate the final best fits by the color dots. 
Notice that for the $B$ filter the difference between the elliptical masks fits 
(blue line) and the irregular mask fit (blue dot) is quite large. 
In a filter where the starburst is more luminous, this result remarks the importance 
of a better constraint of the size and the shape of the mask.

\section{Results}

The results of the 2D fitting of a \mbox{S\'ersic} law to the starburst-free regions
of the galaxy images are presented in Table~\ref{T2}. It is worth 
mentioning that, in all cases after the procedure is completed, stable
parameters were obtained. 
The consistency checks discussed in \S5.4 were applied systematically to all of the galaxies. 
As can be seen in Table~\ref{T2}, the dispersion in $n$ and \mbox{$r_{\rm e}$} between
different filters is always less than 10\% except for I~Zw~123 ($\sim$23\% in $n$). 
These small discrepancies among
different filters are within the uncertainties expected from our simulations and can also be 
explained in terms of the different quality of the data.
For each galaxy, we calculated $P_{\rm area}$. 

We calculate a \mbox{$R_{\rm tran}$} value as the radius of a circular 
mask with the same number of pixels as that used for the best fit. 
In all cases we were able to fit more than 74\% of the galaxy pixels. The ratio
$R_{\rm tran}/R_{\rm e}$ varies between 0.70 (Mrk~370$R$) and 1.73 (I~Zw~123$V$).
To estimate the uncertainties in $n$, \mbox{$r_{\rm e}$} and $m_{\rm tot}$ 
(here $m_{\rm LSB}$) we compute the dispersion between parameters in the 
stability range defined for each frame, as well as the difference between the 
values obtained when fitting with a fixed sky-background and those obtained by setting the 
sky-background level as a free parameter.

\begin{figure*}
\begin{tabular}{c c c}
\includegraphics[width=5.5cm,angle=0]{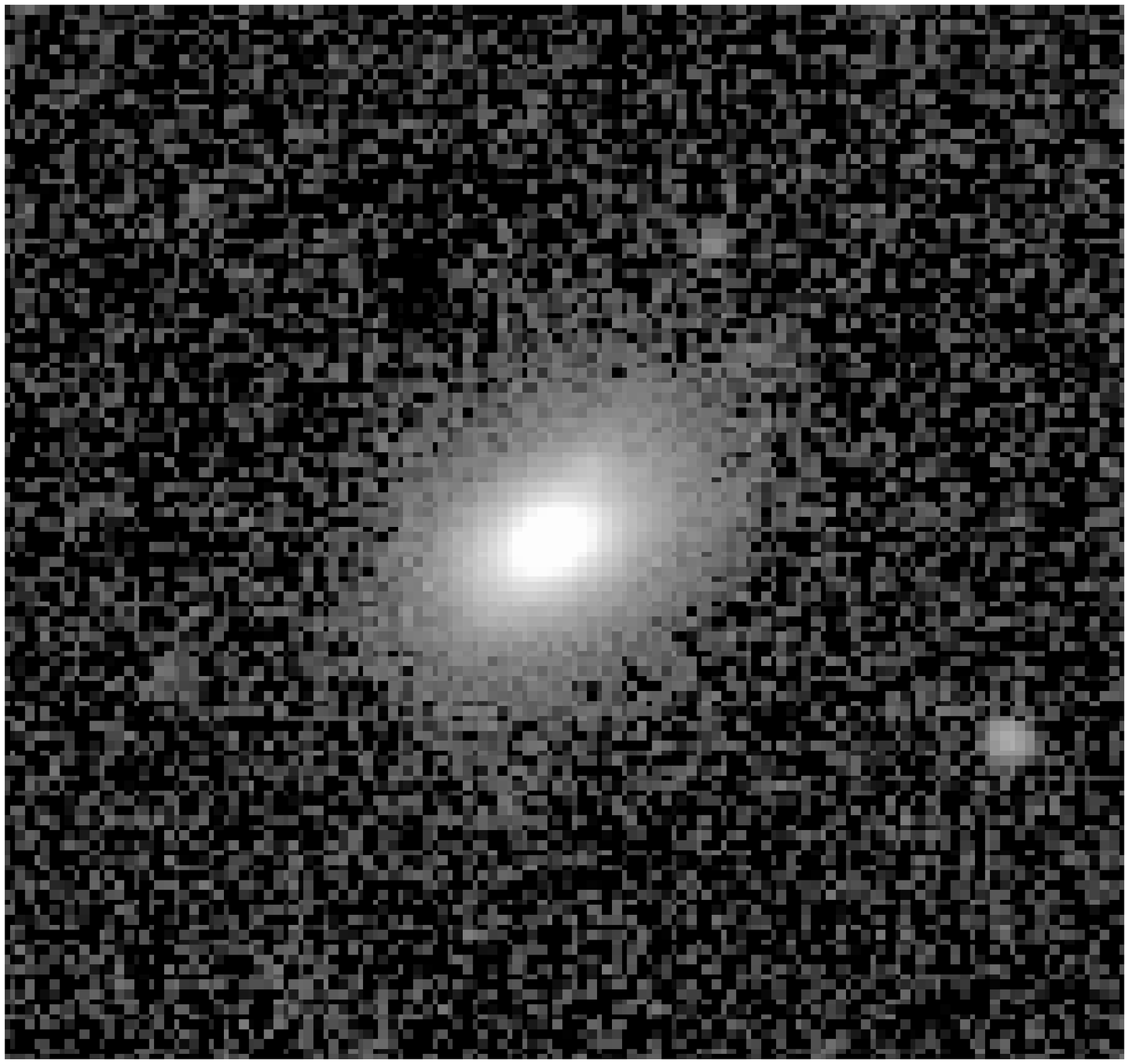}&
\includegraphics[width=5.5cm,angle=0]{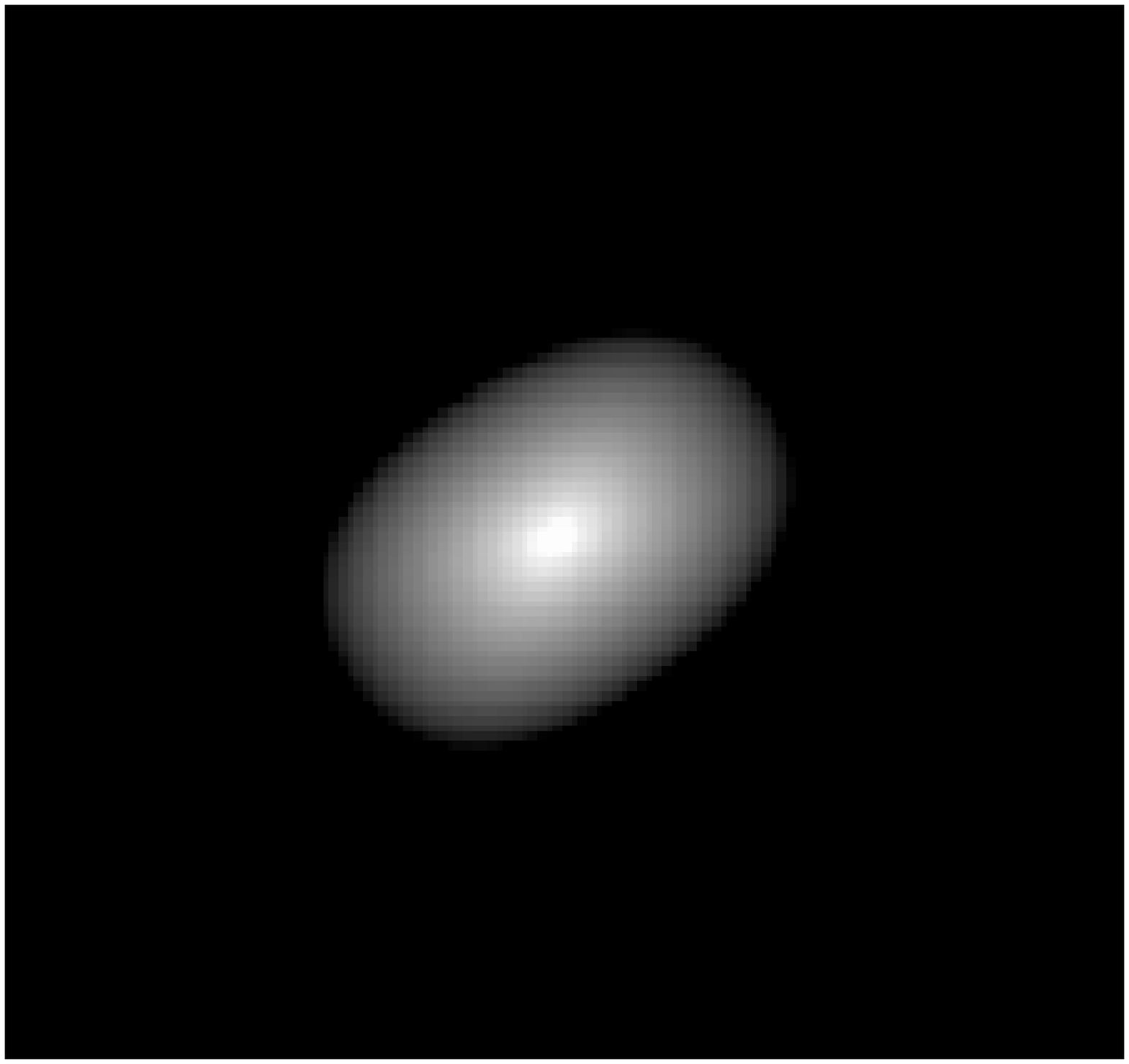}&
\includegraphics[width=5.5cm,angle=0]{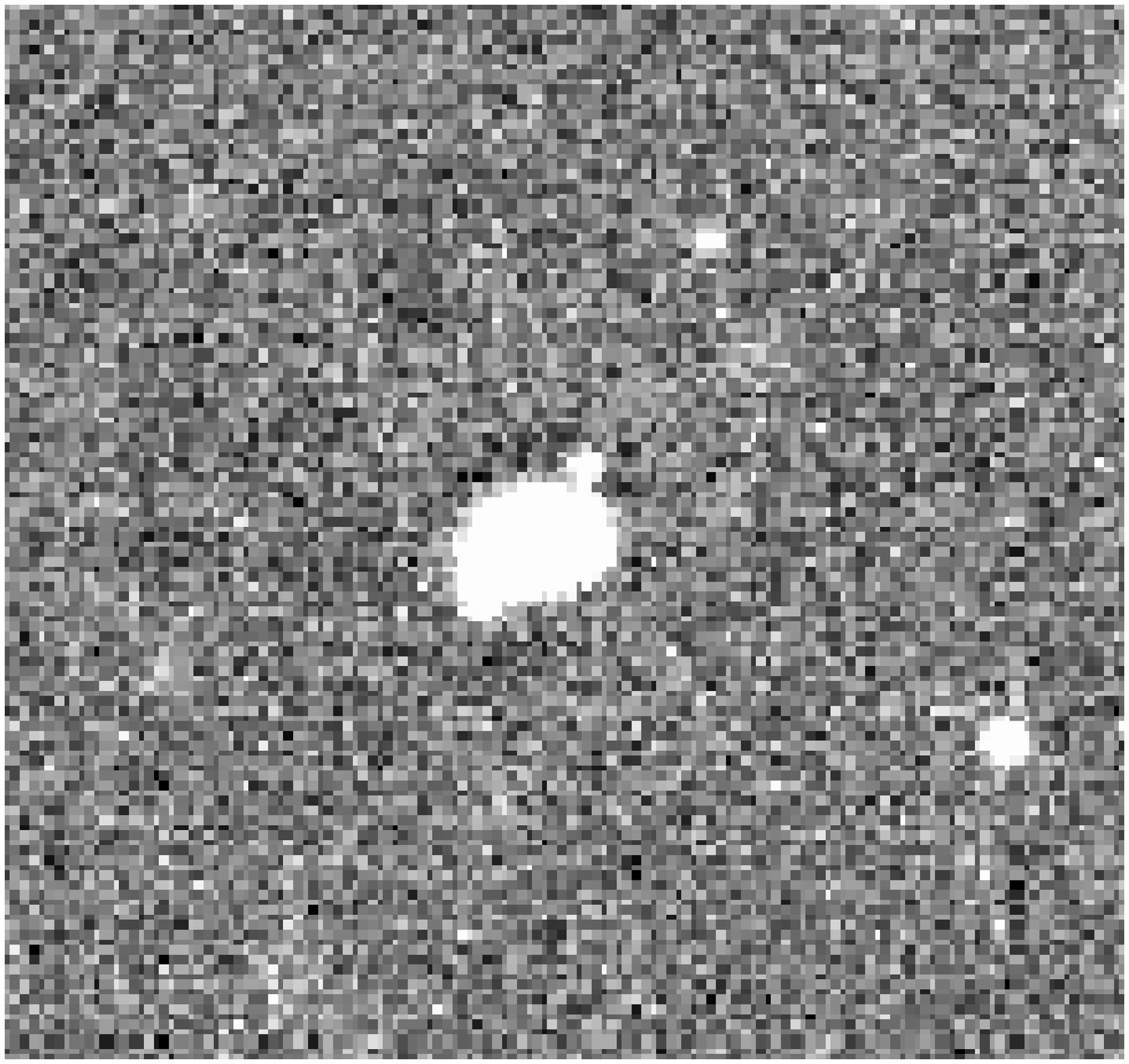}\\
\includegraphics[width=5.5cm,angle=0]{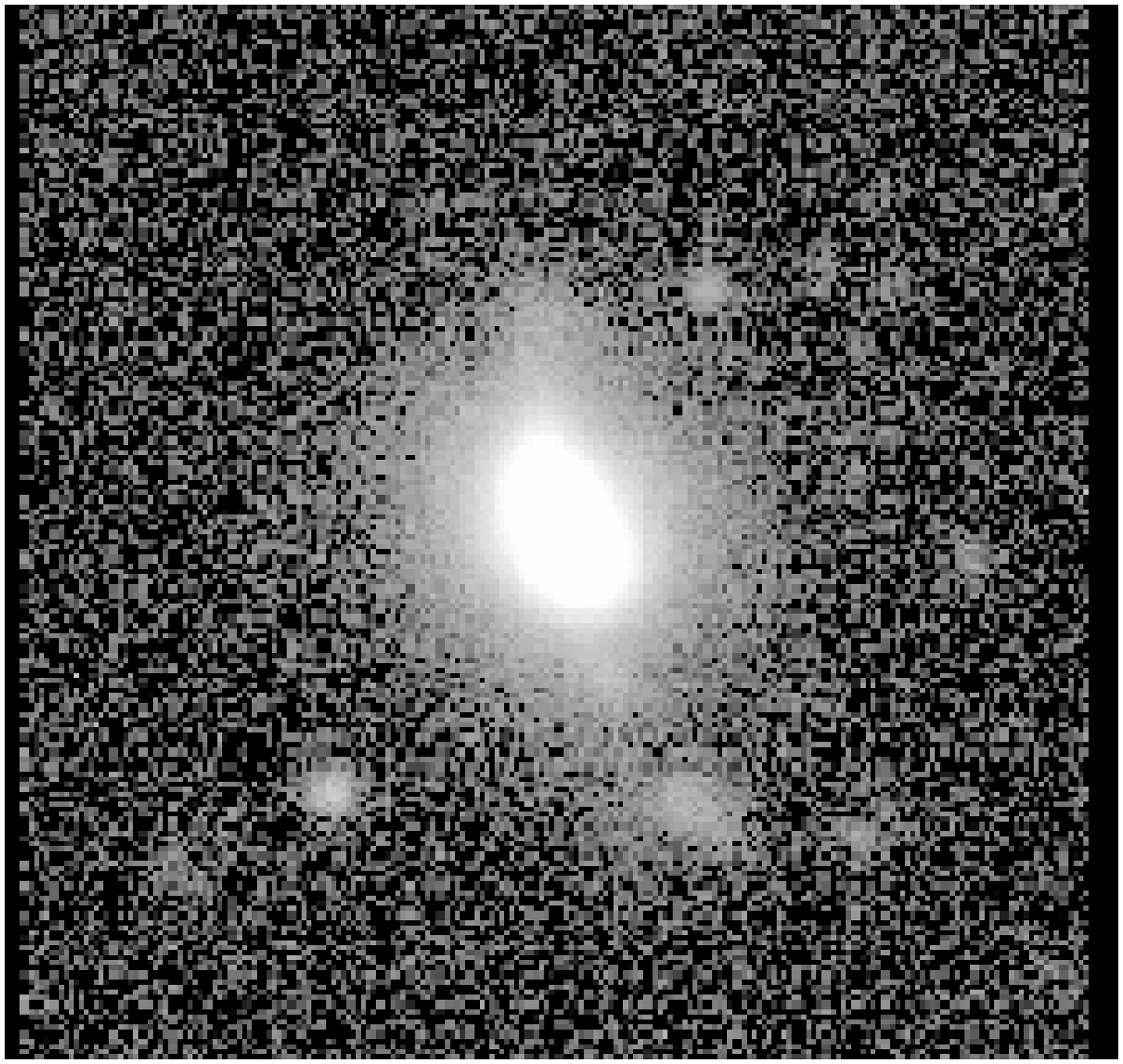}&
\includegraphics[width=5.5cm,angle=0]{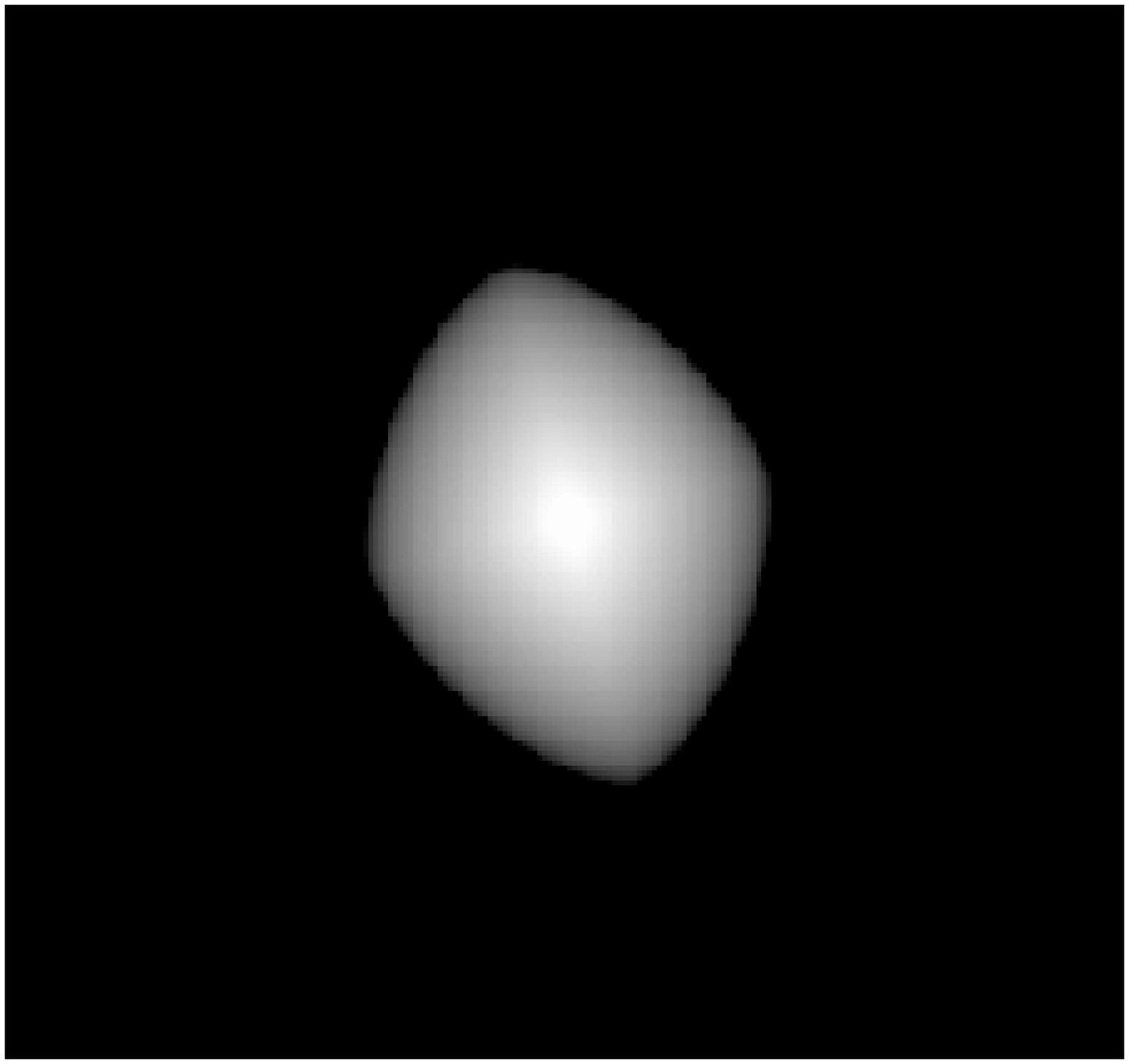}&
\includegraphics[width=5.5cm,angle=0]{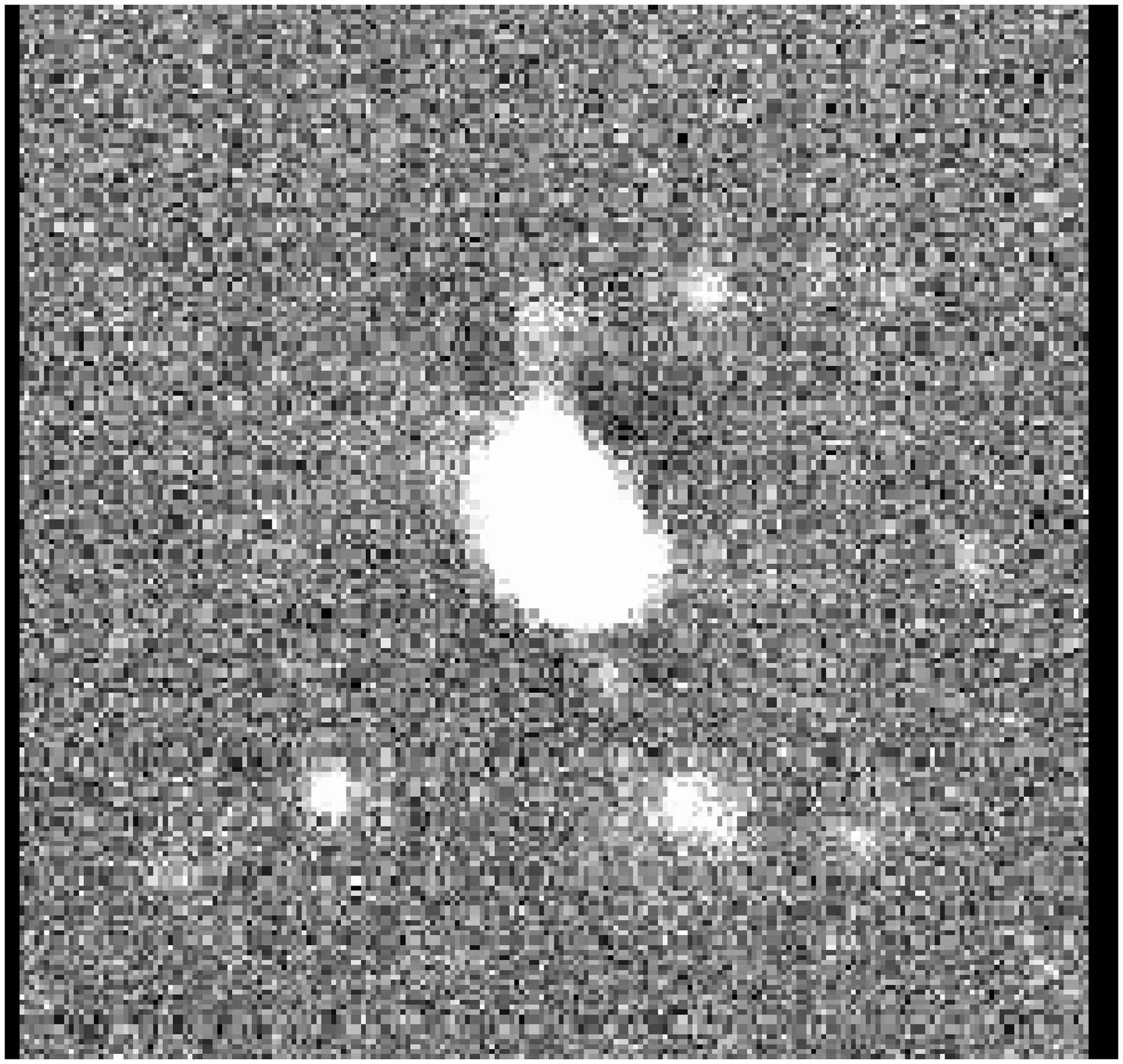}\\
\includegraphics[width=5.5cm,angle=0]{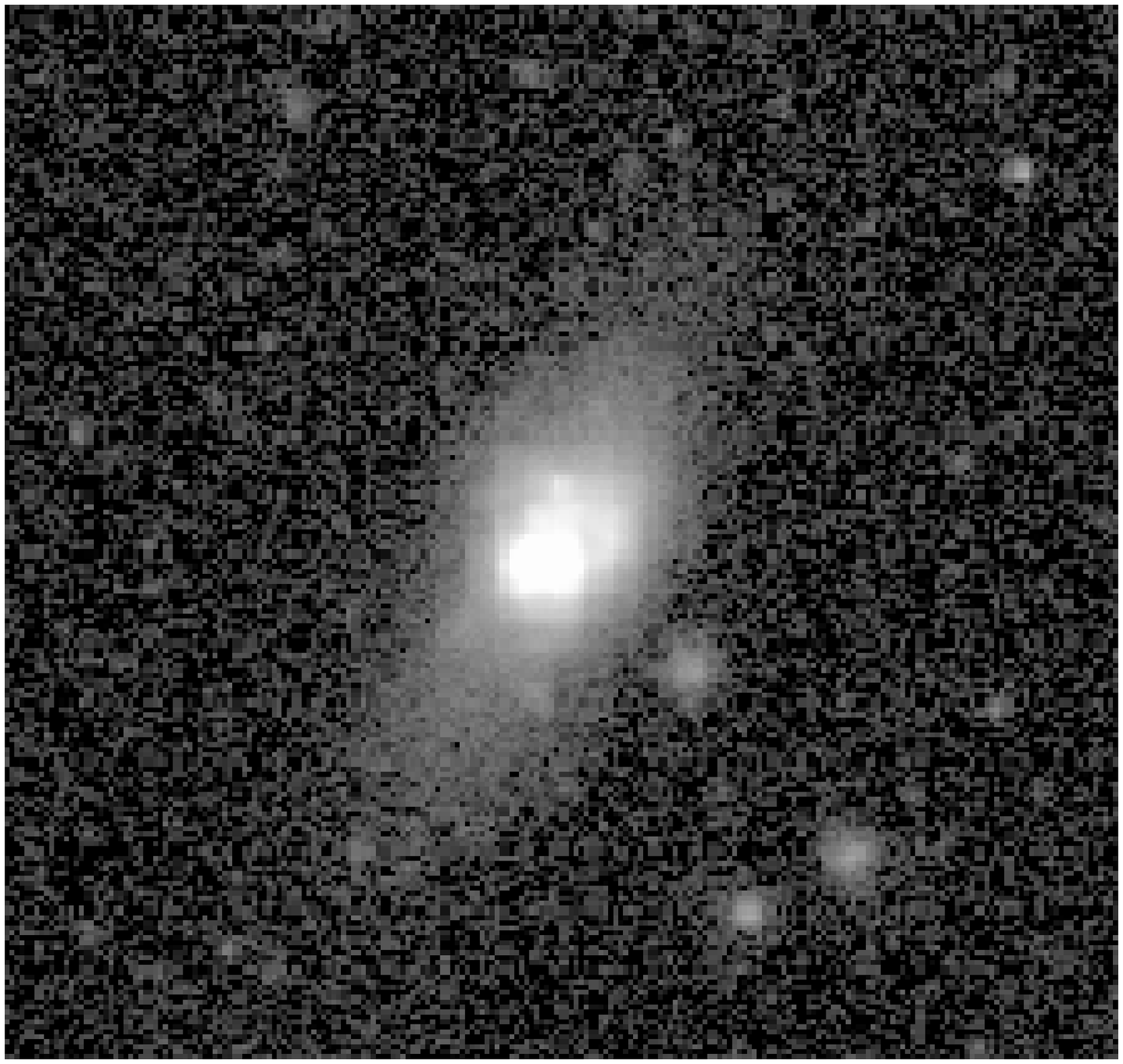}&
\includegraphics[width=5.5cm,angle=0]{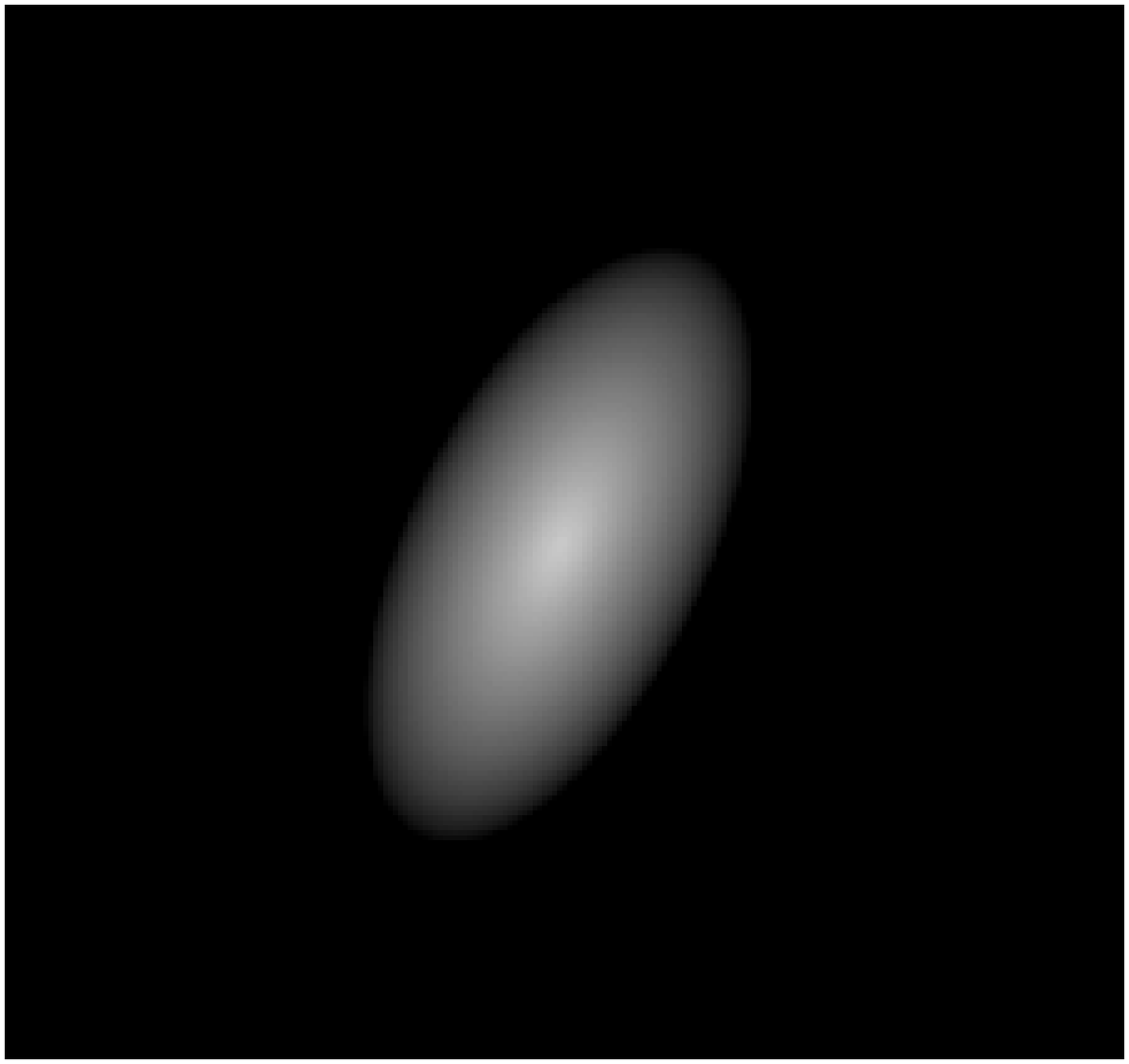}&
\includegraphics[width=5.5cm,angle=0]{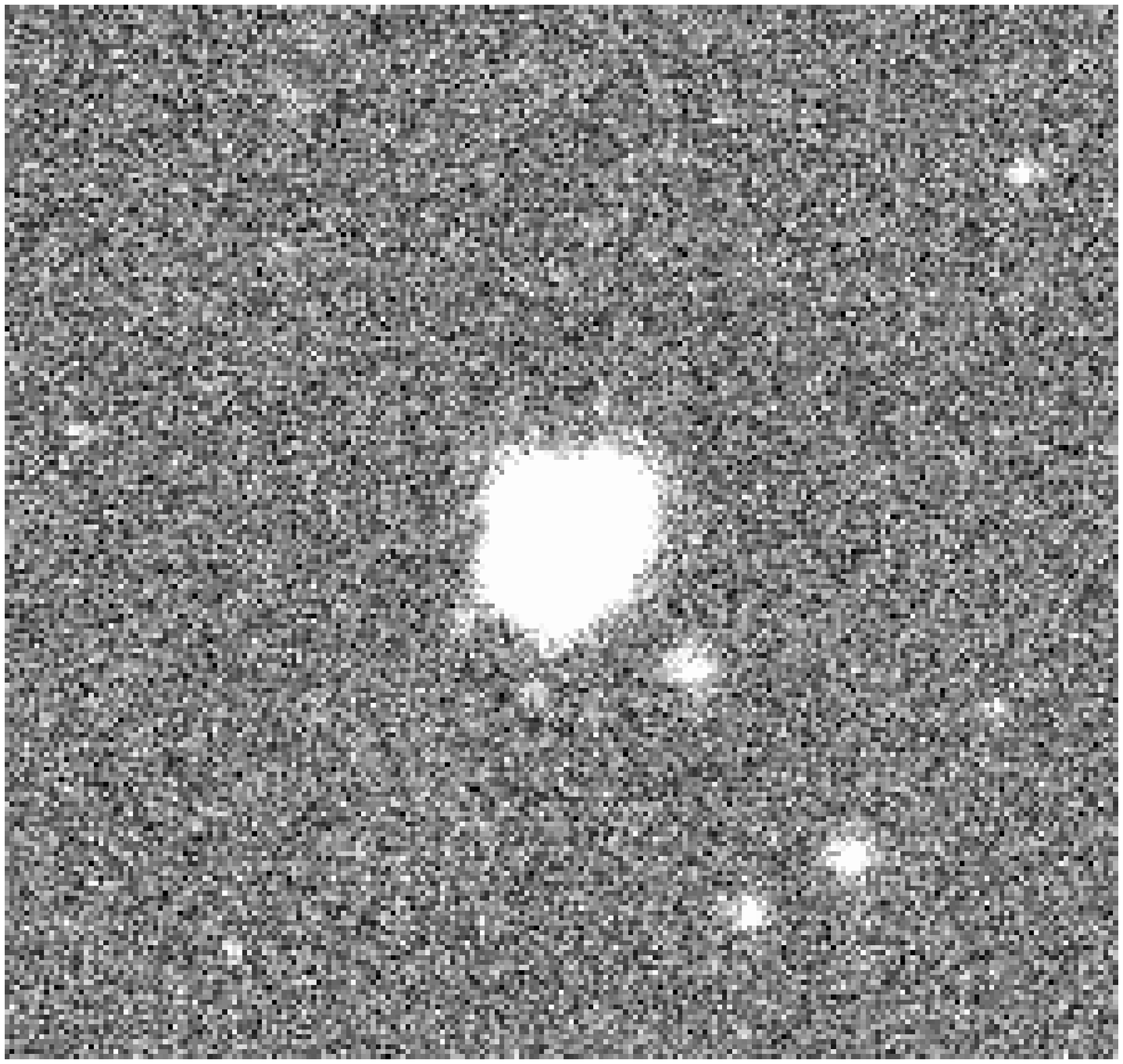}\\
\includegraphics[width=5.5cm,angle=0]{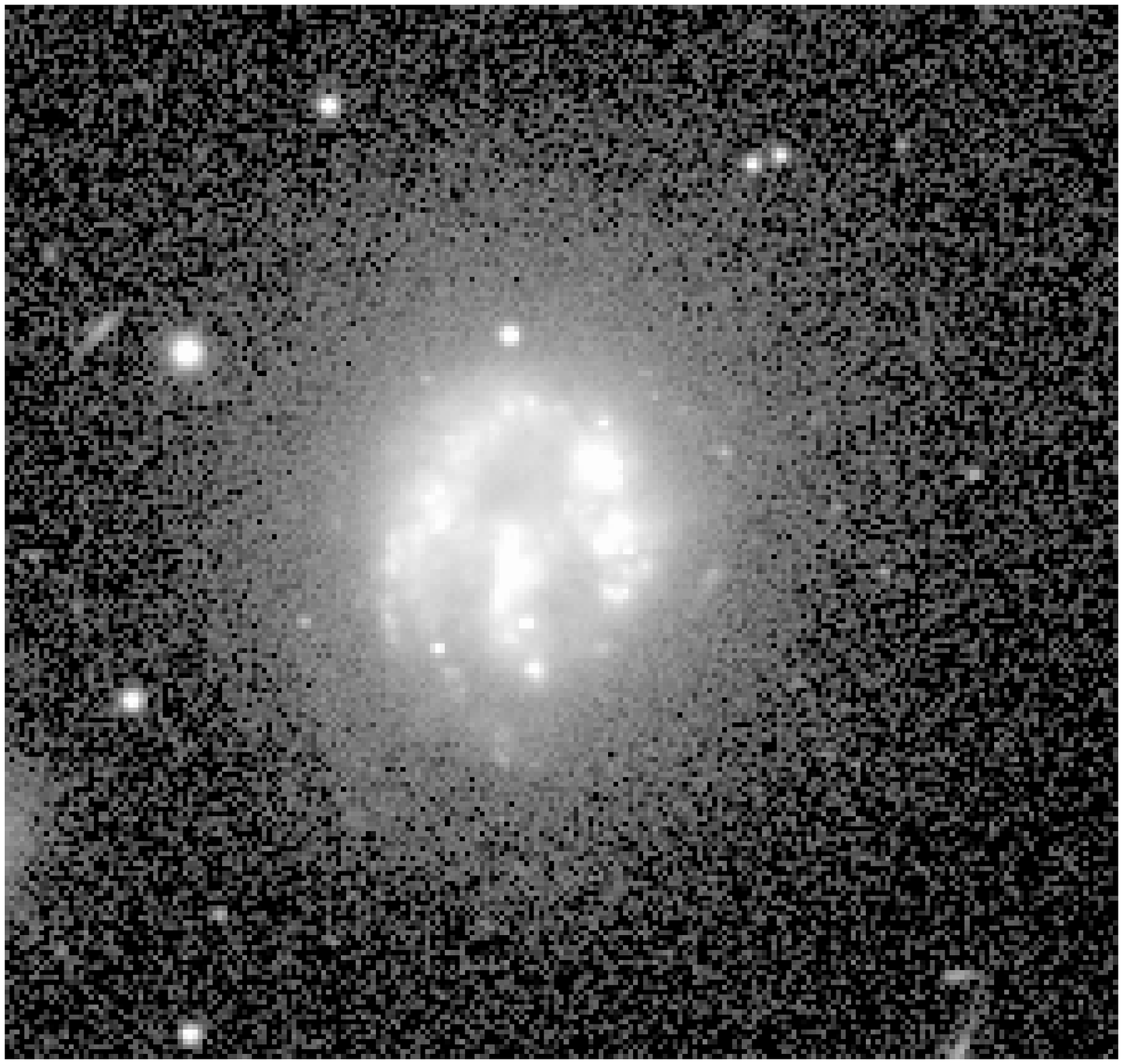}&
\includegraphics[width=5.5cm,angle=0]{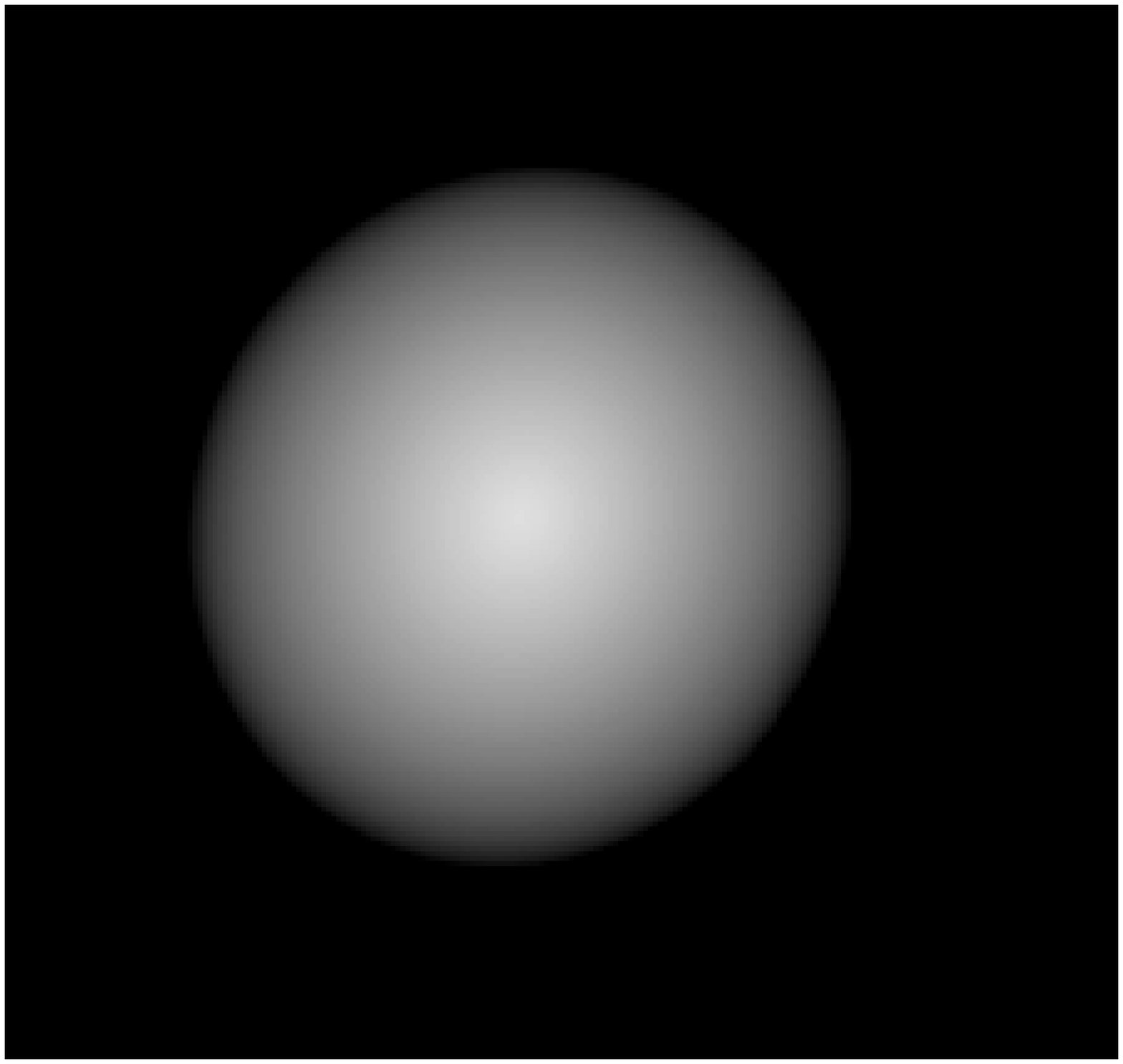}&
\includegraphics[width=5.5cm,angle=0]{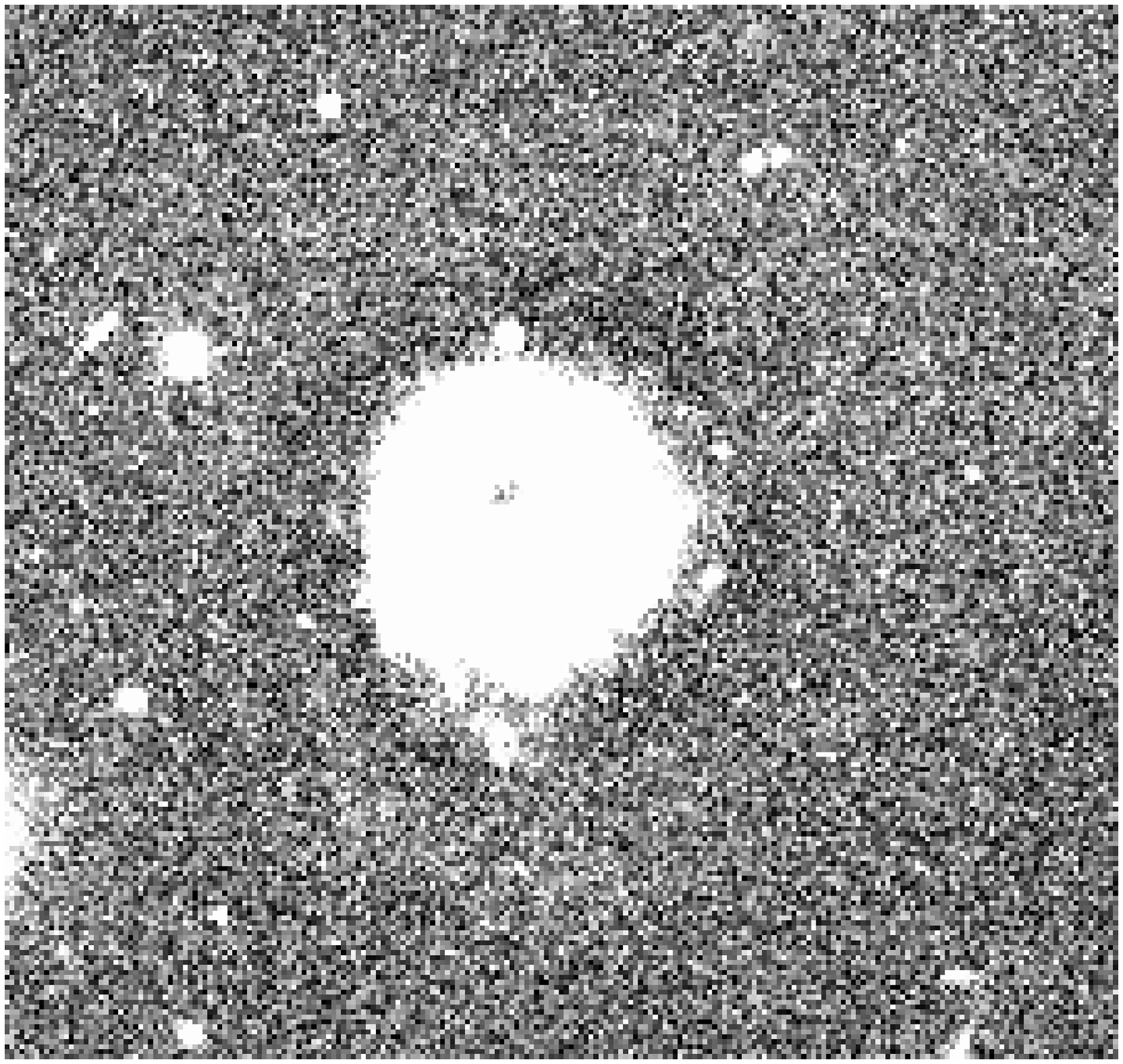}
\end{tabular}
\caption{The galaxy ({\em Left}) and the \mbox{S\'ersic} model 
({\em centre}) in logarithmic intensity grey scale.  
The residual image ({\em right}), in linear grey scale for Tol~127~$B$, Mrk~314~$B$, 
Mrk~36~$B$, and Mrk~86~$B$.}
\label{F8}
\end{figure*}

\begin{figure*}
\begin{tabular}{c c c}
\includegraphics[width=5.5cm,angle=0]{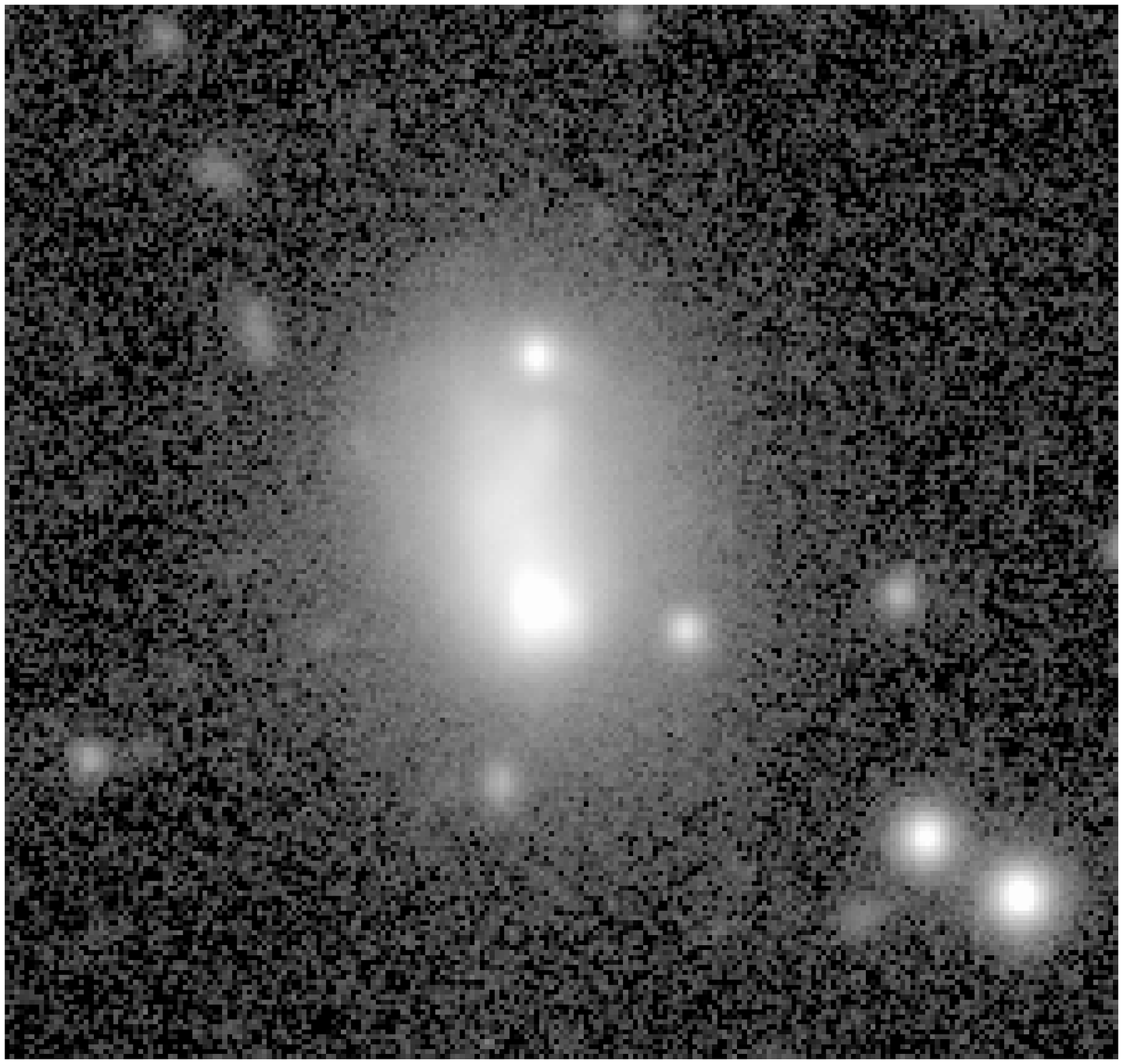}&
\includegraphics[width=5.5cm,angle=0]{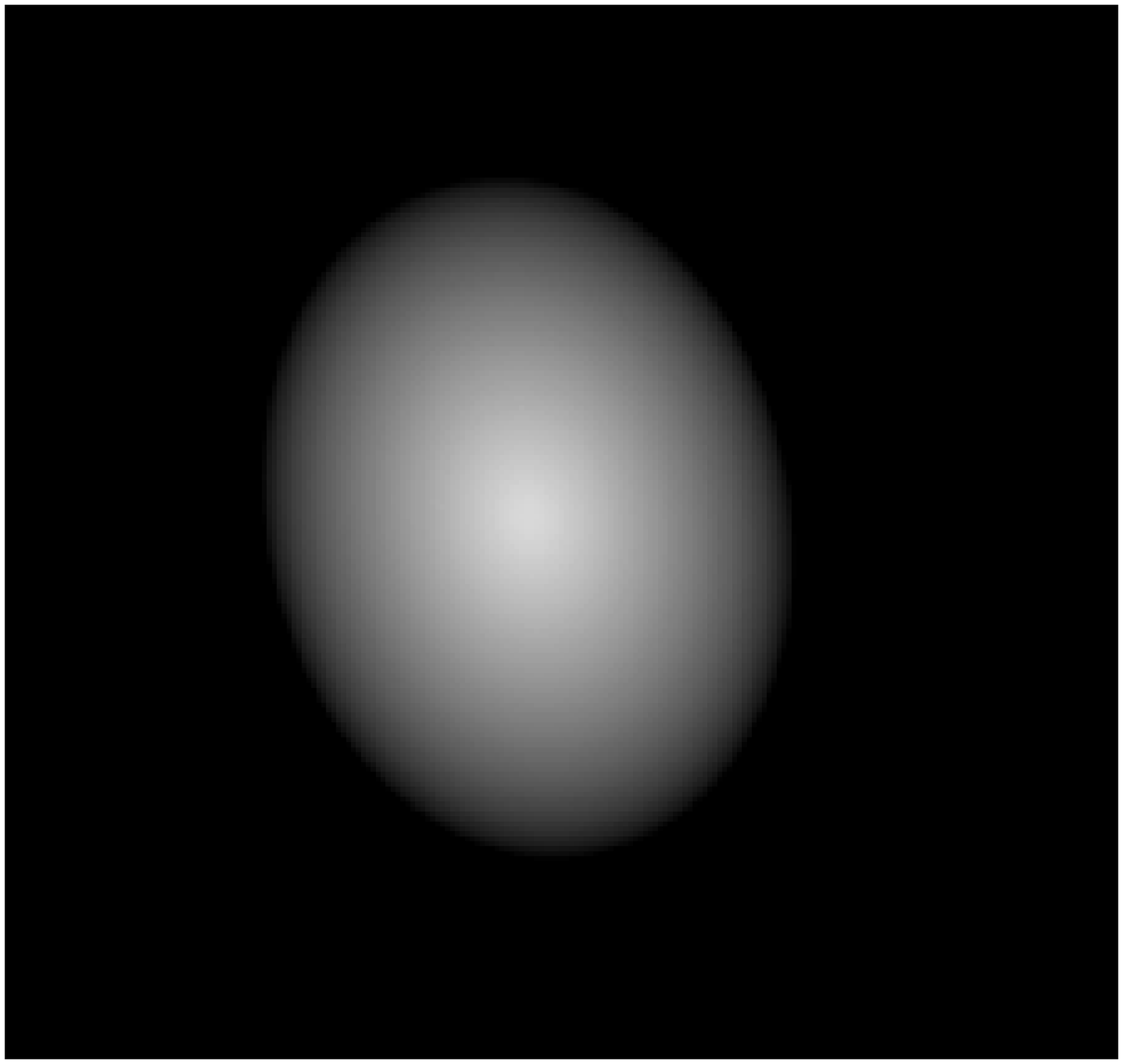}&
\includegraphics[width=5.5cm,angle=0]{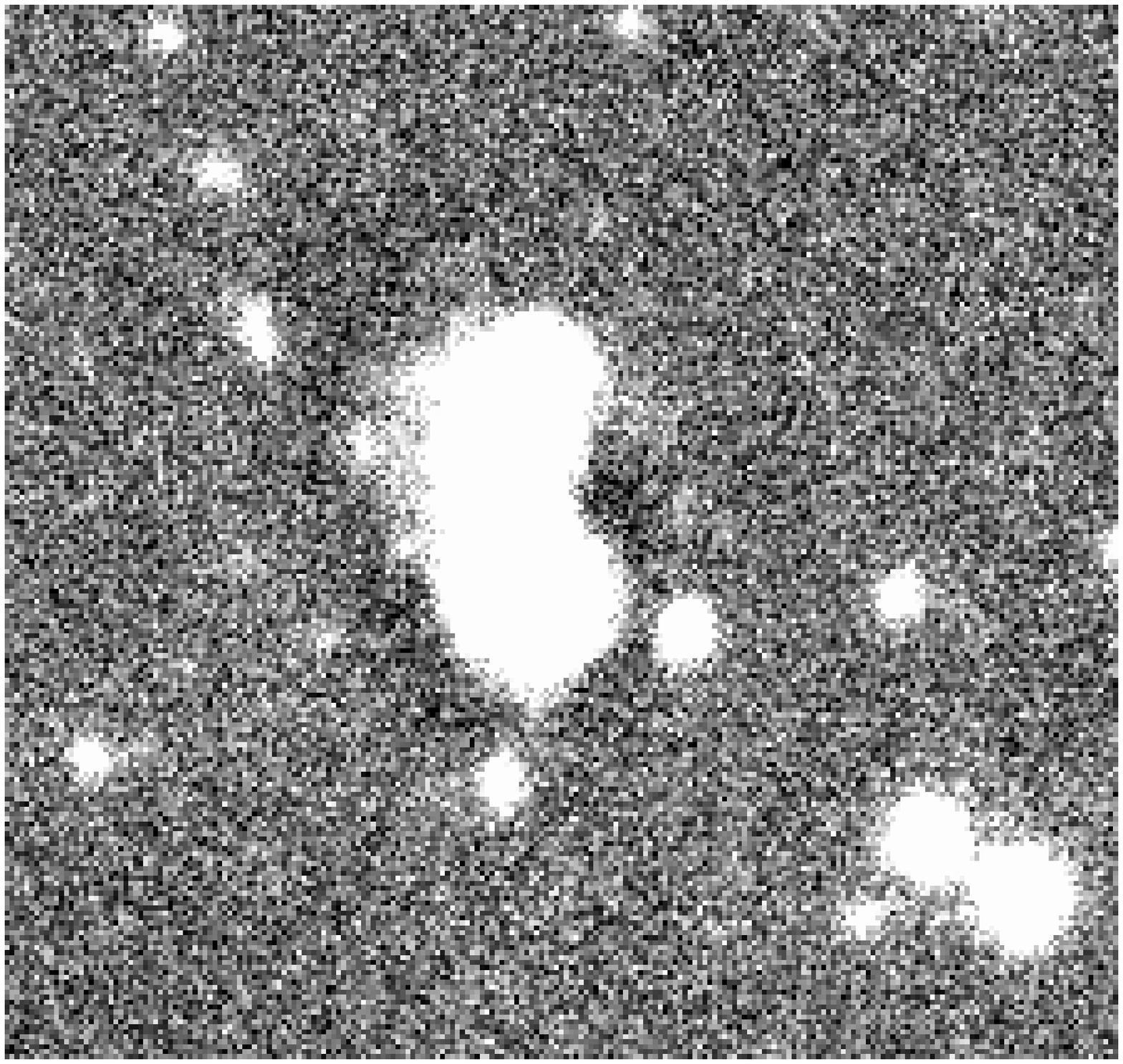}\\
\includegraphics[width=5.5cm,angle=0]{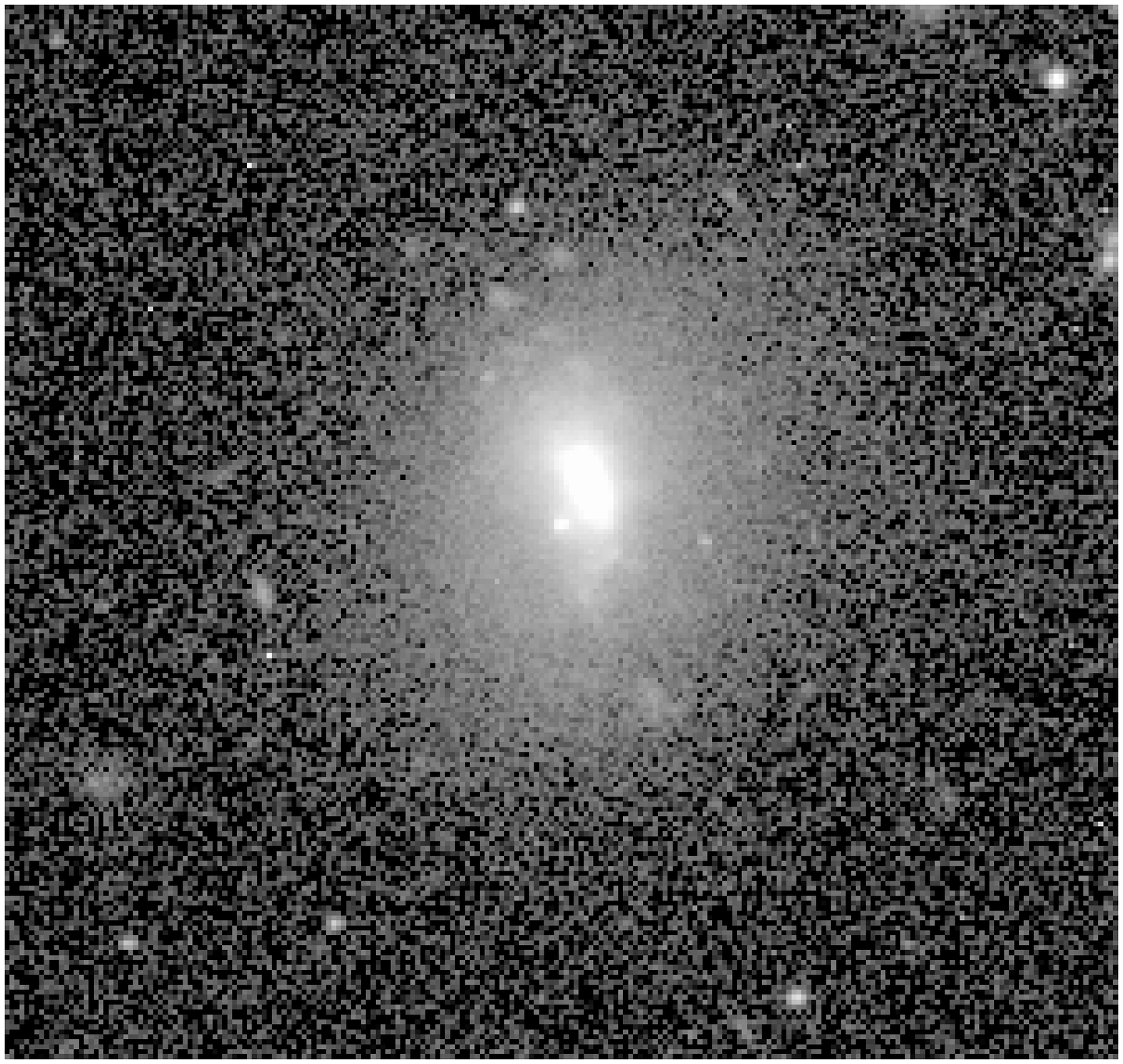}&
\includegraphics[width=5.5cm,angle=0]{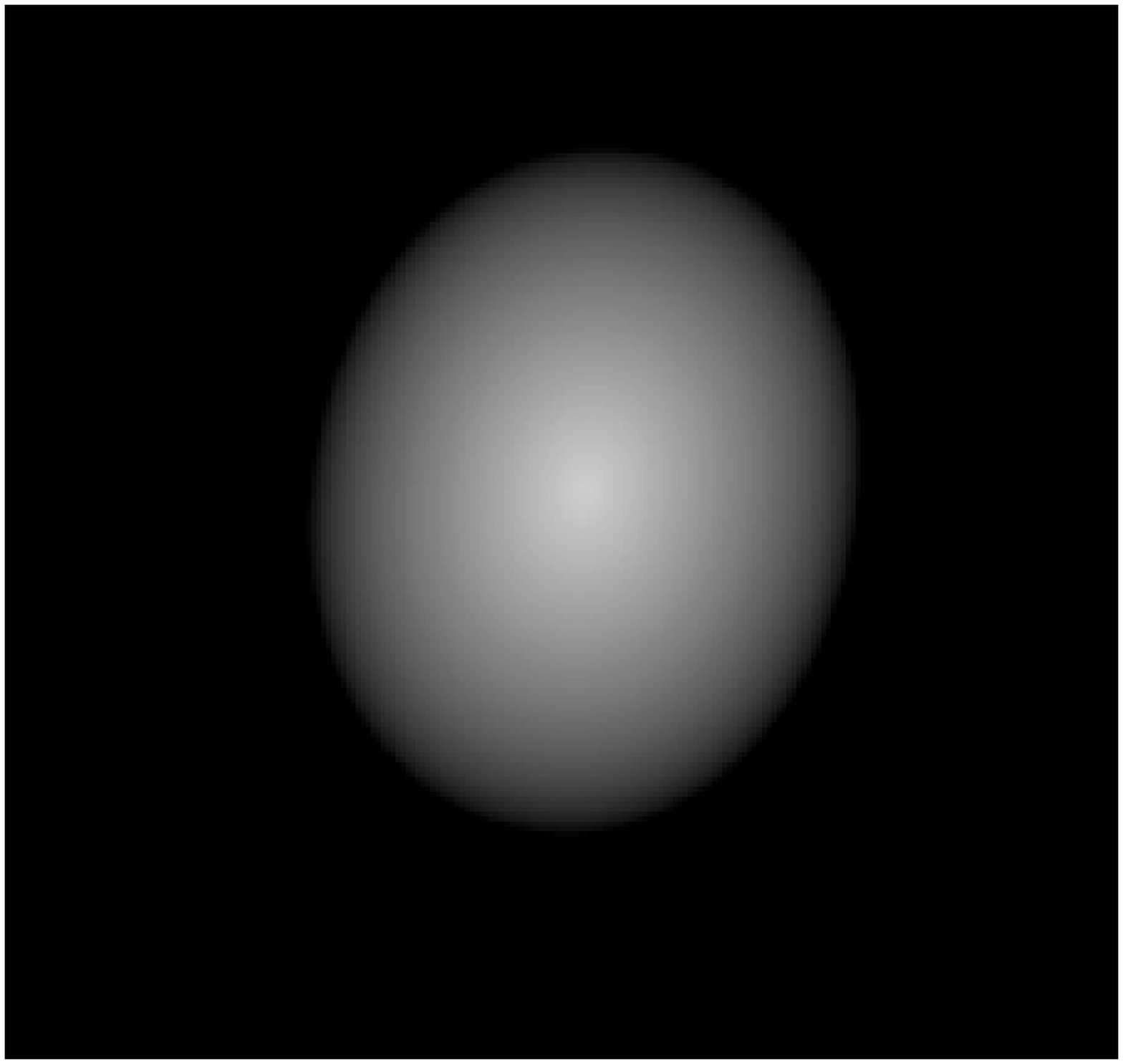}&
\includegraphics[width=5.5cm,angle=0]{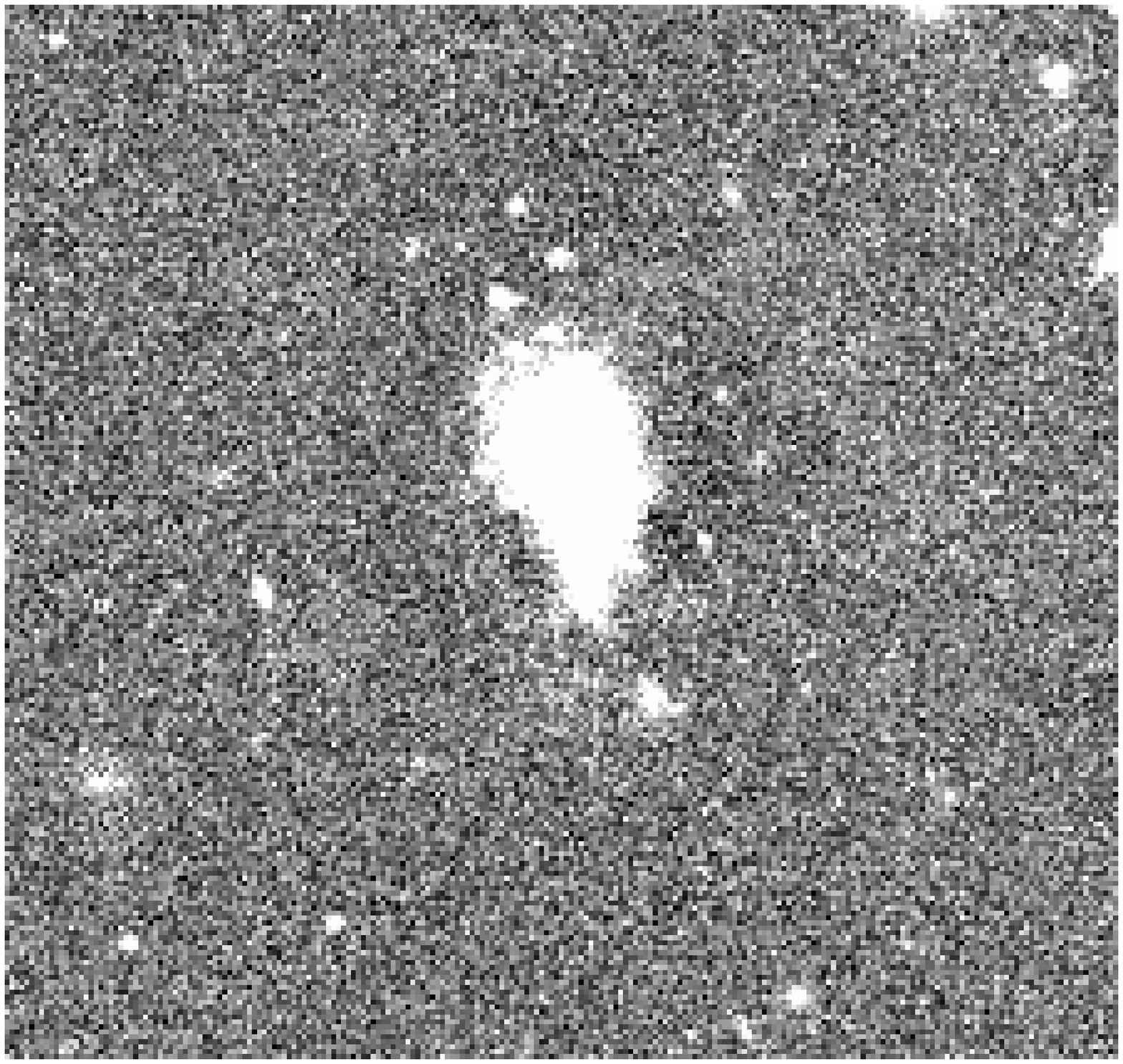}\\
\includegraphics[width=5.5cm,angle=0]{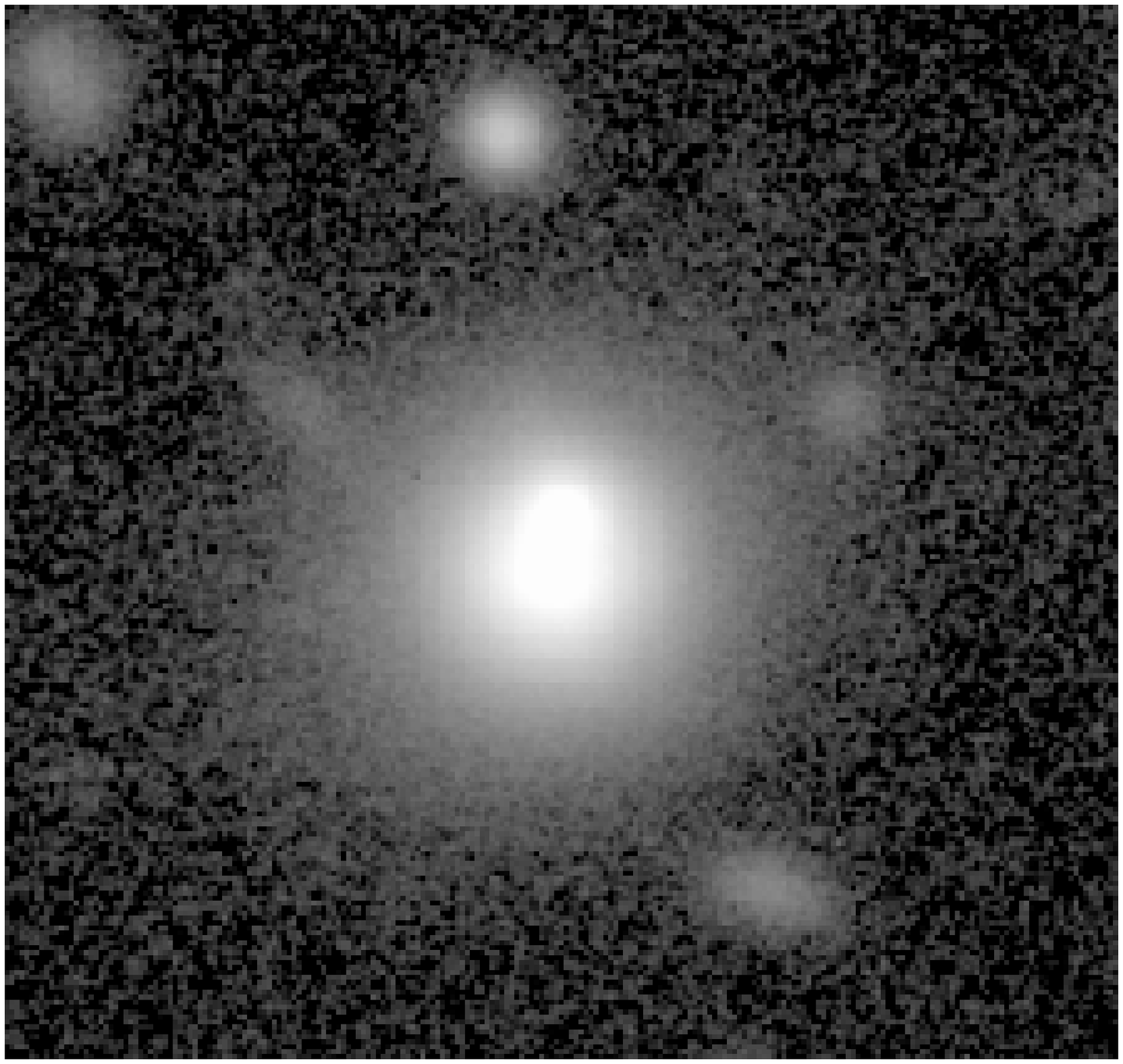}&
\includegraphics[width=5.5cm,angle=0]{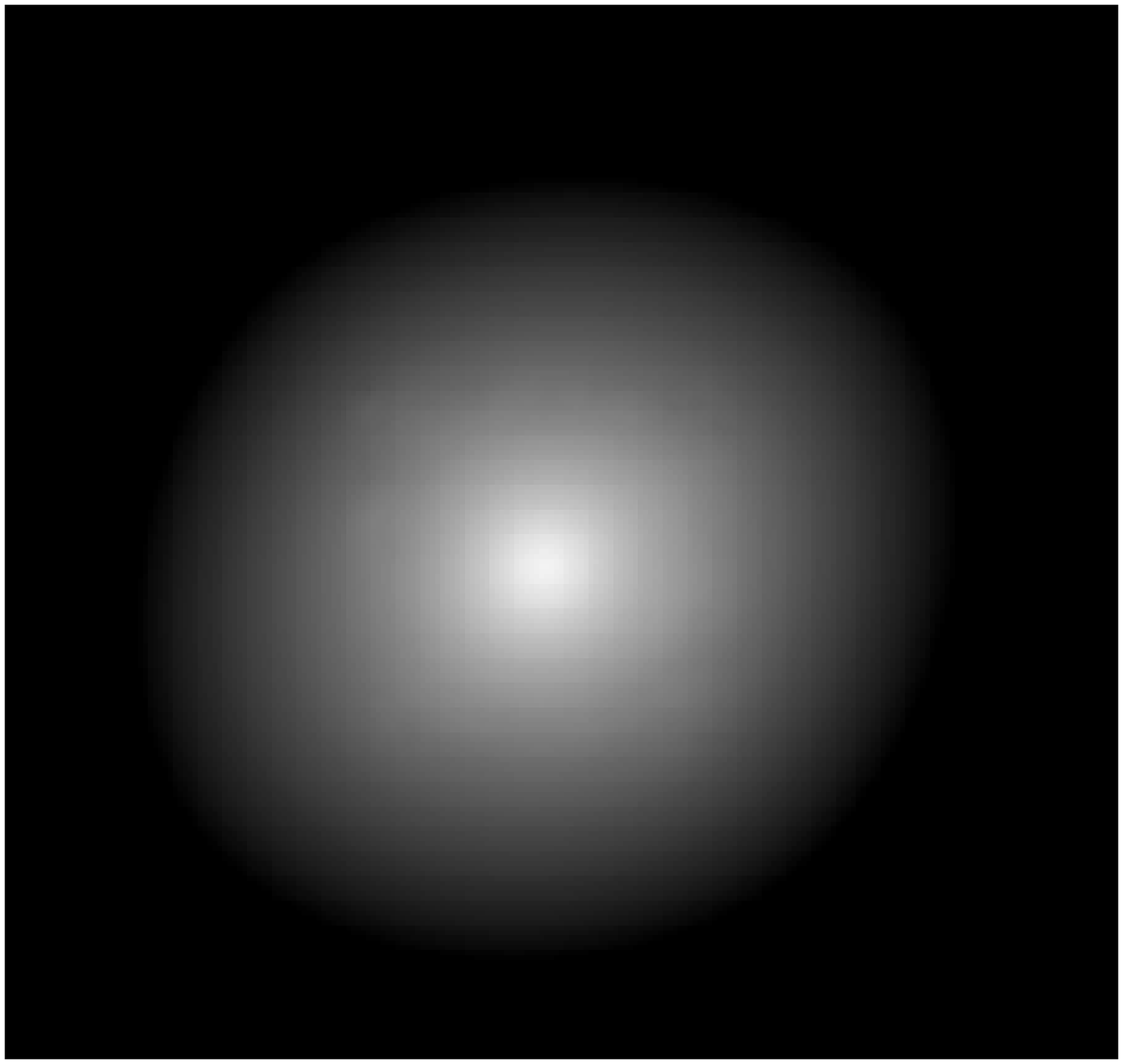}&
\includegraphics[width=5.5cm,angle=0]{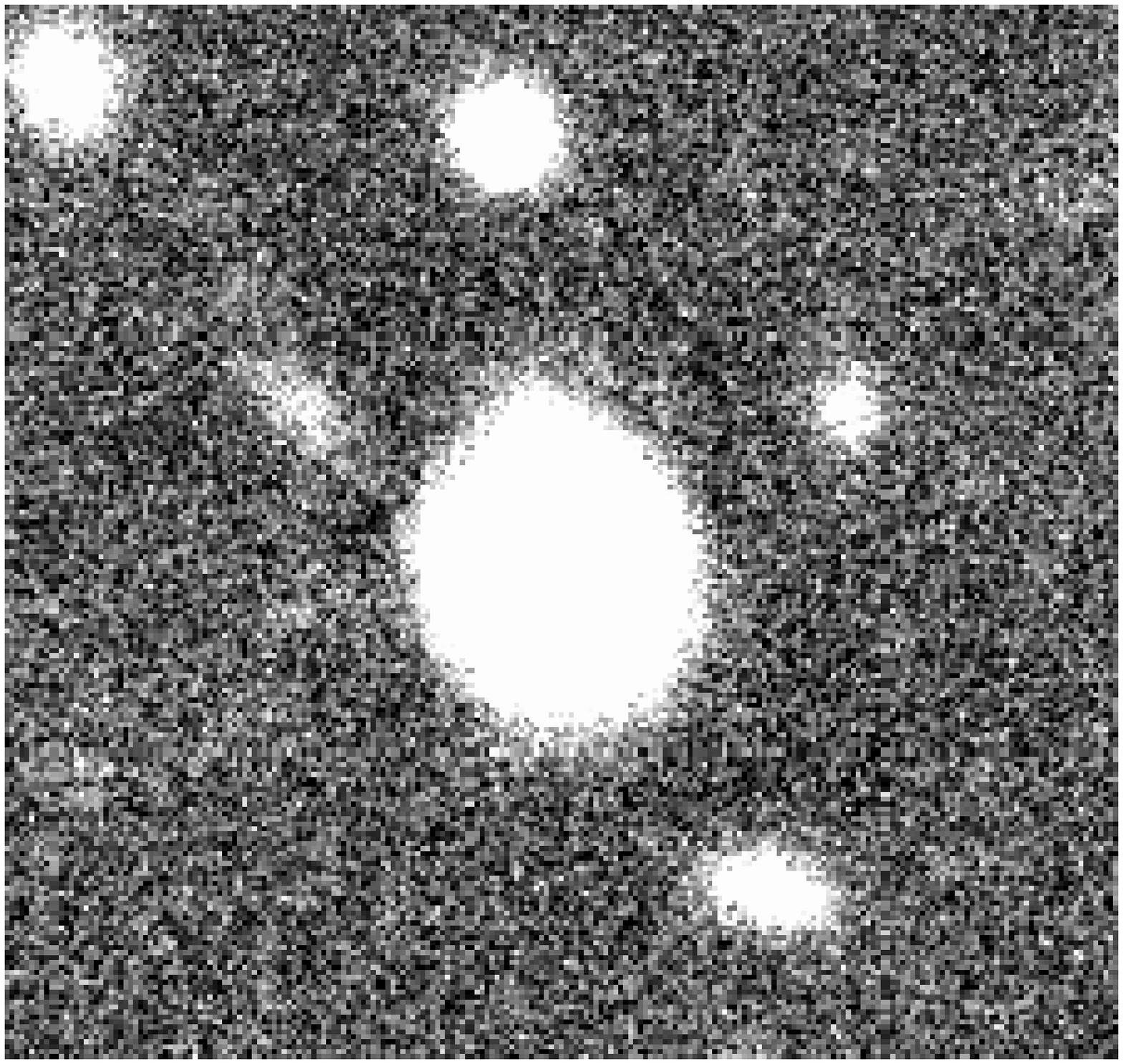}\\
\includegraphics[width=5.5cm,angle=0]{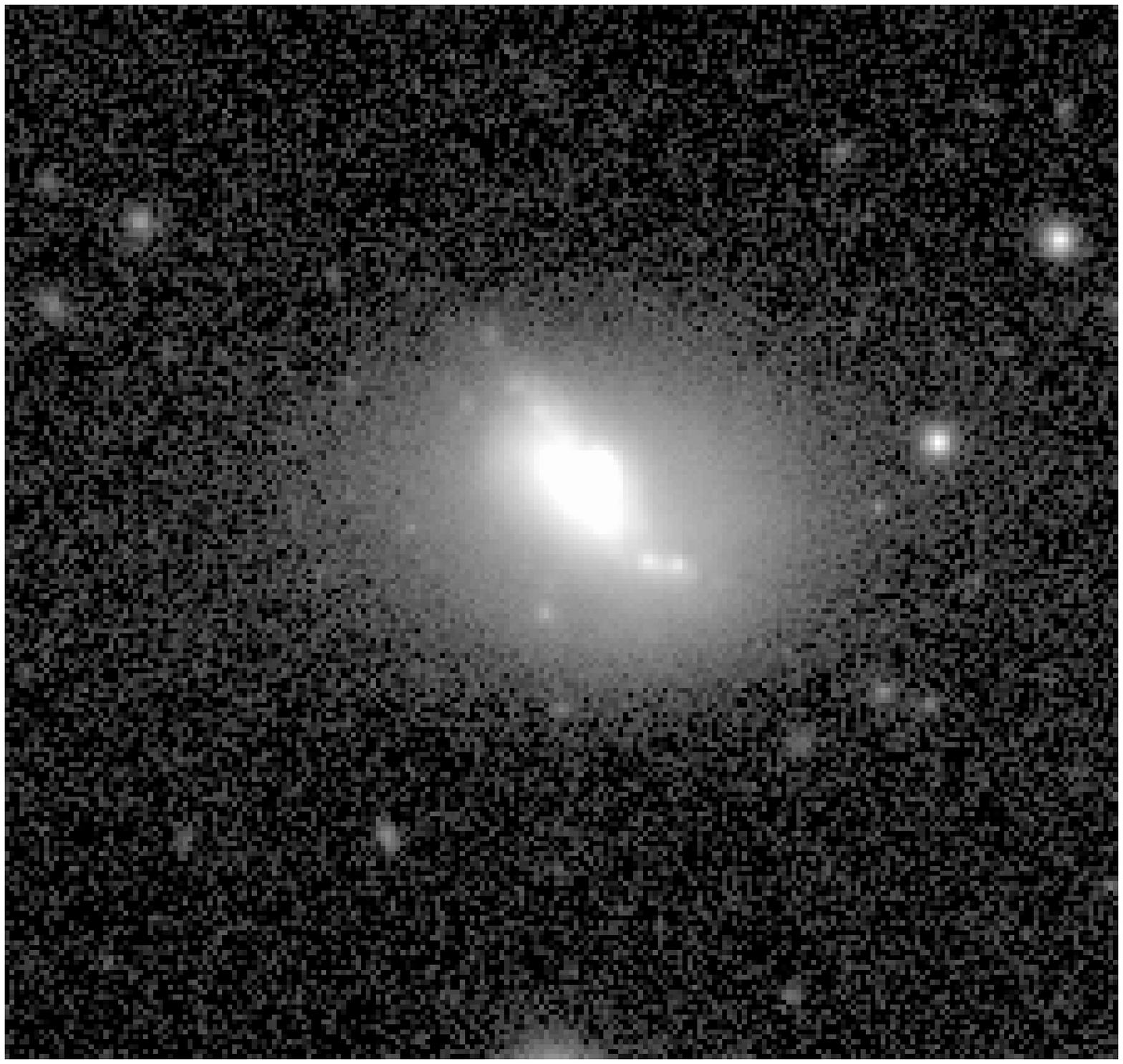}&
\includegraphics[width=5.5cm,angle=0]{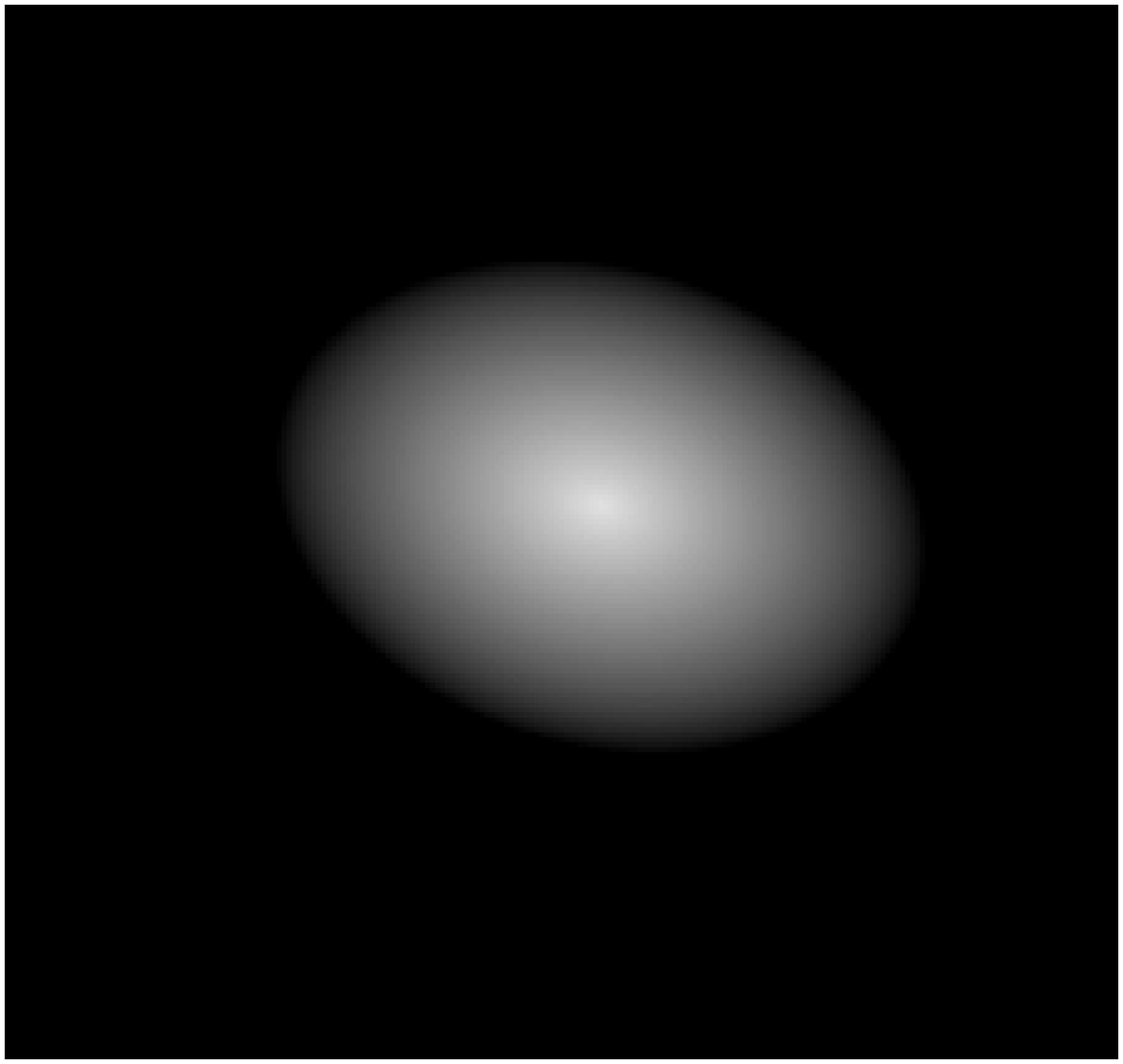}&
\includegraphics[width=5.5cm,angle=0]{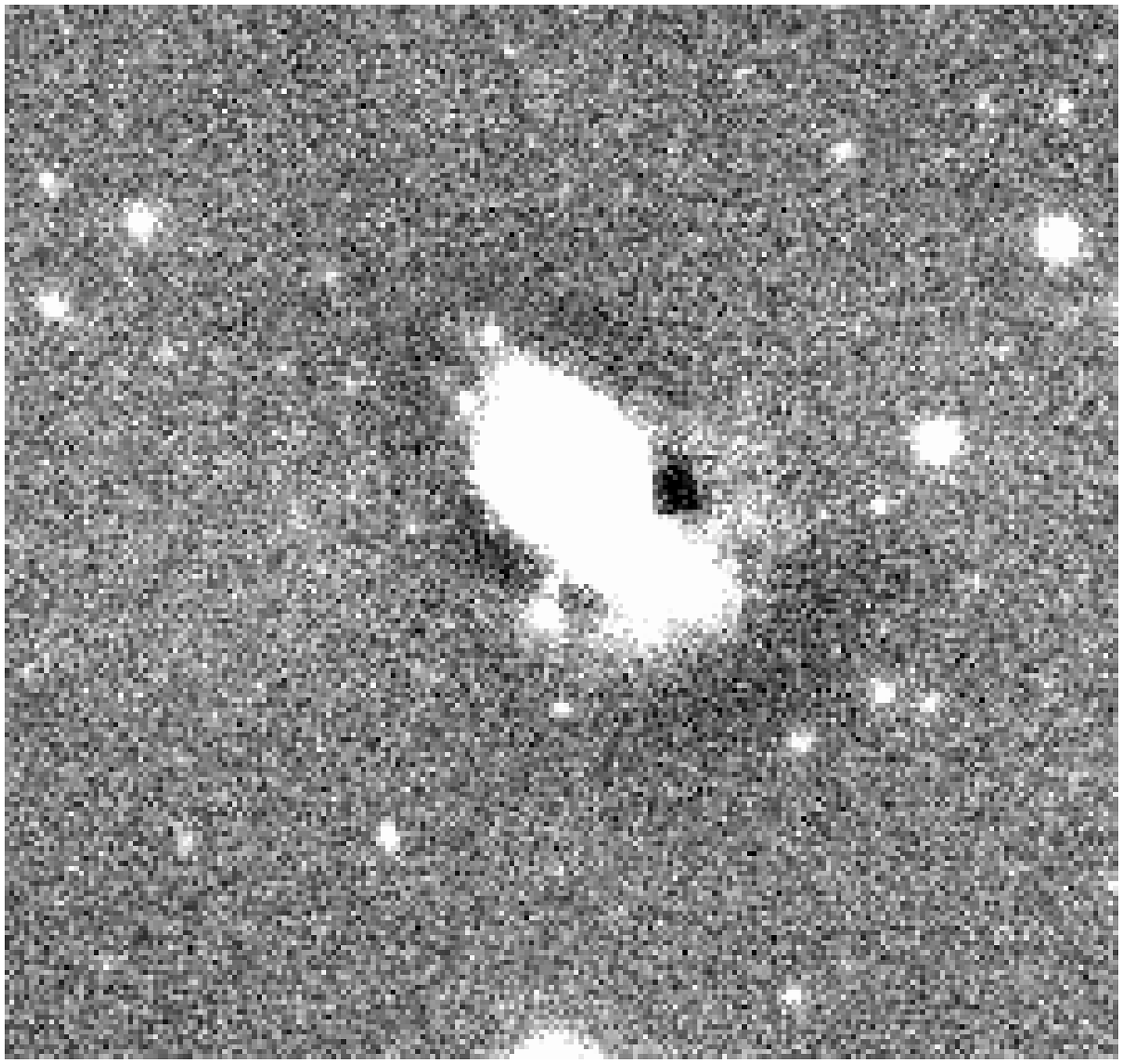}
\end{tabular}
\caption{The galaxy ({\em Left})
and the \mbox{S\'ersic} model ({\em centre}) in logarithmic intensity grey scale.
The residual image ({\em right}) in linear grey scale for Mrk~5~$B$, Mrk~370~$B$, 
I~Zw~123~$B$, and Mrk~35~$B$.}
\label{F9}
\end{figure*}

We discuss our results in the following subsections, comparing them with those 
obtained in Paper~I. In Fig.~\ref{F8} and Fig.~\ref{F9} we show three images for each 
galaxy: (left) the galaxy, (center) the \mbox{S\'ersic} model, and (right) 
the residual image, all in logarithmic intensity grey scale. 

\subsection{Comparison with previous 1D fitting}

This paper confirms that low $n$ \mbox{S\'ersic} models are able to reproduce, 
within the uncertainties, the surface brightness distribution in these three 
pass-bands in all galaxies of the sample. This may indicate that, for 
this sample, the hosts resemble a disc-like --- exponential --- structure.

\begin{figure*}
\begin{tabular}{l l l}
\includegraphics[width=4.65cm,angle=0]{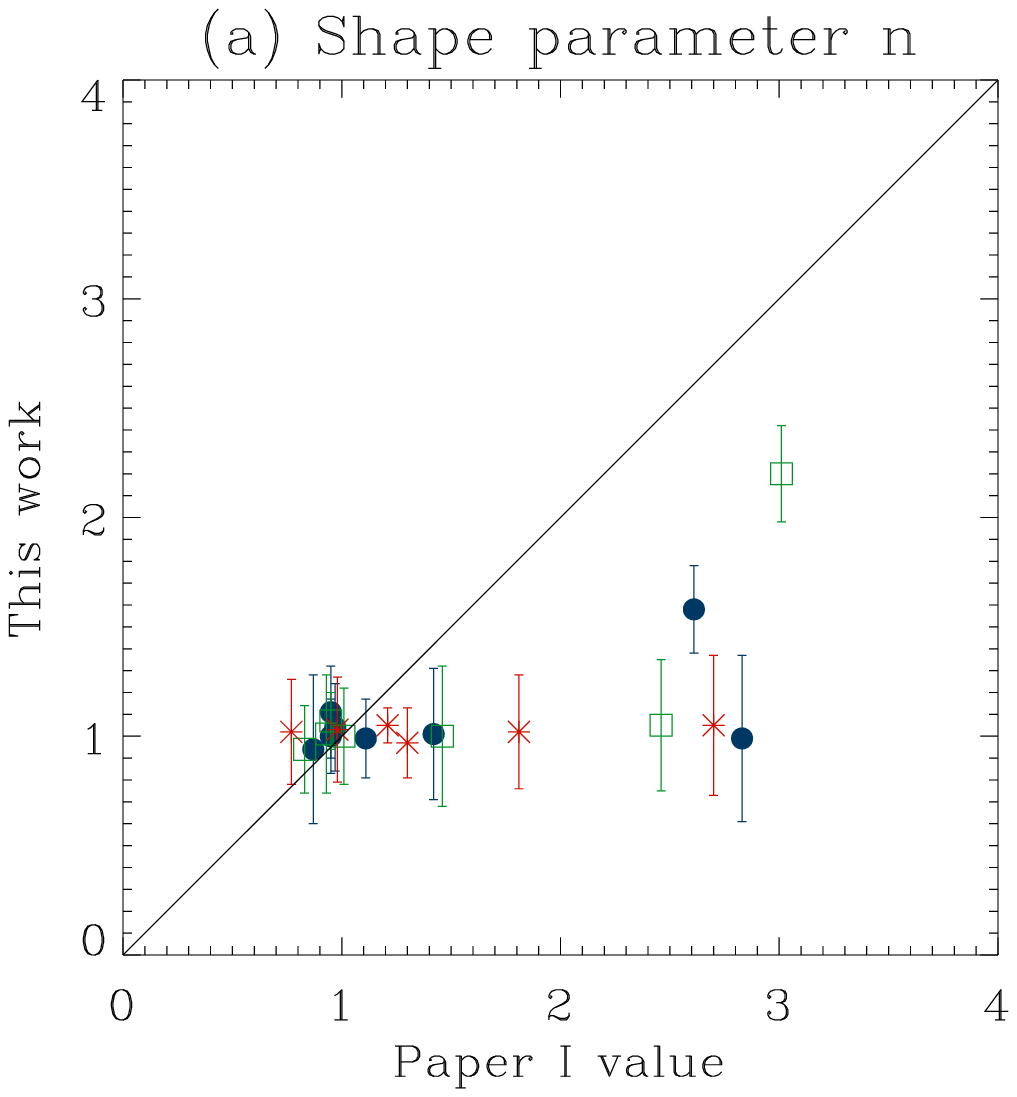}&
\includegraphics[width=4.90cm,angle=0]{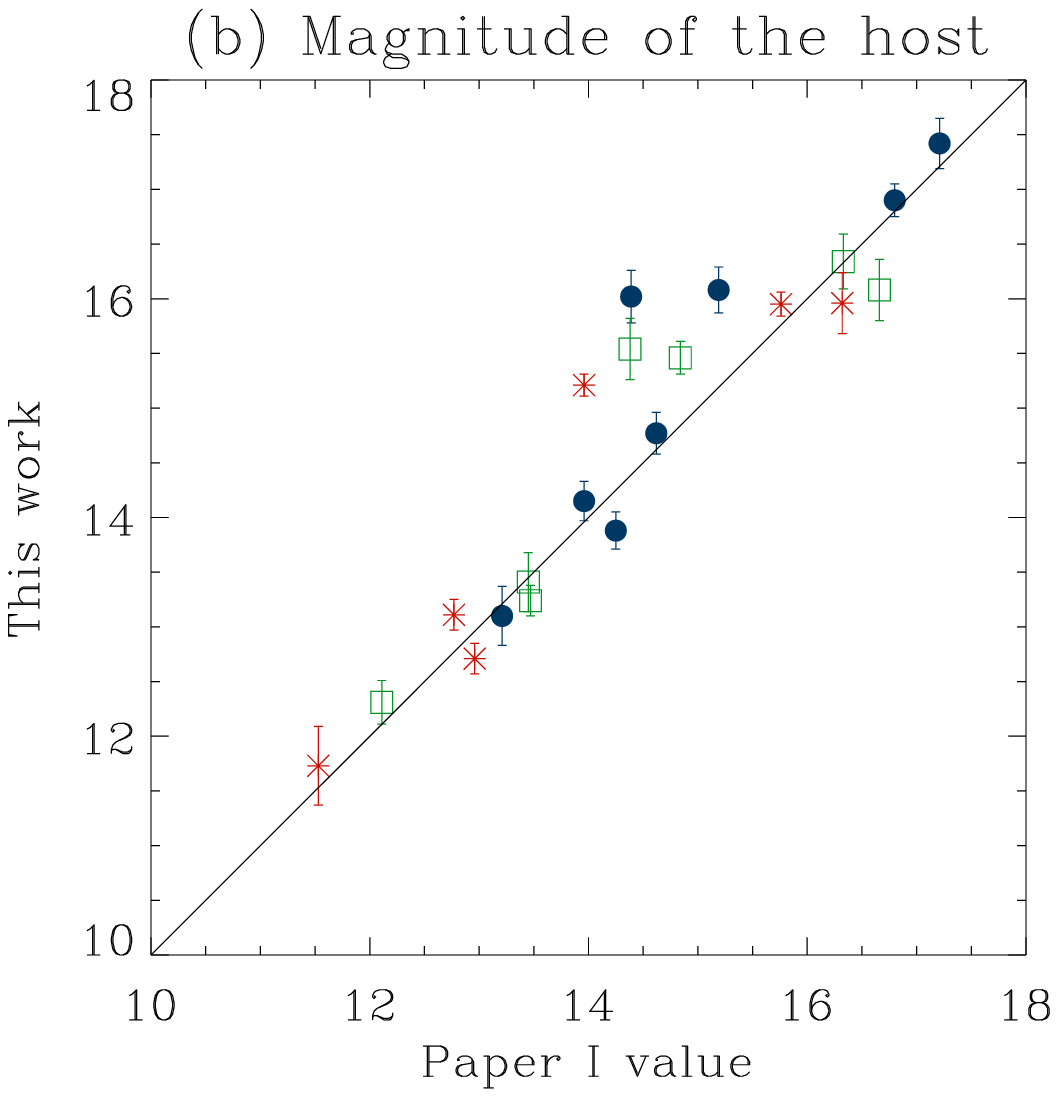}&
\includegraphics[width=5.05cm,angle=0]{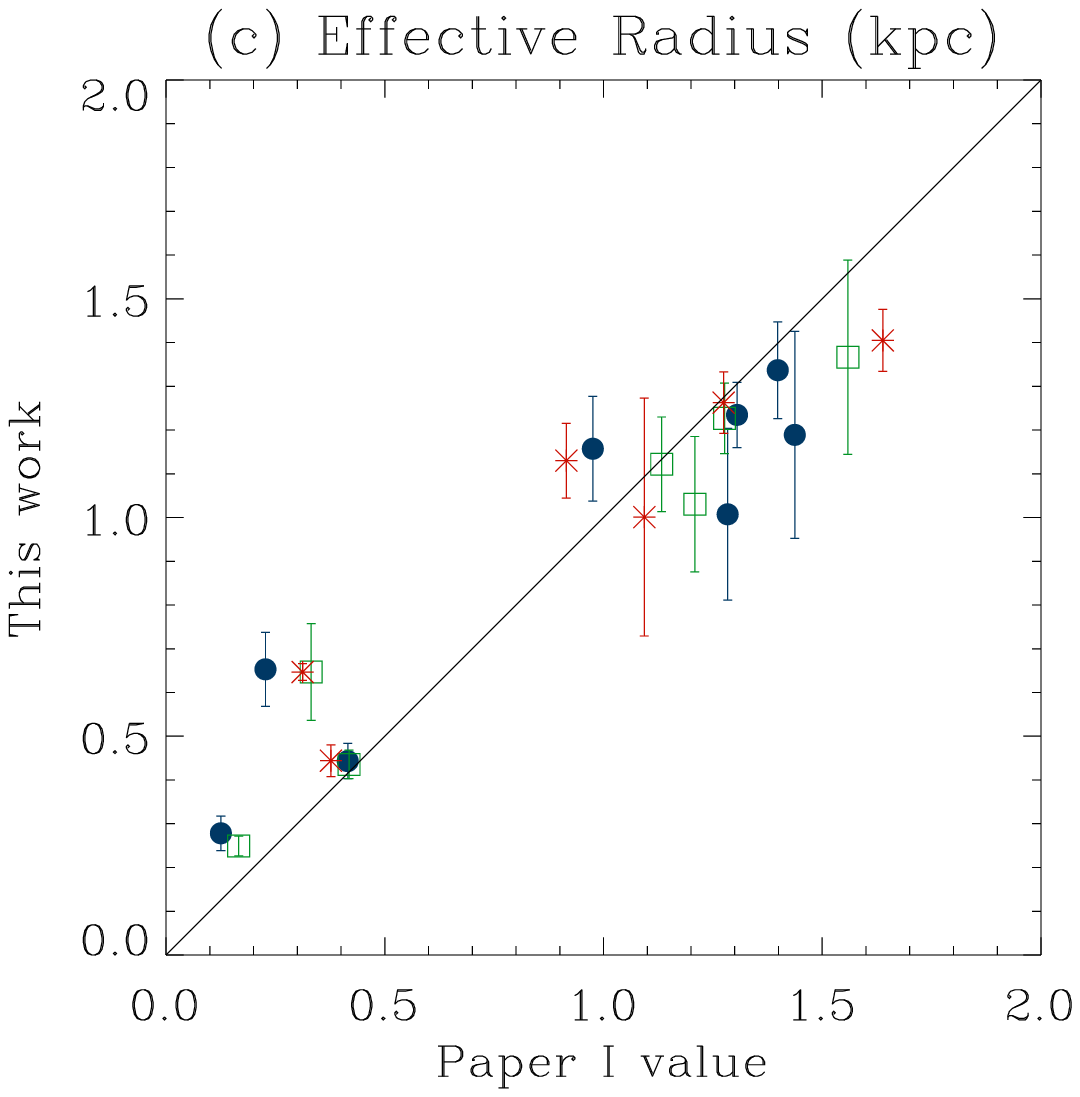}\\
\includegraphics[width=5.25cm,angle=0]{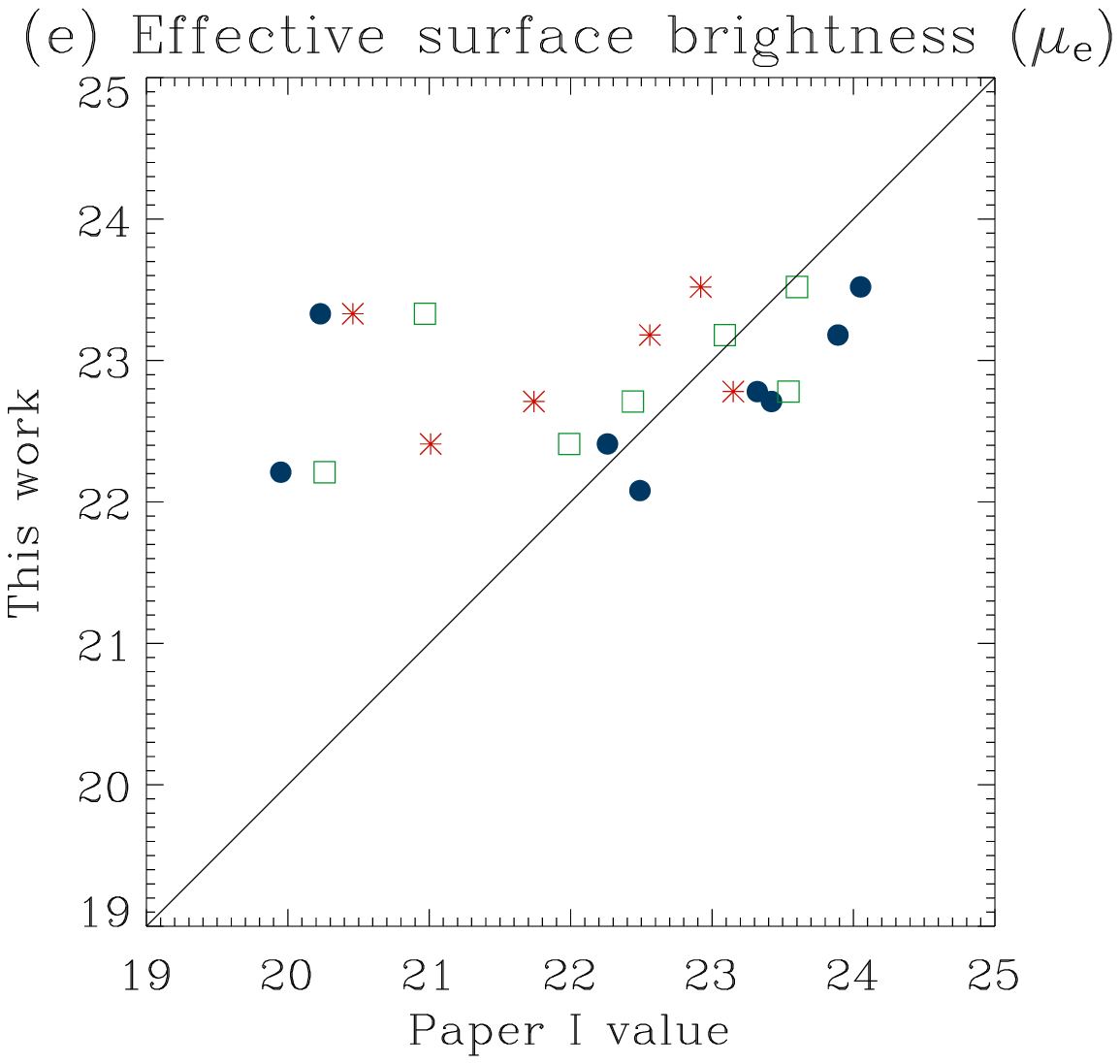}&
\includegraphics[width=5.55cm,angle=0]{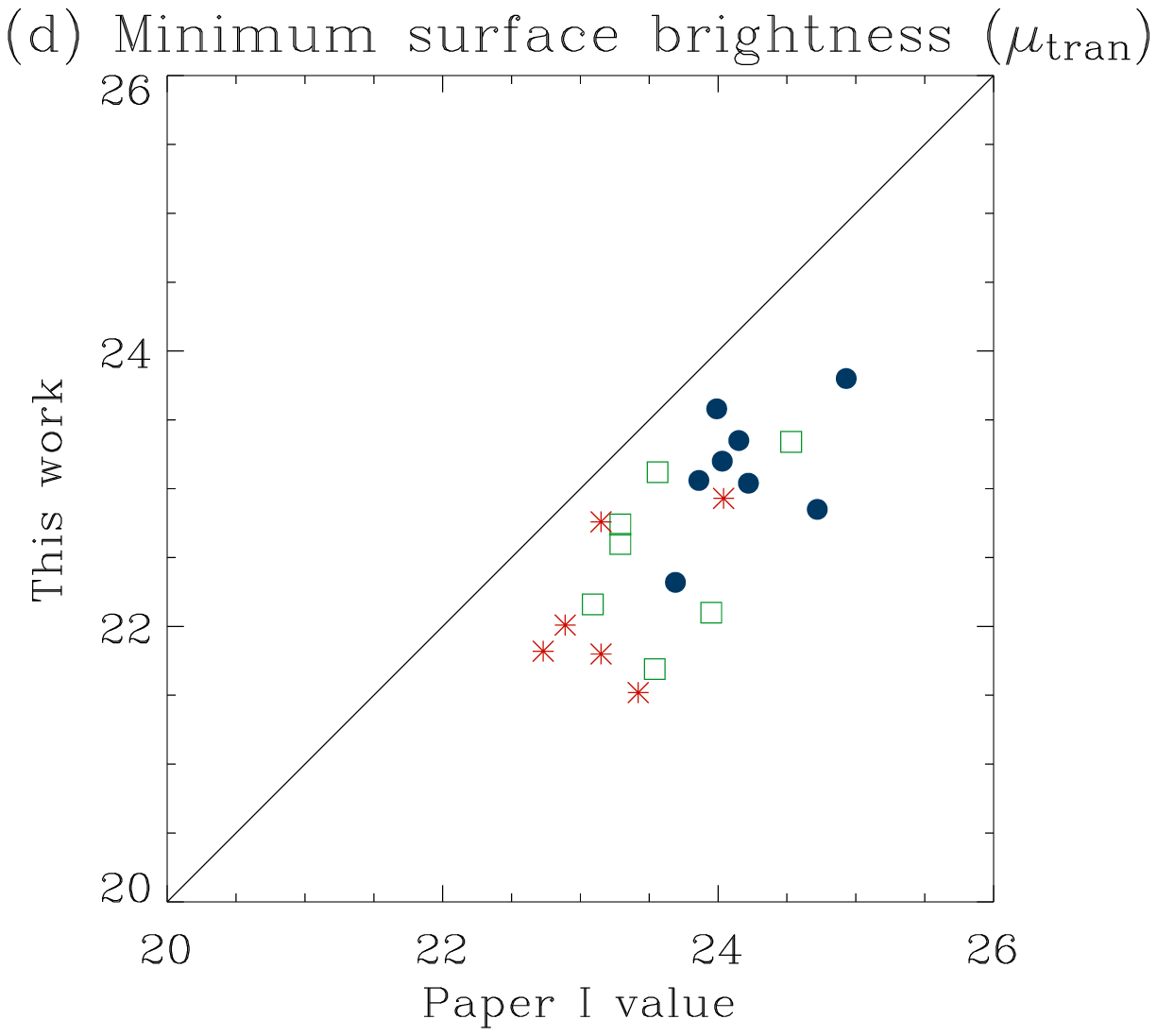}&
\includegraphics[width=4.70cm,angle=0]{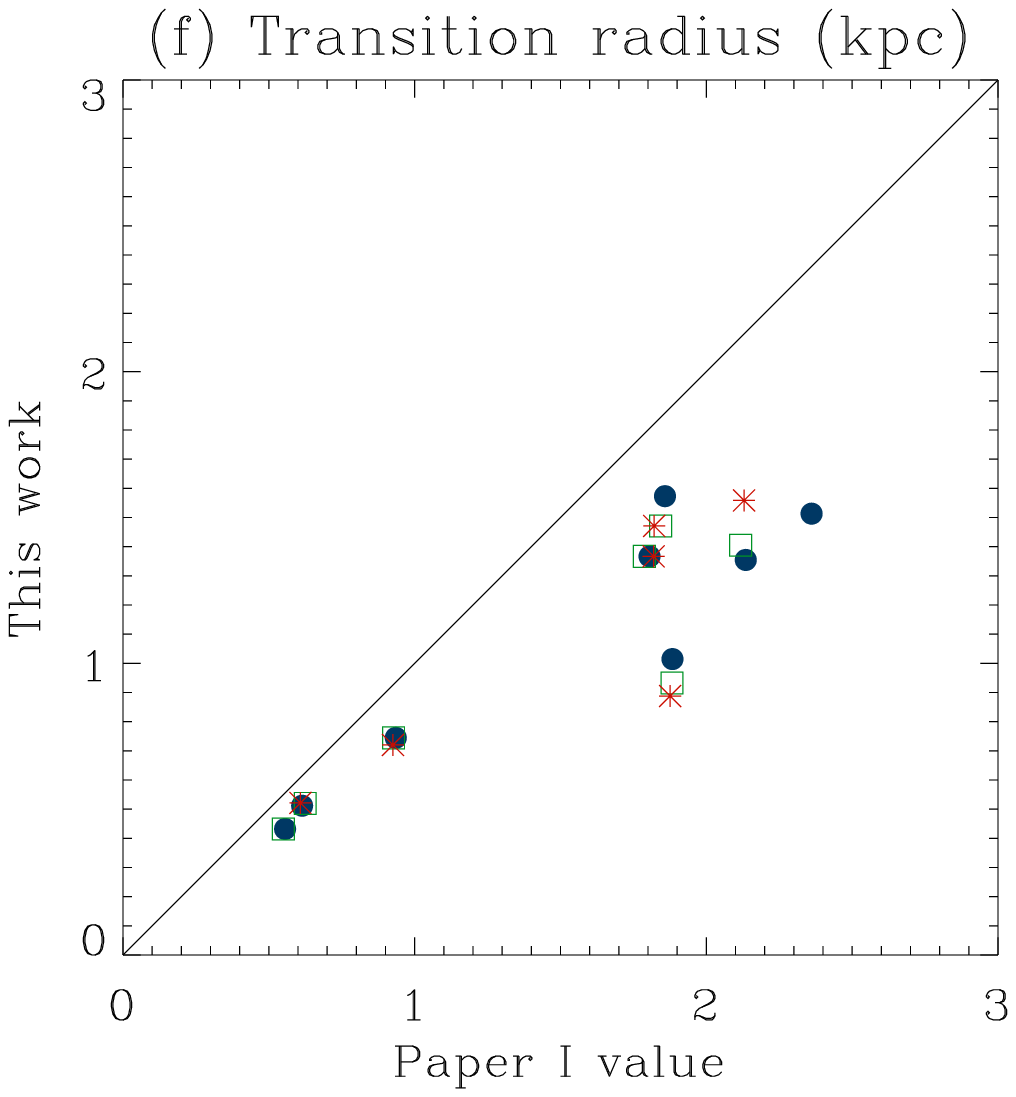}
\end{tabular}
\caption{Comparison between the \mbox{S\'ersic} parameters in the three bands 
(blue dots=$B$, green squares=$V$, and red asterisks=$R$) derived in this work with those 
reported in Paper~I: a) shape parameter $n$, b) LSB host total magnitude, 
c) effective radii (kpc), d) effective surface brightness 
(mag arcsec$^{-2}$), e) minimum surface brightness (mag arcsec$^{-2}$), and f) transition 
radius (kpc).}  
\label{F10}
\end{figure*}

\begin{figure*}
      \centering
      \includegraphics[width=18.0cm,angle=90]{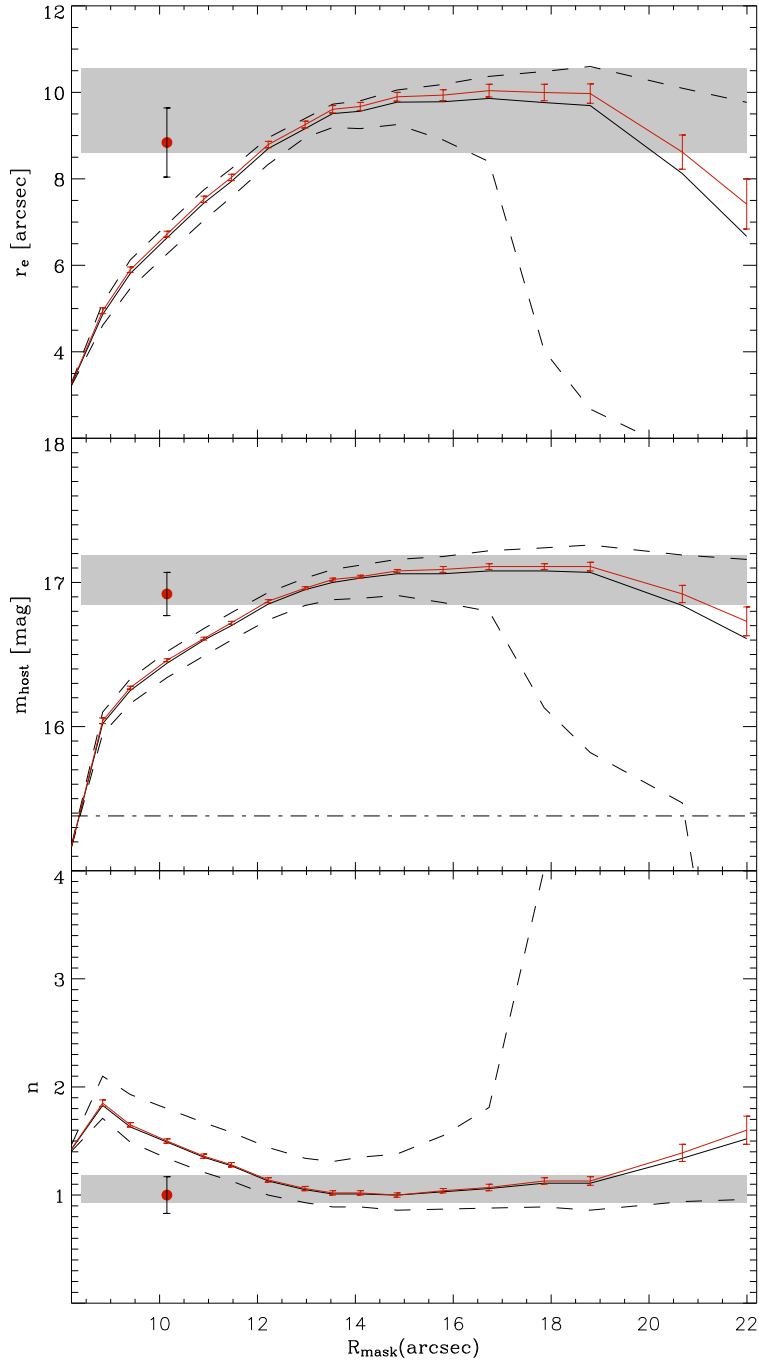}
      \caption{Stability in the 2D fit of Mrk~36 ($B$ band): the parameters $R_{\rm e}$, 
	$m_{\rm tot}$ and $n$, are plotted as a function of $R_{\rm mask}$ in the $B$ band. 
	Red lines indicates fits for which the sky-background was a free parameter.
	Black lines indicates fits with a fixed sky-background level in our best estimation. 
	The error bars (only statistical uncertainties) were put only in the red curve since 
	they are the bigger.
	Dashed lines indicates fits with a fixed sky-background in $<sky> \pm \sigma$, where 
	$\sigma$ is the uncertainty in the sky estimation. Dots are the final fits. 
	The horizontal dashed-dotted line in the $m_{\rm tot}$ plot indicates the magnitude 
	of the whole galaxy (host + starburst).
	Their error bars represent the quadratic sum of uncertainties from both dispersion in 
	the stability range and from the sky subtraction (see text for details). 
	We notice that there is a radial interval,$12\leq R_{\rm mask}\leq 19$, within which 
	the \mbox{S\'ersic} parameters are quite stable in the range indicated by the grey band 
	(with deviations of $\sim 1''$ in $R_{\rm e}$, $\sim 0.17$ in $m_{\rm tot}$, and 
	$\sim 0.14$ in $n$). 
	Our best fits are within the uncertainties and have a transition radius 
	(when $R_{\rm mask}=R_{\rm tran}$) significantly smaller than those obtained by using 
	the elliptical masks.}
      \label{F11}
    \end{figure*}

We compared our results with those of Paper~I.
Fig.~\ref{F10} shows the model parameters, $n$, \mbox{$r_{\rm e}$}, $m_{\rm tot}$, 
and $\mu_{\rm e}$ as well as $R_{\rm tran}$ and the surface 
brightness evaluated in $R_{\rm tran}$, extracted by the 2D technique for each galaxy in
the three bands: $B$ (blue filled dots), $V$ (green squares), and $R$ (red
asterisks), versus the 1D values reported in Paper~I. The diagonal line in
each plot represents a perfect match between the two sets of values. The main
issues are:\\ 
$\bullet$ \ The agreement between the \mbox{S\'ersic} parameters in
both studies is fairly good for five of the eight galaxies of the sample.
However, small differences are found for one galaxy (Mrk~86), while for the 
other two galaxies, Mrk~5 and I~Zw~123, the \mbox{S\'ersic} parameters between the 
two fitting techniques are clearly in disagreement, especially for the
$n$ shape parameter (see Fig.~\ref{F10}a). In Paper~I their host galaxies are
presented as possible spheroidal candidates, being fitted by high $n$ values,
$<n>_{\rm BRV} =2.66$ for Mrk~5 and $<n>_{\rm BV} =2.81$ for I~Zw~123,
while we obtained $<n>_{\rm BVR} =1.03$ for Mrk~5 and $<n>_{\rm BV}=1.89$ 
for I~Zw~123. \\ 
$\bullet$ \ For all the galaxies, we fitted the images successfully with 
smaller $R_{\rm tran}$ than in Paper~I (see Fig.~\ref{F10}f). 
We can fit the images not only with a higher number of pixels --- crucial to 
minimize the uncertainties --- but with a radial range more extended toward the 
centre, still avoiding starburst contamination. 
Here, we reproduce an example presented in Paper~I (see their Fig.~2): 
the stability plots in $n$, $R_{\rm e}$ 
and $m_{\rm tot}$ for Mrk~36 ($B$ band) are shown in Fig.~\ref{F11} (see caption for details).
From the comparison between both techniques, we identify some differences. 
Firstly, the radial range for which the parameters are quite stable is bigger in the 2D case. 
Secondly, the final fit isolates the starburst emission more accurately, thus including 
more high $S/N$ pixels. Finally, the uncertainties are quite smaller than the 1D case.

Furthermore, since it was described in \S5.1, $n$ could be poorly 
constrained when the fit is done in a restricted surface brightness range. 
We can define the surface brightness interval as the difference
between the surface brightness evaluated at $R_{\rm tran}$ ($\mu_{\rm tran}$) and 
the value at $S/N=1$ ($\mu_{\rm s/n=1}$).  
The surface brightness interval is most cases $>4$ mag, and $>3$ mag for the other cases. 
Notice that these values are lower limits since GALFIT fits those pixels with $S/N<1$.   
We list in Table~\ref{T2} both $\mu_{\rm tran}$ and $\mu_{\rm s/n=1}$.

By the 2D fitting technique we can model the starburst-free region more 
easily and with higher accuracy, and improve on the stability of the \mbox{S\'ersic} 
parameters, especially $n$, by reducing the size of the mask and thus providing 
a larger range of surface brightness available for fitting (see Fig~\ref{F10}d).

\subsubsection{Special cases}

A special analysis was devoted to Mrk~5 and I~Zw~123 (see Fig.~\ref{F9}),
whose 1D and 2D parameters show the largest difference (see Fig.~\ref{F10}). \\

\begin{figure*}
      \centering
      \includegraphics[width=15.0cm,angle=0]{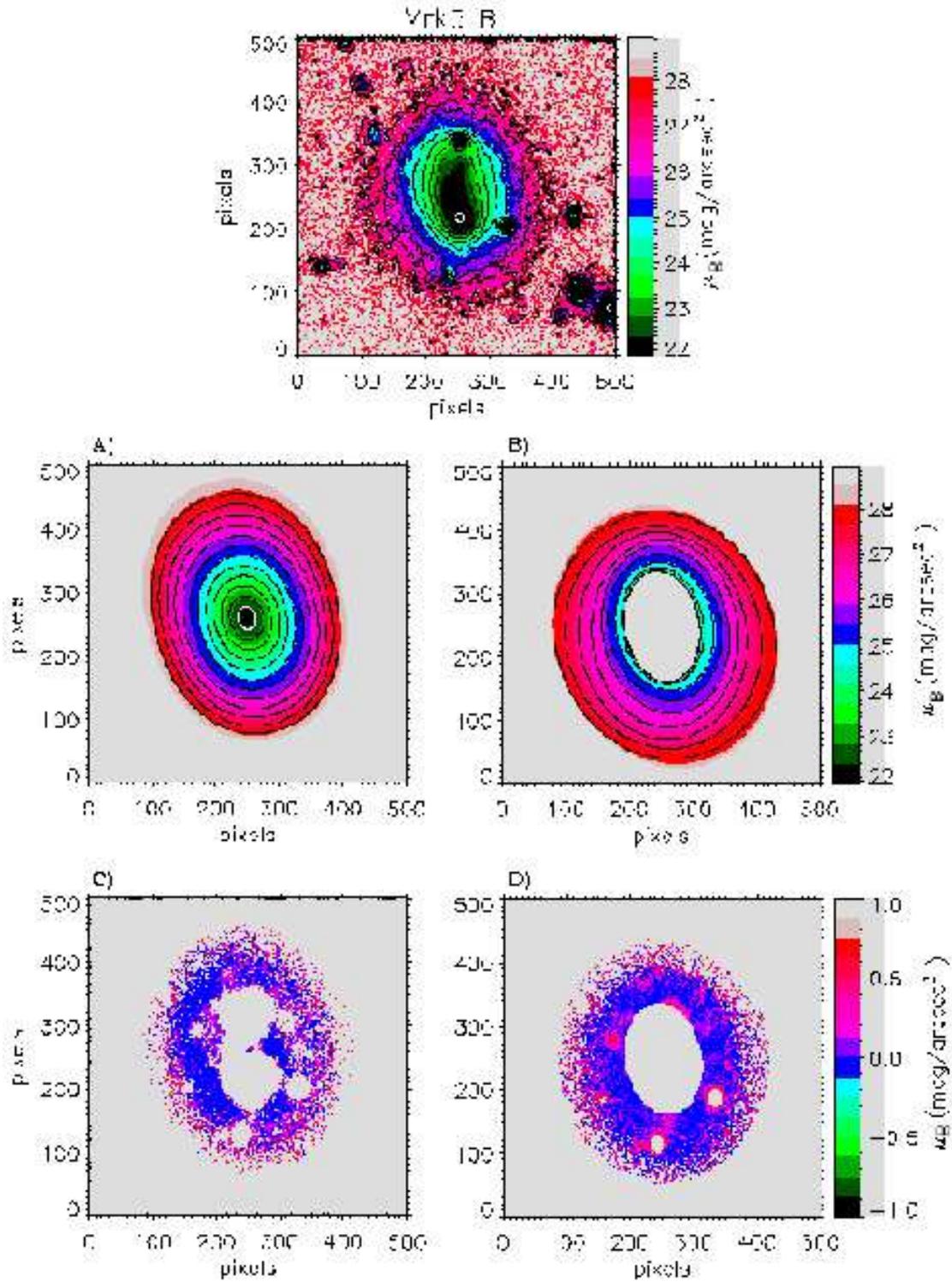}
      \caption{Comparison between the GALFIT 2D model (panels A) and C)) and the 2D model 
	generated by using the {\sl bmodel} task of IRAF from the 1D fit (Paper~I) 
	for Mrk~5 (panels B) and D)). In A) and B) we show the two models, while in C) and D) 
	we show the residuals down to the $1\sigma_{sky}$ level.}
      \label{F12}
    \end{figure*}

\begin{figure*}
     \begin{tabular}{l l l}
       \includegraphics[width=5.25cm,angle=90]{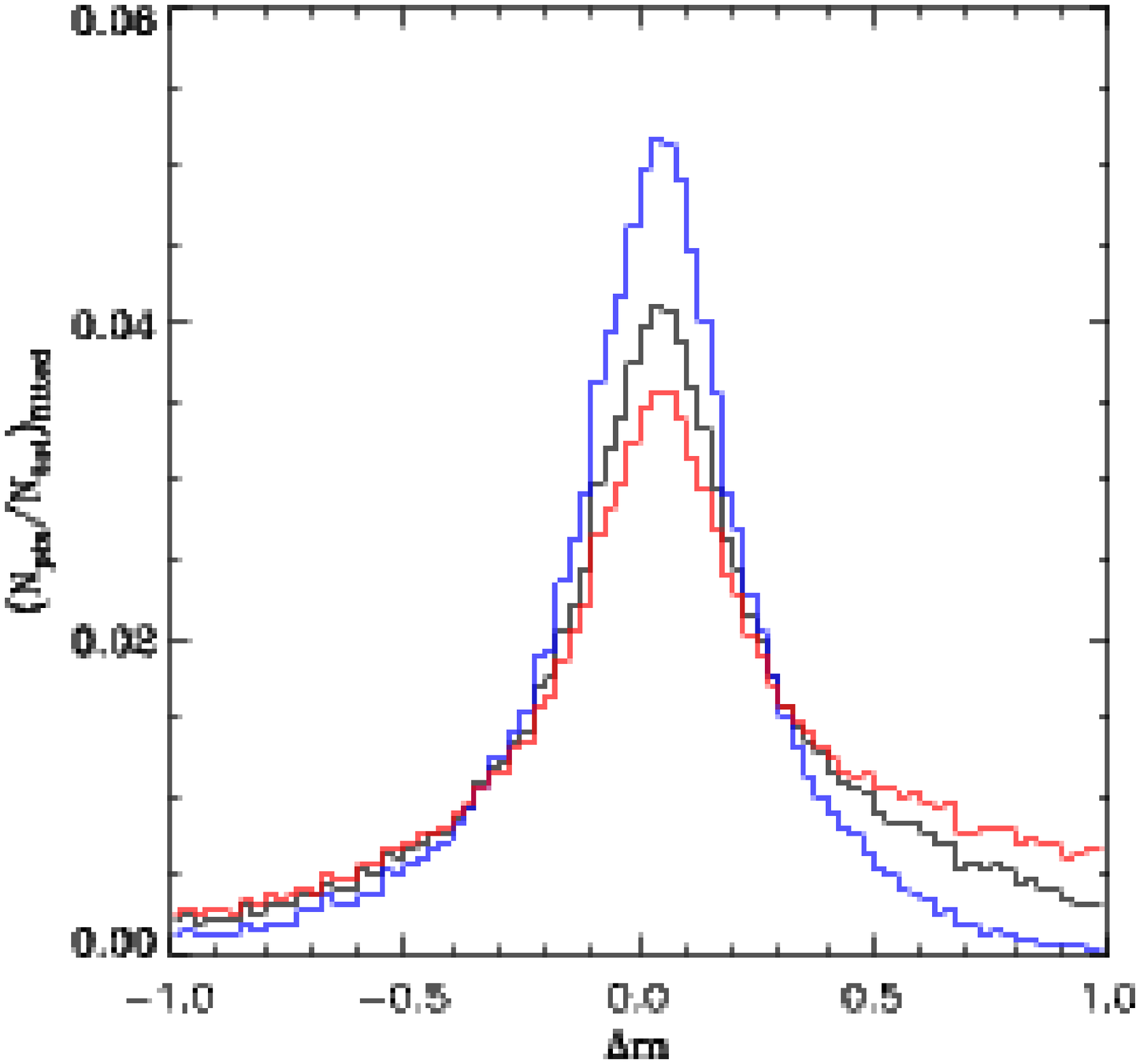}&
       \includegraphics[width=5.25cm,angle=90]{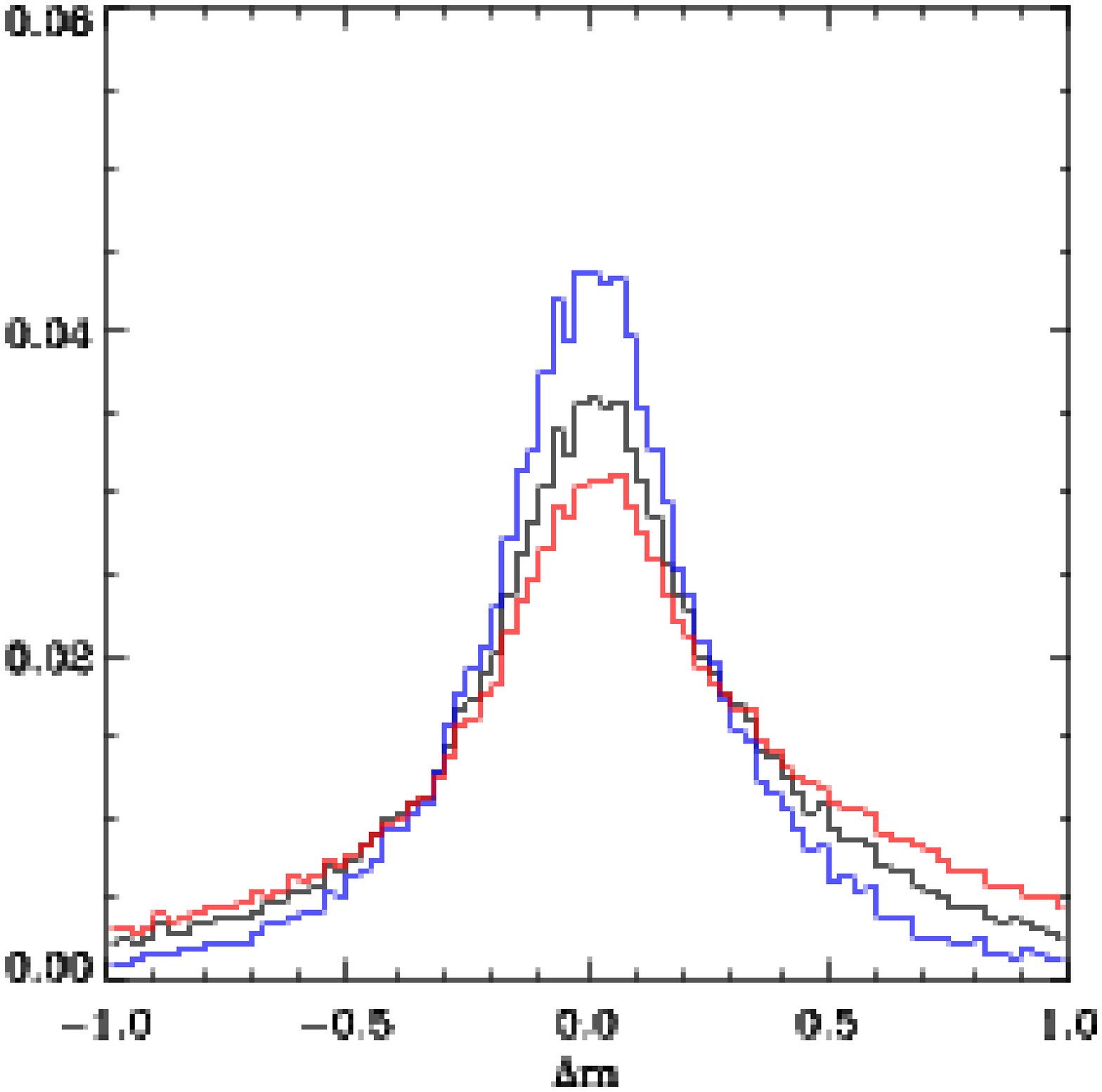}&
       \includegraphics[width=5.25cm,angle=90]{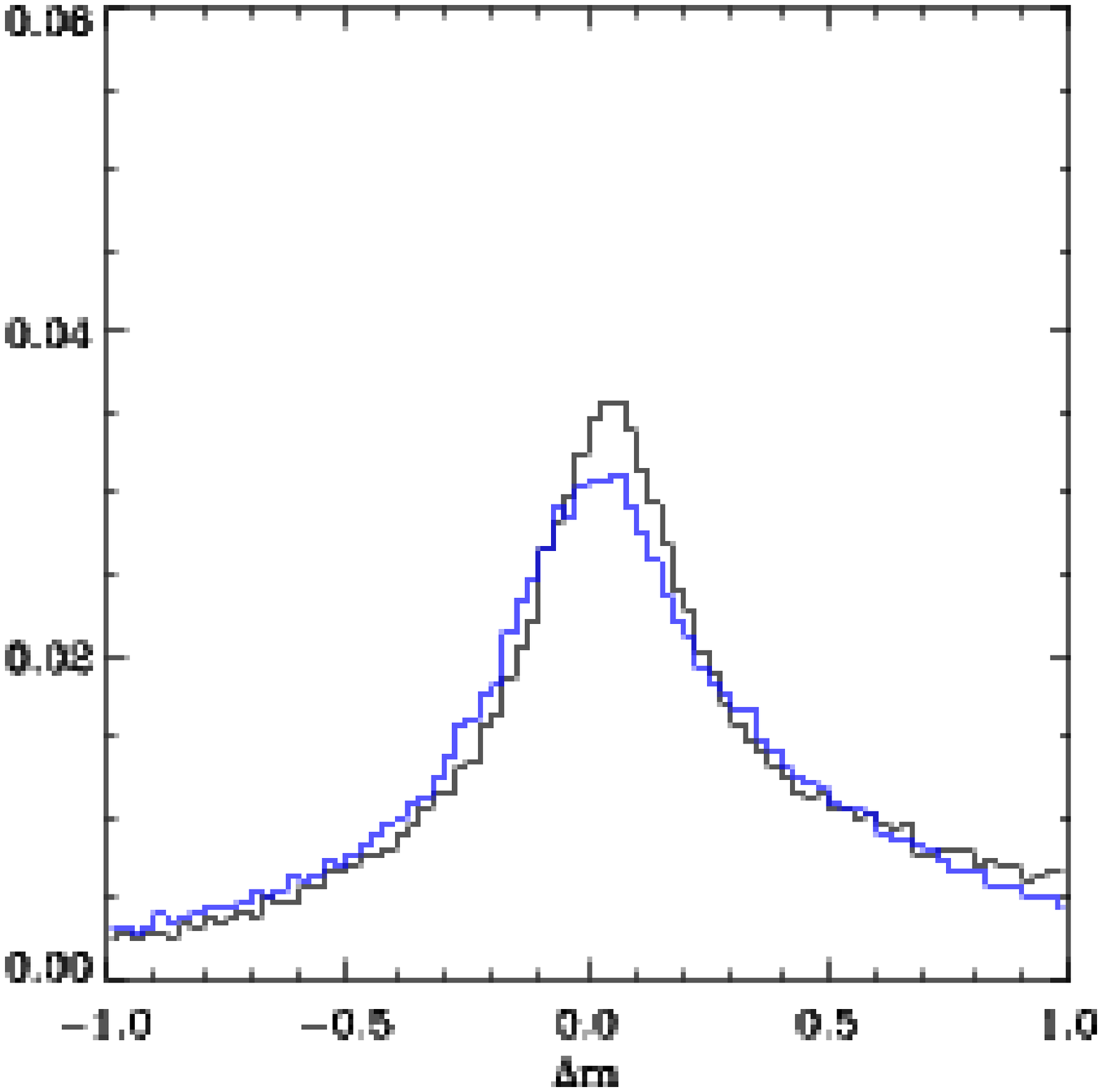}
     \end{tabular}
     \caption{Residual histograms for Mrk~5.
       Comparison between residuals provided from the 2D model and the 2D model 
       generated from the 1D fit. See text in Section~4.3.1 for details.}
     \label{F13}
   \end{figure*}

{\em 1) Mrk~5.$-$}\\
By using the {\sl bmodel} task in IRAF we performed 2D modelling of the radial 
\mbox{S\'ersic} profiles obtained in Paper~I, taking into account only those 
isophotes used in the fit. 
We have subtracted the resulting models from the galaxy image in
order to obtain the residuals and compare them with those given by GALFIT.
Fig.~\ref{F12} gives models and residual images in a surface
brightness colour scale. The regions shown in grey are pixels masked out from
the fit. Quantitatively, both residuals are quite small along the whole LSB
component. Residual histograms for both models are shown in Fig.~\ref{F13}.
Fig.~\ref{F13}(left) presents the GALFIT residual distribution while
Fig.~\ref{F13}(center) shows the residual distribution from the 1D fitting. We
normalize the number of pixels by the total number of pixels above
a given $S/N$ threshold. We present three levels ordered by 
increasing total pixel number: $S/N=2$ (blue), $S/N=1$ (black) and $R_{\rm
max}$ (with $S/N<1$, in red), where $R_{\rm max}$ is the outer limit for the
1D fitting (from Paper~I). In Fig.~\ref{F13}~(right) we show the comparison
between both distributions out to $R_{\rm max}$. 

The residuals show a Gaussian distribution centred on 0. As we can 
observe, by decreasing the $S/N$ threshold the Gaussian becomes broader
as we include more low signal-to-noise pixels. As we can see in 
Fig.~\ref{F12}c, no clear difference is observed between both techniques for 
the same isophote level. 
Also, as we showed in Section~5.2 (see also Table~\ref{T4}), the 
sky background uncertainties in the fit can be well determined. 
Stable solutions are found only among low $n$ values. This way, we are
unable to explain the differences on the \mbox{S\'ersic} parameters from the 1D and 
the 2D techniques.

On the other hand, the inward extrapolation of the best-fitting 1D \mbox{S\'ersic} 
model results in significantly higher luminosity than that observed for 
the radial profile in the central $\sim10$ arcsec (see Fig.~5 in Paper~I). 
Our best-fitting 2D \mbox{S\'ersic} model does not show this characteristic. 
However, if we fit these images with GALFIT, but using the main parameters derived 
from the 1D fit, i.e., fixing $m_{\rm tot}$, $R_{\rm e}$, and $n$, we notice the 
same behaviour again. 
This fit shows a large negative residual feature inside the entire masked 
region, with the LSB component brighter than the actual galaxy light 
distribution. \\

{\em 2) IZw~123. $-$}\\ 
We analysed the possible differences between both results by inspecting the
residuals and all possible uncertainties in the fit, and we found 
possible explanations in terms of both the seeing effects and the 
limited radial interval for fitting. 

The images of IZw~123 have a FWHM of about 2 arcsec. 
In Paper~I we fit the radial profile with 
$r_{\rm e}=2.07$, $n=2.61$, $m_{\rm tot}=15.19$ and $r_{\rm e}=2.74$, 
$n=3.01$, and $m_{\rm tot}=14.84$ in $B$ and $V$ bands respectively. 
In this work $r_{\rm e}$ is 4.59 and 4.11, $n$ 1.58 and 2.20, and 
$m_{\rm tot}$ 16.08 and 15.46, in $B$ and $V$ respectively. 
If we fit the same images in the same conditions but do not use the 
convolution in the models, $r_{\rm e}$ goes down to 2.89 and 2.51, $n$ 
increases to 2.40 and 3.15, while $m_{\rm tot}$ decreases to 15.60 and 14.86, 
in $B$ and $V$ respectively. 
These differences show that seeing convolution of the model is important 
when fitting small galaxies observed in poor seeing conditions. 

\begin{figure*}
     \centering
     \includegraphics[width=13.75cm,angle=90]{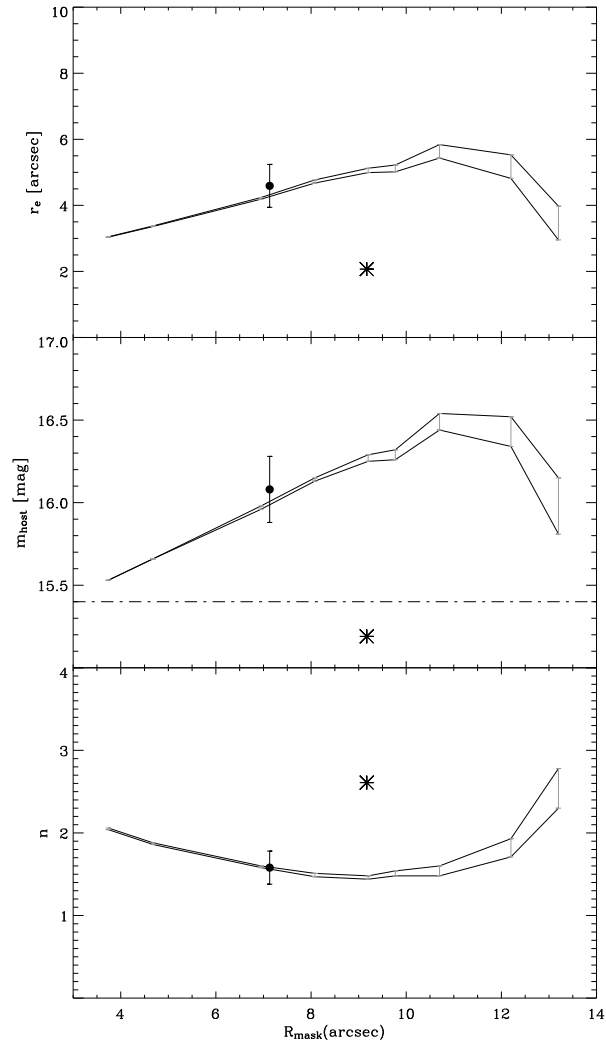}
     \caption{I~Zw~123 $B$-band: the free model parameters $R_{\rm e}$, $m_{\rm tot}$ 
       and $n$, are plotted as a function of $R_{\rm mask}$. Dots are the final best 2D fits. 
       Their $R_{\rm mask}$ value correspond to the radius beyond which starburst emission is 
       practically absent, $R_{\rm tran}$.  
       Results from Paper~I for each parameter are also indicated by asterisks. 
       The horizontal dashed-dotted line in the $m_{\rm tot}$ plot indicates the magnitude of 
       the whole galaxy (host + starburst).}
     \label{F14}
   \end{figure*}

On the other hand, I~Zw~123 has the largest $R_{\rm tran}/R_{\rm e}$ ratio, 
i.e., it has the starburst emission with the largest spatial extent, measured in 
terms of the host galaxy's structural parameters. If it also has a steep 
\mbox{S\'ersic} profile and taking into account the results shown in the 
simulations (see \S4), the resulting parameters could be very sensitive to 
exactly excluding the starburst emission and to the limited sampled area (see also \S5.1).
Figure~\ref{F14} presents the three model parameters,
$n$, \mbox{$r_{\rm e}$} and $m_{\rm tot}$, as a function of $R_{\rm mask}$ in the $B$ band. 
There is only a narrow radial interval, $7\lesssim R_{\rm mask}\lesssim11$, 
for which the parameters became reasonably stable, especially for $n$. Dots with error bars 
indicates the final best fits, where $R_{\rm mask}=R_{\rm tran}$, 
while asterisks indicates the results from Paper~I.
The horizontal dashed-dotted line in the $m_{\rm tot}$ plot indicates the 
magnitude of the whole galaxy (host + starburst); host luminosities brighter than that 
are physically meaningless.

Comparing the results from both techniques, we propose our best-fitting 2D 
\mbox{S\'ersic} model as the more physically reliable solution, for the outer 
regions of the LSB host.

\section{Summary and final conclusions}

We have presented a new two-dimensional fitting methodology devoted to the 
characterization of the LSB component in BCD galaxies.
The technique is based on the GALFIT algorithm developed by 
\cite{2002AJ....124..266P}, which permits to fit the whole galaxy image after 
suitably masking out the pixels contaminated by the starburst.

We described the different steps in our methodology. 
We carried out a set of fits using synthetic galaxies in order to validate 
the method and to determine its robustness and estimate the associated 
uncertainties. 
The method was applied to the eight BCD galaxies already analysed in Paper~I. 
We fit the same dataset, consisting of deep $BVR$ images, using a \mbox{S\'ersic} 
law. Finally, we analysed our results in terms of the possible error sources 
and we performed a set of consistency checks comparing the results with those 
of Paper~I. We paid special attention to Mrk~5 and I~Zw~123. 

The main results of this work can be summarized as follows:

\begin{enumerate}
\item Our simulations test the robustness of the methodology, and show that
the uncertainties in recovering stable \mbox{S\'ersic} parameters can be measured in
terms of the percentage of how many pixels are fitted relative to the total
number of pixels above a given signal-to-noise ratio.
Low $n$ synthetic galaxies are characterized by more stable \mbox{S\'ersic}
parameters and smaller uncertainties compared to high $n$ synthetic galaxies.
The shape parameter, $n$, is the most affected by the uncertainties. 
The 2D \mbox{S\'ersic} parameters of galaxies with $R_{\rm tran}/R_{\rm e}$ 
ratios between 0.5 and 2.5 are recovered within typical deviations of about 10\% 
and 20\% for low and high $n$ galaxies respectively, if we use --- at least --- the 50\%
of the galaxy pixels with $S/N>1$.
\item All the sampled galaxies show generally stable fits when using more 
than 74\% of the pixels with $S/N>$1.
\mbox{S\'ersic} indexes $n$ and effective radii for all the objects show good
agreement (within the uncertainties) in the three bands. 
The eight galaxies present a red LSB component and low $n$ values, suggesting 
that the LSB hosts of this sample of BCDs share similar structural 
properties. Seven of them have near exponential light distributions (very 
close to 1), while the other one has $n$ values close to 2. 
The uncertainties, estimated from the dispersion in the stability range and 
those from sky subtraction, are in all cases but one lower than 30\%. 
\item The mean advantage of the 2D technique over the 1D technique is that we can 
maximize the fitted portion of the galaxy. Especially in those galaxies with
irregular starburst we can accomplish this by using masks that follow
the actual shape of the starburst emission.
Furthermore, we have been able to put constraints to the sky background uncertainties, 
which play an important role in the \mbox{S\'ersic} parameter errors. 
Also, we have paid attention to the seeing effects when fitting a small galaxy.
\end{enumerate}
     
Because of the difficulties of the problem, it is indispensable to work with a
homogeneous data set, using the same reduction process and deriving the 
structural parameters with a well defined methodology, taking into account 
all the possible error sources. 
This two-dimensional technique can provide an important improvement 
in this kind of analysis.

\begin{acknowledgements}
Based on observations made at the Nordic Optical Telescope (NOT), 
in the Spanish Observatory del Roque de los Muchachos of the Instituto de 
Astrof\'{\i}sica de Canarias and on observations taken at the German/Spanish
 Calar Alto Observatory. This work has been partially funded by the Spanish
DGCyT, grant AYA2004-08260-C03-01. Thanks are given by R.O.A. to Adriana Caldiz 
and M. Dobrin\v{c}i\'{c} for revising the English. 
R.O.A. thanks R. S{\'a}nchez-Janssen for very stimulating discussions and helpful comments. 
We are sincerely grateful to  Dr. Richer for his very detailed comments, suggestions 
and time which helped us to improve the quality of the manuscript.    
\end{acknowledgements}

\clearpage
\begin{table*}
\caption{The galaxy sample}
\label{T1}
\centering
\footnotesize{
\begin{tabular}{l l c c c c c c }
\noalign{\smallskip}
\hline\hline
\noalign{\smallskip}
Galaxy & Other names & RA(J2000) & Dec.(J2000) & D (Mpc) & $A_{B}$ (mag)  & $m_{B}$ (mag) & $M_{B}$ (mag)\\
(1)     &    (2)        &     (3)     &    (4)      &   (5)   &     (6)       &     (7)    &   (8)  \\
\noalign{\smallskip}
\hline
\noalign{\smallskip}
Tololo~0127-397 &                       & 01 29 15.8 &$-39$ 30 37&    61.0 & 0.067  & 16.18 & $-17.75$\\[3pt]
Mrk~370         & NGC~1036, UGC~02160   & 02 40 29.0 & 19 17 50  &    10.9 & 0.399  & 13.19 & $-16.99$\\[3pt]
Mrk~5           & UGCA~130              & 06 42 15.5 & 75 37 33  &    14.0 & 0.364  & 15.22 & $-15.51$\\[3pt]
Mrk~86          & NGC~2537, UGC~04274   & 08 13 14.7 & 45 59 26  &     8.1 & 0.232  & 12.18 & $-17.37$\\[3pt]
Mrk~35          & Haro~3, NGC~3353      & 10 45 22.4 & 55 57 37  &    15.6 & 0.031  & 13.18 & $-17.79$\\[3pt]
Mrk~36          & Haro~4, UGCA~225      & 11 04 58.5 & 29 08 22  &    10.4 & 0.131  & 15.25 & $-14.84$\\[3pt]
I~Zw~123        & UGCA~410, Mrk~487     & 15 37 04.2 & 55 15 48  &    12.5 & 0.062  & 15.40 & $-15.09$\\[3pt]
Mrk~314         & NGC~7468, UGC~12329   & 23 02 59.2 & 16 36 19  &    28.9 & 0.383  & 13.78 & $-18.53$\\[3pt]
\noalign{\smallskip}
\hline
\hline
\end{tabular}
}
\begin{list}{}{}
\item Notes.$-$ Columns: 
(1) name of the galaxy; 
(2) other designations of the galaxies; 
(3) right ascension in hours, minutes and seconds; 
(4) declination in degrees, arcminutes and arcseconds; 
(5) distance, computed assuming a Hubble flow with a Hubble constant 
$H_{0}$ = 75 km s$^{-1}$ Mpc$^{-1}$, after correcting recession velocities 
relative to the centroid of the local group for Virgocentric infall; 
(6) extinction coefficient in the $B$ band, from \cite{schlegel98}; 
(7) asymptotic magnitude in the $B$ band, from C01b. Notice that the 
asymptotic magnitudes listed in C01b were corrected for Galactic extinction 
following \cite{BurstHeil84}; here they have been recomputed using the 
\cite{schlegel98} extinction values; 
(8) absolute magnitude, obtained from the $B$ asymptotic magnitudes using 
the distances tabulated in (5). 
\end{list}
\end{table*}
\begin{sidewaystable*}
\begin{minipage}[t][180mm]{\textwidth}
\caption{Parameters for the LSB host galaxy derived from the \mbox{S\'ersic} fit}
\label{T2}
\centering
\scriptsize{
\begin{tabular}{l l c c c c c c c c c c c c c c c c} 
\hline\hline             
\noalign{\smallskip}
Galaxy & Filter & $P_{\rm area}$(\%) & $PA$ & $q(=b/a)$ & $c$ & \mbox{$R_{\rm tran}$} & \mbox{$R_{\rm tran}^{\rm 1D}$}
& $n$ & $\sigma_{n}$ & \mbox{$r_{\rm e}$} & $\sigma_{r_{\rm e}}$ & m$_{\rm LSB}$ & $\sigma_{m_{\rm LSB}}$ & $M_{\rm LSB}$ & $\mu_{e}$& $\mu_{\rm tran}$& $\mu_{\rm s/n=1}$\\
\noalign{\smallskip}
   (1)  &   (2)  &   (3)  &   (4)   &   (5)   &   (6)  & (7) & (8) &  (9)  &  (10) & (11)  & (12)&  (13)& (14) & (15) & (16) & (17)& (18) \\
\noalign{\smallskip}
\hline
\noalign{\smallskip}
Tol~0127-397  & $B$ & 13.2 & $-$53.8 & 0.69 & 0.23&4.58 & 7.22 &1.11$\pm$0.19&0.21&  4.02$\pm$0.80&0.15&17.42$\pm$0.19&0.13& $-$16.51 & 22.78& 23.04& 27.06\\[3pt]
              & $V$ & 14.2 & $-$50.5 & 0.67 & 0.08&4.75 & 7.16 &0.94$\pm$0.15&0.20&  4.62$\pm$0.72&0.75&16.08$\pm$0.20&0.20& $-$17.85 & 21.64& 21.69& 25.95\\
              & $R$ & 14.7 & $-$50.8 & 0.66 & 0.08&5.27 & 7.20 &1.02$\pm$0.24&0.20&  4.75$\pm$0.24&0.20&15.96$\pm$0.21&0.19& $-$17.97 & 21.60& 21.80& 25.98\\\hline
Mrk~370       & $B$ & 15.0 & $-$14.8 & 0.79 & 0.09&19.20& 35.65&1.04$\pm$0.08&0.20& 23.36$\pm$1.00&1.41&13.88$\pm$0.13&0.12& $-$16.31 & 23.18& 22.85& 26.42\\
              & $V$ & 15.4 & $-$18.4 & 0.81 & 0.09&17.65& 35.62&1.00$\pm$0.10&0.22& 23.21$\pm$1.54&0.70&13.24$\pm$0.10&0.11& $-$16.95 & 22.53& 22.10& 25.63\\
              & $R$ & 10.4 & $-$14.0 & 0.80 & 0.08&16.80& 35.50&1.03$\pm$0.14&0.14& 23.89$\pm$1.33&1.18&12.71$\pm$0.10&0.10& $-$17.48 & 22.07& 21.52& 25.49\\\hline
Mrk~5         & $B$ & 10.9 &  13.1 & 0.76 & 0.10&10.97& 13.78&0.99$\pm$0.15&0.38&  9.62$\pm$1.25&0.61&16.02$\pm$0.20&0.14& $-$14.71 & 23.33& 23.58& 28.20\\
              & $V$ & 9.6  &  13.2 & 0.75 & 0.07&10.97& 13.67&1.05$\pm$0.14&0.30&  9.53$\pm$0.43&1.63&15.54$\pm$0.08&0.27& $-$15.19 & 22.84& 23.12& 27.56\\
              & $R$ & 9.0  &  12.5 & 0.78 &$-$0.01&10.61& 13.64&1.05$\pm$0.25&0.32&  9.53$\pm$0.28&0.16&15.21$\pm$0.08&0.06& $-$15.52 & 22.55& 22.76& 27.39\\\hline
Mrk~86        & $B$ & 18.9 & $-$21.2 & 0.91 & 0.00&34.80& 45.97&0.94$\pm$0.34&0.03& 25.65$\pm$5.00&0.21&13.10$\pm$0.27&0.02& $-$16.44 & 22.71& 23.35& 27.30\\
              & $V$ & 16.0 & $-$23.7 & 0.92 & 0.01&34.80& 45.52&1.00$\pm$0.32&0.05& 26.24$\pm$3.94&0.95&12.31$\pm$0.20&0.05& $-$17.23 & 22.01& 22.60& 26.44\\
              & $R$ & 14.6 & $-$21.3 & 0.90 & 0.00&34.80& 46.33&1.02$\pm$0.26&0.25& 25.49$\pm$6.92&5.36&11.73$\pm$0.30&0.21& $-$17.81 & 21.35& 22.01& 26.15\\\hline
Mrk~35        & $B$ & 20.2 &  74.0 & 0.72 &$-$0.02&20.80& 24.57&0.99$\pm$0.18&0.11& 15.30$\pm$1.58&0.07&14.15$\pm$0.18&0.05& $-$16.82 & 22.41& 23.06& 27.50\\
              & $V$ & 17.6 &  77.3 & 0.71 & 0.03  &19.45& 24.37&1.01$\pm$0.27&0.23& 14.83$\pm$1.43&1.12&13.41$\pm$0.15&0.23& $-$17.56 & 21.59& 22.16& 27.71\\
              & $R$ & 18.5 &  77.7 & 0.70 &$-$0.01&19.45& 24.08&0.97$\pm$0.16&0.08& 14.94$\pm$1.13&0.28&13.11$\pm$0.12&0.08& $-$17.86 & 21.28& 21.82& 27.88\\\hline
Mrk~36        & $B$ & 26.5 & $-$25.9 & 0.47 &$-$0.00&10.15 & 12.18&1.00$\pm$0.04&0.17&  8.79$\pm$0.80&0.18&16.92$\pm$0.15&0.04& $-$13.17 & 23.52& 23.80& 27.76\\
              & $V$ & 26.0 & $-$24.4 & 0.48 &$-$0.12&10.34 & 12.39&1.07$\pm$0.13&0.12&  8.64$\pm$0.65&0.40&16.34$\pm$0.25&0.05& $-$13.72 & 22.98& 23.34& 26.41\\
              & $R$ & 26.5 & $-$23.8 & 0.49 &$-$0.10&10.34 & 12.07&1.05$\pm$0.08&0.06&  8.81$\pm$0.72&0.03&15.95$\pm$0.10&0.04& $-$14.14 & 22.62& 22.93& 26.84\\\hline
I~Zw~123      & $B$ & 17.5 & $-$51.1 & 0.92 &$-$0.26&7.13 & 9.17 &1.58$\pm$0.20&0.18&  4.59$\pm$0.65&0.50&16.08$\pm$0.20&0.06& $-$14.40 & 22.21& 23.20& 27.59\\
              & $V$ & 20.0 & $-$52.9 & 0.89 &$-$0.20&7.13 & 9.07 &2.20$\pm$0.17&0.22&  4.11$\pm$0.37&0.15&15.46$\pm$0.14&0.04& $-$15.02 & 21.48& 22.74& 26.68\\\hline
Mrk~314       & $B$ & 13.5 &  14.8   & 0.77 &$-$0.46&10.80& 16.85&1.01$\pm$0.17&0.30&  9.54$\pm$0.79&0.70&14.77$\pm$0.11&0.15& $-$17.53 & 22.08& 22.32& 25.95\\
\noalign{\smallskip}
\hline
\hline
\end{tabular}
}
\begin{list}{}{}
\item Notes.$-$ Columns 
(1) name of the galaxy; 
(2) filter; 
(3) relative percentage area (see definition in 3.1.4); 
(4) position angle; 
(5) ellipticity; 
(6) boxy/discy shape parameter; 
(7) 2D transition radius (\arcsec); 
(8) 1D transition radius (\arcsec); 
(9) 2D fit \mbox{S\'ersic} shape parameter; 
(10) uncertainty in the shape parameter due to sky subtraction errors; 
(11) effective radius (\arcsec); 
(12) uncertainty in effective radius (\arcsec) due to sky subtraction errors; 
(13) total magnitude of the LSB host (mag); 
(14) uncertainty in total magnitude of the LSB host (mag) due to 
sky-subtraction errors; 
(15) absolute magnitude of the host derived from the 2D fit; 
(16) effective surface brightness (mag arcsec$^{-2}$) ;
(17) surface brightness of the 2D model at its transition radius (mag arcsec$^{-2}$); 
(18) surface brightness of the 2D model at its $S/N=1$ level (mag arcsec$^{-2}$). \\
Uncertainties in columns 10, 12 and 14, are estimated as described in 
Section~4.
\end{list}
\vfill
\end{minipage}
\end{sidewaystable*}
\begin{table*}
\centering
Sensitivity of \mbox{S\'ersic} parameters of Mrk~5~($B$) to changes in the sky 
background
\begin{tabular}{l l c c c c c c}
\noalign{\smallskip}
\hline\hline
\noalign{\smallskip}
Sky background & \mbox{$R_{\rm tran}$} & $P_{\rm area}$ (\%) & $n$ & \mbox{$r_{\rm e}$}  & $\mu_e$  & $m_{\rm tot}$  &  ${\chi^2}_{\nu}$ \\
\noalign{\smallskip}
\hline
\noalign{\smallskip}
 Correct sky &  {\bf 10.97}  &  {\bf 10.9}  & {\bf 0.99} & {\bf 9.62} & {\bf 23.63} &{\bf 16.02} & {\bf 2.232} \\[2pt]
 Sky over-subtracted &  10.97  & 10.9  & 0.88 & 9.23 & 23.54  & 16.08 & 3.093\\[2pt]
 Sky undersusbtracted&  10.97  & 10.9  & 1.17 & 10.47 & 23.83  & 15.97 & 3.013  \\[2pt]
\noalign{\smallskip}
\hline
\hline
\end{tabular}
\caption{Columns: 
(2) transition radius (arcsec); 
(3) masked area relative to $1\sigma_{sky}$ of the galaxy; 
(4) \mbox{S\'ersic} shape parameter; 
(5) effective radius (\arcsec); 
(6) effective surface brightness (mag arcsec$^{-2}$);
(7) total magnitude of the LSB component derived form the fit (mag); 
(8) $\chi^{2}_\nu$. 
In bold the best fitting parameters.}
\label{T4}
\end{table*}

\end{document}